\documentclass[fleqn,usenatbib]{mnras}

\usepackage{mathptmx}
\usepackage{txfonts}

\usepackage[T1]{fontenc}
\usepackage{ae,aecompl}

\usepackage{graphicx}	

\usepackage{float}

\newcommand{\beq}{\begin{equation}}
\newcommand{\eeq}{\end{equation}}
\newcommand{\beqa}{\begin{eqnarray}}
\newcommand{\eeqa}{\end{eqnarray}}

\newcommand{\NH}{N_{\rm{H}}}
\newcommand{\NHI}{N_{\rm{HI}}}
\newcommand{\NHM}{N_{\rm{H}_2}}

\newcommand{\NHIMW}{N_{\rm{HI,Gal}}}
\newcommand{\NHlos}{N_{\rm{H,los}}}

\newcommand{\SHI}{\Sigma_{\rm{HI}}}
\newcommand{\SHM}{\Sigma_{\rm{H}_2}}
\newcommand{\SHMS}{\Sigma^{*}_{\rm{H}_2}}
\newcommand{\SSFR}{\Sigma_{\rm{SFR}}}

\newcommand{\SGMC}{\Sigma_{\rm{H}_2,GMC}}
\newcommand{\fGMC}{f_{\rm{GMC}}}

\newcommand{\Tbb}{T_{\rm{bb}}}
\newcommand{\Tin}{T_{\rm{in}}}

\newcommand{\Cpl}{C_{\rm{PL}}}
\newcommand{\Cbb}{C_{\rm{BB}}}
\newcommand{\Cdiskbb}{C_{\rm{DISKBB}}}

\newcommand{\Mbh}{M_{\rm{BH}}}

\newcommand{\Funabs}{F_{\rm{unabs}}}

\newcommand{\Fobs}{F_{\rm{X,obs}}}
\newcommand{\Fobsmean}{\langle F_{\rm{X,obs}}\rangle}

\newcommand{\Lx}{L_{\rm{X}}}
\newcommand{\Ledd}{L_{\rm{E}}}

\newcommand{\Lunabs}{L_{\rm{X,unabs}}}
\newcommand{\Lunabsi}{L_{\rm{X,unabs},i}}
\newcommand{\Lobs}{L_{\rm{X,obs}}}
\newcommand{\Lobsi}{L_{\rm{X,obs},i}}
\newcommand{\LSunabs}{L_{\rm{SX,unabs}}}

\newcommand{\LSobs}{L_{\rm{SX,obs}}}

\newcommand{\Lmean}{\langle{L}_{\rm{X,unabs}}\rangle}

\newcommand{\Fsoft}{F_{0.25-2,{\rm unabs}}}
\newcommand{\Ftot}{F_{0.25-8,{\rm unabs}}}

\newcommand{\Spec}{{\rm Spec}}
\newcommand{\Speci}{{\rm Spec}_i}

\title[HMXB luminosity function]{Bright end of the luminosity
  function of high-mass X-ray binaries: contributions of hard, soft
  and supersoft sources}
 
\author[S. Sazonov and
  I. Khabibullin]{S. Sazonov$^{1,2}$\thanks{E-mail:
    sazonov@iki.rssi.ru} and I. Khabibullin$^{1,3}$\\
$^{1}$Space Research Institute, Russian Academy of Sciences,
  Profsoyuznaya 84/32, 117997 Moscow, Russia\\ 
$^{2}$Moscow Institute of Physics and Technology, Institutsky per. 9,
  141700 Dolgoprudny, Russia \\
$^{3}$Max Planck Institute for Astrophysics,
  Karl-Schwarzschild-Strasse 1, D-85741 Garching, Germany
}

\begin{document}
\label{firstpage}
\pagerange{\pageref{firstpage}--\pageref{lastpage}}
\maketitle

\begin{abstract}
  Using a spectral analysis of bright {\sl Chandra} X-ray sources
  located in 27 nearby galaxies and maps of star-formation rate (SFR)
  and ISM surface densities for these galaxies, we constructed the
  intrinsic X-ray luminosity function (XLF) of luminous high-mass
  X-ray binaries (HMXBs), taking into account absorption effects and
  the diversity of HMXB spectra. The XLF per unit SFR can be described
  by a power law $dN/d\log\Lunabs\approx 2.0(\Lunabs/10^{39}\,{\rm
    erg~s}^{-1})^{-0.6}$ $(M_\odot$~yr$^{-1})^{-1}$ from
  $\Lunabs=10^{38}$ to $10^{40.5}$~erg~s$^{-1}$, where $\Lunabs$ is
  the unabsorbed luminosity at 0.25--8~keV. The intrinsic number of
  luminous HMXBs per unit SFR is a factor of $\sim 2.3$ larger than
  the observed number reported before. The intrinsic XLF is composed
  of hard, soft and supersoft sources (defined here as those with the
  0.25--2~keV to 0.25--8~keV flux ratio of $<0.6$, 0.6--0.95 and
  $>0.95$, respectively) in $\sim $ 2:1:1 proportion. We also
  constructed the intrinsic HMXB XLF in the soft X-ray band
  (0.25--2~keV). Here, the numbers of hard, soft and supersoft
  sources prove to be nearly equal. The cumulative present-day
  0.25--2~keV emissivity of HMXBs with luminosities between $10^{38}$
  and $10^{40.5}$~erg~s$^{-1}$ is $\sim 5\times 10^{39}$~erg~s$^{-1}$
  $(M_\odot$~yr$^{-1})^{-1}$, which may be relevant for studying the
  X-ray preheating of the early Universe.
\end{abstract}


\section{Introduction}
\label{s:intro}

One of the indicators of recent star formation activity in galaxies is
the total number and cumulative luminosity of high-mass X-ray binaries
(HMXBs, e.g. \citealt{grietal03,lehetal10}). The X-ray luminosity 
function (XLF) of HMXBs in nearby galaxies can be described by a power
law $dN/d\Lx\propto\Lx^{-1.6}$ across more than four decades in
luminosity (at $\sim 0.5$--8~keV energies) from $\Lx\sim{\rm
  a~few}\times 10^{35}$ to $\sim 10^{40}$~erg~s$^{-1}$
\citep{minetal12}, with some indication of a cutoff at $\Lx\gtrsim
10^{40}$~erg~s$^{-1}$. This XLF thus smoothly connects low-luminosity
HMXBs with ultraluminous X-ray sources (ULXs, usually defined as
off-nuclear point-like objects with $\Lx\gtrsim
10^{39}$~erg~s$^{-1}$). It is possible that some ULXs, especially the
most luminous ones, contain a black hole of intermediate
($>100$~$M_\odot$) rather than stellar mass, but the compact object is
nevertheless expected to be accreting matter from a massive donor
star in a close orbit, hence such systems may still be called HMXBs
in a broad sense.

The $\sim 1.6$ slope of the HMXB XLF implies that more than 80\% of
the total X-ray emission from point sources in actively
star-forming galaxies is produced by HMXBs with $\Lx>
10^{38}$~erg~s$^{-1}$, with ULXs alone contributing more than
half. The X-ray output of the young stellar population is thus 
strongly dominated by neutron stars and black holes accreting at near-
and supercritical accretion rates. Such objects are very interesting
but rare, and their population properties can be disclosed by
accurately measuring the bright end of the HMXB XLF. 

Knowledge of the shape and normalisation (i.e. number of objects 
per unit star-formation rate, SFR) of the HMXB XLF, as well as its
possible dependence on other galactic properties such as metallicity
(e.g. \citealt{linetal10,douetal15}), is also important for
studying the 'cosmic dawn', when X-ray binaries belonging to the
first generations of (massive) stars and their remnants might have
(together with other mechanisms, such as supernova cosmic-ray heating,
\citealt{sazsun15}) significantly preheated the Universe
(e.g. \citealt{venetal01,madetal04,ricost04,miretal11}) before it was
reionised by UV radiation from subsequent generations of stars and
quasars. 

Previous statistical studies of HMXBs were usually aimed at
constructing an {\sl observed} XLF, and paid litte attention to survey
selection effects that may arise due to the diversity of HMXB X-ray
spectral shapes and the presence of X-ray absorbing gas along the line
of sight. This is a reasonable approach as long as one is dealing with
prevalent low- and medium-luminosity HMXBs with
$\Lx<10^{38}$~erg~s$^{-1}$, since most of such objects are X-ray
pulsars with characteristic hard X-ray spectra, with the bulk of the
luminosity emitted above 2~keV (see \citealt{waletal15} for a recent
review). However, the situation is different for luminous
($\Lx>10^{38}$~erg~s$^{-1}$) HMXBs, as explained below. 

First, the vast majority of such high-luminosity HMXBs are powered by
Roche lobe overflow rather than by wind, with a significant and
increasing with luminosity fraction of systems with a black hole as
the compact object. In contrast to neutron-star HMXBs, whose X-ray
luminosity is dominated by hard X-ray emission from magnetically and
radiatively supported accretion columns on the surface of the neutron
star \citep{bassun76}, black-hole X-ray binaries (both low- and
high-mass ones) are known to experience dramatic spectral state
transitions when the accretion rate is between $\sim 1$\% and $\sim
100$\% of the Eddington one and tend to be in a so-called high/soft
state (sometimes also in a 'very high' state) at subcritical accretion
rates (see \citealt{donetal07} for a review). 

In the soft state, most of the X-ray emission originates in a standard
thin accretion disk \citep{shasun73} and has a multicolour black-body
spectrum with characteristic temperature $\Tin\sim
(\Mbh/10M_\odot)^{-1/4}(\Lx/\Ledd)^{1/4}$ keV, where $\Mbh$ is the 
black hole mass and $\Ledd\approx 1.3\times
10^{39}(M/10\Mbh)$~erg~s$^{-1}$ is the corresponding Eddington
luminosity. A photon limited X-ray survey performed at 0.5--8~keV
energies might have a better chance to detect such a source
than one of the same luminosity but in a hard spectral state (due to
the larger number of photons per unit X-ray flux in the former case)
in the absence of absorption along the line of sight to the
source. However, propagation through the interstellar medium (ISM) in
the Galaxy and especially in the host galaxy may significantly
suppress the X-ray flux from a source in a soft state and reverse the
selection effect in favour of those in hard states. 

Secondly, most ULXs are probably stellar-mass black holes accreting
matter from a massive donor at a supercritical rate, although a
non-negligible fraction of such systems may contain a neutron star
rather than a black hole, as witnessed by the well-known pulsars
LMC~X-4 and SMC~X-1 with $\Lx\lesssim 10^{39}$~erg~s$^{-1}$ and the
recently discovered pulsars M82~X-2 \citep{bacetal14} and NGC 7793 P13
\citep{fueetal16,isretal16} with $\sim 10^{40}$~erg~s$^{-1}$. In this
case, accretion onto the black hole proceeds through a thick
disk. When viewed through the funnel of the disk with a strong wind
outflowing from it, such a system is perceived as a ULX and exhibits a
characteristic hard spectrum with a downturn above $\sim 5$--10 keV
(e.g. \citealt{sazetal14,waletal14,ranetal15,krisaz16}). However,
if the same system is observed from outside the funnel, through
the optically thick wind, only reprocessed emission from the central
regions can be visible, which, depending on the viewing direction and
accretion rate, may emerge i) in the soft X-ray band or ii) at even
longer wavelengths
(e.g. \citealt{pouetal07,midetal15,urqsor16,guetal16,fenetal16}). So-called
ultraluminous supersoft X-ray sources (ULSs) -- rare objects in nearby
galaxies with $\Lx\gtrsim 10^{39}$~erg~s$^{-1}$ and thermal spectra 
with colour temperature $\Tbb\lesssim 0.1$~keV \citep{diskon03} -- may
represent case i), while case ii) is probably realised in the Galactic
microquasar SS~433 (\citealt{fabetal15}, see also a relevant
discussion in \citealt{khasaz16}). The recent discovery of SS~433-like
baryonic jets in a ULS in the M81 galaxy \citep{liuetal15} provides
additional support to this picture.

There may also exist ULSs containing an IMBH accreting in subcritical
regime via a standard geometrically thin disk with low temperature
$\Tin\lesssim 0.3$~keV (since $\Tin\propto \Mbh^{-1/2}$ for a given
$\Lx$). 

Measured X-ray fluxes of ULSs are very sensitive to line-of-sight
absorption, and X-ray surveys may miss such objects despite their high
intrinsic luminosity, due to the presence of interstellar gas in their
host galaxies \citep{diskon03}. This raises a question: how many ULSs
are there in nature compared to ULXs? The answer could provide strong
constraints on models of supercritical accretion.

The purpose of this work is to construct an {\sl intrinsic}
(i.e. unabsorbed) XLF of HMXBs at $\Lx>10^{38}$~erg~s$^{-1}$ and study 
its composition in hard and soft X-ray sources. To
this end, we use a catalogue of X-ray sources detected by the {\sl
  Chandra X-ray observatory} in nearby (within 15~Mpc) galaxies. Since 
its launch in 1999, {\sl Chandra} has systematically observed many
local galaxies and detected thousands of X-ray sources in them. On
the other hand, recent UV, infrared, 21~cm and CO surveys have
provided high-quality maps of the SFR and atomic and molecular
ISM in the same galaxies. We combine this information with the {\sl
  Chandra} data to infer the shape and normalisation of the bright end
of the intrinsic XLF of HMXBs.  

Our study consists of three stages. First, we perform a spectral
analysis of bright {\sl Chandra} X-ray sources to infer their
intrinsic luminosities, $\Lunabs$, in the 0.25--8~keV energy band (and
the photoabsorption columns, $\NH$, affecting the measured
spectra). We then use the H, H$_2$ and SFR maps of the studied
galaxies to study how the total SFR 'probed' by {\sl Chandra} depends
on $\Lunabs$ and source spectral type. Our underlying
assumption is that the number of HMXBs follows star formation activity
in galaxies. At the final stage, we build an intrinsic XLF of HMXBs by
dividing the numbers of {\sl Chandra} sources of different spectral
types in specified $\Lunabs$ bins by the corresponding SFR ($\Lunabs$)
functions and adding together the contributions of hard and soft sources.

Making allowance for line-of-sight absorption effects is especially
important in the soft X-ray range. We thus also construct an intrinsic
HMXB XLF in the 0.25--2~keV band. We use this soft XLF elsewhere
\citep{sazkha17} for estimating the X-ray preheating of the early
Universe.

\section{Galaxy sample}
\label{s:galaxies}

This study is based on galaxies from The HI Nearby Galaxy Survey
(THINGS, \citealt{waletal08}), which is a high spatial resolution
($\sim 6''$) 21~cm survey of nearby galaxies performed by the Very Large
Array. The original THINGS catalogue comprises 34 galaxies 
located at distances $2<D<15$~Mpc and north of $\delta=-33^\circ$,
covering a wide range of Hubble types, star-formation rates, absolute
luminosities and metallicities.

{\sl Chandra} has observed all THINGS galaxies except for
NGC~2366. Apart from this object, we also excluded from consideration
i) NGC~3077, because of problems with determination of its
SFR\footnote{A bright star contaminates the {\sl GALEX} UV image.},
and ii) five dwarf galaxies, DDO~53, DDO~154, Ho~I, M81~DwA and
M81~DwB. The total SFR is likely less than 0.1~$M_\odot$~yr$^{-1}$ for
NGC~3077 and is below 0.01~$M_\odot$~yr$^{-1}$ for each of the
excluded dwarf galaxies \citep{leretal08,waletal08}, hence the
combined contribution of these galaxies to the HMXB population 
of the THINGS sample is expected to be negligible\footnote{We
  nevertheless searched, according to our criteria defined below, for
  bright {\sl Chandra} X-ray sources in these excluded galaxies and
  found none.}. 

\begin{table*}
  \caption{Galaxy sample
    \label{tab:galaxies}
  }
  \begin{tabular}{|l|r|r|r|r|l|c|r|r|}
    \hline
    \multicolumn{1}{|c|}{Galaxy} &
    \multicolumn{1}{c|}{$D$} &
    \multicolumn{1}{c|}{$i$} &
    \multicolumn{1}{c|}{P.A.} &
    \multicolumn{1}{c|}{$R_{25}$} &
    \multicolumn{1}{c|}{Type} &
    \multicolumn{1}{c|}{SFR} &
    \multicolumn{1}{c|}{log $M_\ast$} &
    \multicolumn{1}{c|}{$N_{\rm H,Gal}$} \\  
  & \multicolumn{1}{c}{Mpc} & \multicolumn{1}{c}{$^\circ$} &
    \multicolumn{1}{c}{$^\circ$} & \multicolumn{1}{c}{kpc} & & 
    \multicolumn{1}{c}{$M_\odot$~yr$^{-1}$} (Ref.) &
    \multicolumn{1}{c}{$M_\odot$} (Ref.) & $10^{20}$~cm$^{-2}$\\ 
    \hline
  HO II & 3.4 & 41 & 177 & 3.3 & Irr & 0.05 (1) & 8.3 (1) & 3.6\\
  IC 2574 & 4.0 & 53 & 56 & 7.5 & Irr & 0.07 (1) & 8.7 (1) & 2.2\\
  NGC 628 & 7.3 & 7 & 20 & 10.4 & Sc & 0.81 (1) & 10.1 (1) & 4.5\\
  NGC 925 & 9.2 & 66 & 287 & 14.3 & SBcd & 0.56 (1) & 9.9 (1) & 5.8\\
  NGC 1569 & 2.0 & 63 & 112 & 1.2 & Irr & 0.11 (2) & 9.1 (3) & 21.6\\
  NGC 2403 & 3.2 & 63 & 124 & 7.4 & SBc & 0.38 (1) & 9.7 (1) & 4.2\\
  NGC 2841 & 14.1 & 74 & 153 & 14.2 & Sb & 0.74 (1) & 10.8 (1) & 1.4\\
  NGC 2903 & 8.9 & 65 & 204 & 15.2 & SBb & 2.00 (2) & 10.5 (3) & 3.1\\
  NGC 2976 & 3.6 & 65 & 335 & 3.8 & Sc & 0.09 (1) & 9.1 (1) & 5.1\\
  NGC 3031 & 3.6 & 59 & 330 & 11.2 & Sab & 0.34 (2) & 10.6 (3) & 5.1\\
  NGC 3184 & 11.1 & 16 & 179 & 12.0 & SBc & 0.90 (1) & 10.3 (1) & 1.1\\
  NGC 3198 & 13.8 & 72 & 215 & 13.0 & SBc & 0.93 (1) & 10.1 (1) & 0.9\\
  NGC 3351 & 10.1 & 41 & 192 & 10.6 & SBb & 0.94 (1) & 10.4 (1) & 2.5\\
  NGC 3521 & 10.7 & 73 & 340 & 12.9 & SBb & 2.10 (1) & 10.7 (1) & 3.7\\
  NGC 3621 & 6.6 & 65 & 345 & 9.4 & SBcd & 0.39 (2) & 10.0 (3) & 6.8\\
  NGC 3627 & 9.3 & 62 & 173 & 13.8 & SBb & 2.22 (1) & 10.6 (1) & 2.1\\
  NGC 4214 & 2.9 & 44 & 65 & 2.9 & Irr & 0.11 (1) & 8.8 (1) & 1.9\\
  NGC 4449 & 4.2 & 60 & 230 & 2.9 & Irr & 0.37 (1) & 9.3 (1) & 1.4\\
  NGC 4736 & 4.7 & 41 & 296 & 5.3 & Sab & 0.48 (1) & 10.3 (1) & 1.3\\
  NGC 4826 & 7.5 & 65 & 121 & 11.4 & Sab & 0.51 (2) & 10.2 (3) & 2.9\\
  NGC 5055 & 10.1 & 59 & 102 & 17.3 & Sbc & 2.12 (1) & 10.8 (1) & 1.3\\
  NGC 5194 & 8.0 & 42 & 172 & 9.0 & Sbc & 3.13 (1) & 10.6 (1) & 1.8\\
  NGC 5236 & 4.5 & 24 & 225 & 10.1 & SBc & 3.06 (2) & 10.6 (3) & 4.0\\
  NGC 5457 & 7.4 & 18 & 39 & 25.8 & SBc & 3.34 (2) & 10.5 (3) & 1.5\\
  NGC 6946 & 5.9 & 33 & 243 & 9.9 & SBc & 3.24 (1) & 10.5 (1) & 18.9\\
  NGC 7331 & 14.7 & 76 & 168 & 19.5 & Sbc & 2.99 (1) & 10.9 (1) & 6.3\\
  NGC 7793 & 3.9 & 50 & 290 & 5.9 & Scd & 0.24 (1) & 9.5 (1) & 1.2\\
  \hline
\end{tabular}

\textbf{References:} (1) \cite{leretal08}; (2) this work, within
$R_{25}$; (3) derived using $L_K$ from \cite{karetal13}.

\end{table*}

We have chosen the THINGS galaxy sample for this study not only
because of the availability of homogeneous HI data but also because
there exist similarly high-quality and sufficiently extended
H$_2$ and SFR maps for these galaxies. The former are provided by the
Heterodyne Receiver Array CO Line Extragalactic Survey (HERACLES,
\citealt{leretal09}), an atlas of CO $J=2\to 1$ emission with $11''$
angular resolution obtained with the IRAM 30-m telescope, while the
latter can be readily constructed from {\sl Galaxy Evolution Explorer}
({\sl GALEX}) far-UV (1350--1750~\AA) maps and {\sl Spitzer}/MIPS
24~$\mu$m maps (mostly from the {\sl Spitzer} Infrared 
Nearby Galaxy Survey, SINGS, \citealt{kenetal03}). The angular
resolution (FWHM) of these UV and IR maps is $4''$ and $6''$, respectively
(see \citealt{bigetal08}). {\sl GALEX} and {\sl Spitzer} data are  
available for all of our galaxies, whereas HERACLES data exist 
for all but 7~objects: NGC~1569, NGC~3031, NGC~3621, NGC~4449,
NGC~4826, NGC~5236 and NGC~7793. As a result, we use HERACLES H$_2$
maps for most of our galaxies and construct synthetic H$_2$ maps from
{\sl GALEX}/{\sl Spitzer} SFR maps for the 7 galaxies listed above,
using a well-established correlation between SFR and H$_2$
surface density (see Section~\ref{s:galaxy_profiles} below). 

Our sample thus consists of 27 nearby galaxies. Table~\ref{tab:galaxies}
provides key information on these objects. Distances ($D$),
inclinations ($i$), position angles (P.A.) and optical sizes ($B$-band
isophotal radii at 25~mag~arcsec$^{-2}$, $R_{25}$) are adopted from
the THINGS description paper \citep{waletal08}. Galaxy types are taken
from the HyperLeda
database\footnote{http://leda.univ-lyon1.fr/}. Estimates of the total 
SFR and stellar masses ($M_\ast$) are adopted from \citealt{leretal08},
except for 7 galaxies absent from their sample. For these, we have
added to Table~\ref{tab:galaxies} our own estimates of the SFR
(obtained, similar to \citealt{leretal08}, using {\sl GALEX} and {\sl
  Spitzer} maps, see Section~\ref{s:galaxy_profiles}) and
$M_\ast$. The latter were calculated from the $K$-band luminosities
given in the Updated Nearby Galaxy Catalogue \citep{karetal13}, assuming
a mass-to-light ratio $M_\ast/L_K=0.5M_\odot/L_{\odot,K}$, the same as used by
\citet{leretal08} for the other galaxies. We need total stellar masses
to estimate (in Section~\ref{s:lmxb}) the contribution of
low-mass X-ray binaries (LMXBs) to the derived XLFs. Finally,
Table~\ref{tab:galaxies} includes information on the Galactic HI
column density in the direction of the galaxies \citep{kaletal05}.  

The sample consists mainly of spiral galaxies, although it also
includes 5 smaller, irregular galaxies. The total SFRs range
from $\sim 0.05$~$M_\odot$~yr$^{-1}$ (Ho~II) to $\sim 3$~$M_\odot$~yr$^{-1}$
(NGC~5194, NGC~5236, NGC~5457, NGC~6946 and NGC~7331). The combined
SFR of all 27 galaxies is $\sim 32$~$M_\odot$~yr$^{-1}$.

\section{X-ray sources}
\label{s:raw_sample}

We used the recently published catalogue of {\sl Chandra}/ACIS
X-ray point sources by \citet{wanetal16}. This is the largest publicly
available catalogue of {\sl Chandra} sources, based on the
{\sl Chandra} data archive covering $\sim 14$ years of
observations. Importantly, the sources in this catalogue had been
selected based on their detection in the 0.3--8~keV energy band, which
reduces bias against soft X-ray sources compared to catalogues based
on source detection in harder X-ray bands. 

Most of the sources in the \cite{wanetal16} catalogue are weak
detections, with just $\sim 10$ counts in an individual ACIS
observation. Since our goal is to estimate unabsorbed X-ray
luminosities of sources, which requires measurement of the
line-of-sight absorption column from their spectra, we decided to use
only sources with at least 100 counts in some observation. A number of
galaxies in our sample (Table~\ref{tab:galaxies}) have been observed
more than once by {\sl Chandra}, and the \cite{wanetal16} catalogue
provides information on individual observations for a given source. In
what follows, we use the ACIS observation with the highest counts for
a given source. However, in Section~\ref{s:variable} we estimate the
influence of this selection procedure on our results using information
from all existing {\sl Chandra} observations for our sources.

We cross-correlated the \cite{wanetal16} catalogue with the positions of
the galaxies in our sample and selected X-ray sources located within
the 25~mag~arcsec$^{-2}$ isophotes of the galaxies, i.e. at radii
$R<R_{25}$ in the plane of the galaxy\footnote{Note that M51b
  (NGC~5195) and Holmberg~IX, the well-known satellites of NGC~5194
  (the Whirlpool Galaxy) and NGC~3031 (M81), respectively, do not fall
  into their $R_{25}$ regions and hence are not studied here.}, but
outside $0.05 R_{25}$. With this choice of the outer boundary we wish
to include most of the X-ray sources actually belonging to the galaxy
(since usually more than 90\% of the total SFR is contained within
$R_{25}$) and minimise contamination by background active galactic
nuclei (AGN). The $R>0.05R_{25}$ condition is introduced to avoid  
complications related to possible supermassive black hole activity in
the nuclei of the studied galaxies and because galactic central
regions are often characterized by highest ISM column densities, which
is difficult to properly take into account in our analysis. Also,
although source confusion plays a minor role in our study (see below)
due to the excellent angular resolution of {\sl Chandra}, it may
become problematic in the nuclear regions of the galaxies.

We additionally filtered out sources that are reported by
\cite{wanetal16} to be partially confused with another source in
a particular {\sl Chandra} observation. Sometimes, the same source may
be reported to be unconfused in another observation, but we
nevertheless excluded all such sources from consideration to have 
better confidence in the results of our X-ray spectral analysis. We
also excluded several sources for which the total counts in an X-ray
spectrum that we extracted (see below) differed by more than a factor
of 2 from the value reported by \cite{wanetal16} for the same {\sl
  Chandra} observation. In all such cases, there proves to be another
X-ray source in the vicinity, so this discrepancy is yet 
another manifestation of source confusion. In total, 34 sources ($\sim
10$\% of all sources) with 100 or more counts have been filtered out. 

The remaining sample consists of 330 sources with at least 100 counts,
located between $0.05R_{25}$ and $R_{25}$ in the 27 studied galaxies. 

\subsection{Spectral modeling}
\label{s:spectra}

We used all available archival {\sl Chandra}/ACIS data to extract
X-ray spectra for the selected sources. Data search and reduction was
performed by means of standard tools from the {\sc ciao} (v.~4.8)
package. Namely, we took advantage of the {\it specextract} thread
with source and background extraction regions defined as follows: the
source region is the $6''$-radius circle around the source, while the
background region is the $18''$-radius circle centred on the source
from which all source regions (for the source under consideration and
any other sources) are excised. The technique works well sufficiently
far from the crowded central regions of the investigated galaxies,
which we have already excluded by imposing the condition
$R>0.05R_{25}\sim 10''$--$30''$. We verified (using {\sc ciao} {\it
  psfsize\_srcs} tool) that the $6''$-radius extraction region
encompasses at least 2/3 of the source photons even for the most
off-axis ($\lesssim 9$ arcmin) cases in our sample and more than 90\%
of the source photons for 97\% of a subsample that is used for
construction of XLFs below.

Using the extracted spectra along with the corresponding response and
aperture correction functions, we performed spectral analysis and
measured source fluxes in the 0.25--8~keV energy band, which is only
slightly different from the 0.3--8~keV band used by \cite{wanetal16}
for source detection. We extended the energy range down to 0.25 keV to
minimise loss of information for sources with the softest spectra. 

The purpose of our spectral analysis was to i) characterize the
spectral hardness/softness of sources, and ii) to determine their
intrinsic (i.e. corrected for line-of-sight absorption) X-ray fluxes
and luminosities. These tasks are not independent and in fact can only
be solved together, since we do not know the positions of our sources
with respect to the ISM in their host galaxies along the line of
sight. Therefore, we need to determine both the intrinsic spectral
shape and the line-of-sight column density, $\NH$, for a given source
from its X-ray spectrum. This is a notoriously difficult task,
especially for very soft sources (see a relevant discussion in
\citealt{devetal07}), even despite the fact that we pre-selected
sources with at least 100 counts.

After a number of trials we decided to take the following approach. We
fitted the spectrum of each source, using {\sc xspec} \citep{arnaud96},
by three alternative  models: 1) absorbed power law (hereafter, PL),
2) absorbed black-body spectrum (hereafter, BB) and 3) absorbed
multicolour disk emission (hereafter, DISKBB) -- {\it wabs(powerlaw)},
{\it wabs(bbody)} and {\it wabs(diskbb)} in {\sc xspec}
terminology. These models are similar in the sense that each has just
3 parameters: the absorption column density, $\NH$, a parameter
describing the intrinsic spectral shape (slope $\Gamma$ for PL,
temperature $\Tbb$ for BB and the inner disk temperature $\Tin$ for
DISKBB) and the normalisation (we actually took advantage of the {\sc
  xspec} convolution model {\it cflux} which calculates the unabsorbed
flux in a given energy band, 0.25--8~keV in our case). We allowed
$\Gamma$ to range between 0 and 10, $\Tbb$ and $\Tin$ from 0.01 to
10~keV, and $\NH$ from $10^{20}$ to $10^{24}$~cm$^{-2}$. 

As regards the absorption model, we also tried to use {\it tbabs}
\citep{wiletal00} instead of {\it wabs} \citep{mormcc83}, but found
the results to be relatively insensitive to this (using {\it tbabs} in
conjunction with the abundance set of \citealt{wiletal00} results in
$\sim 20$\% higher $\NH$ estimates), while {\it wabs} provides a 
strong advantage in terms of computational time. We also note that
differences in $\NH$ of the same order may arise from variations in
the ISM metallicity between different galaxies or within a given
galaxy, none of which is taken into account in our analysis.

We used $C$-statistics ({\it cstat} in {\sc xspec}, \citealt{cash79})
for comparing the results of spectral fitting by the different
models. For very soft sources, we applied an additional criterion (see
below) to finally select the best model. 

Our choice of spectral models is not arbitrary. Both DISKBB and PL
models are commonly used for spectral analysis of X-ray binaries in
their different spectral states, whereas BB often adequately describes
the spectra of supersoft sources such as ULSs. Theoretically, DISKBB
spectra can be produced by standard accretion disks, PL spectra can
naturally arise in coronae of accretion disks and jets, and BB spectra
can emerge from supercritical accretion disks with winds. Furthemore,
PL should fit well the spectra of AGN, which may contaminate our
sample of sources. 

However, our single-component modeling is certainly oversimplified for
some spectra, especially for brighter sources with $\gtrsim 1000$
counts. We nevertheless consider this approach adequate
for the purposes of this study, first because it is practically
impossible to reliably deduce more than 3 spectral parameters for the
vast majority of our sources, and also because even for the spectra of
some brighter sources that seem to require fitting by a multicomponent
model (e.g. a combination of PL and BB) it usually turns out (as
demonstrated below) that one of these components clearly dominates and
its parameters can be inferred with reasonable accuracy by fitting the
corresponding single model to the data.

\subsection{Comparison of different spectral models}
\label{s:spectral_models}

\begin{figure}
\centering
\includegraphics[width=\columnwidth,viewport=30 200 560 710]{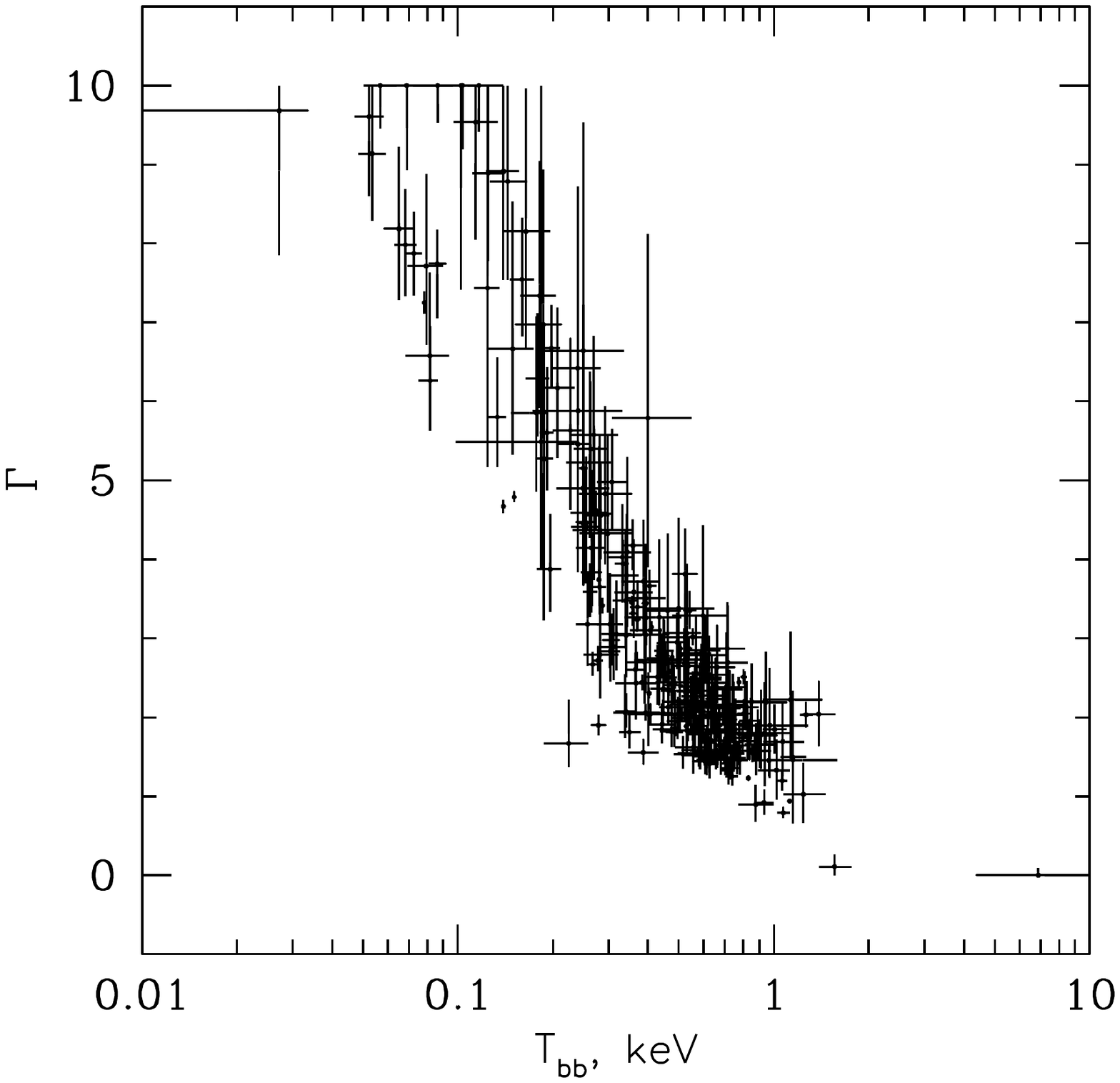}
\includegraphics[width=\columnwidth,viewport=30 200 560 710]{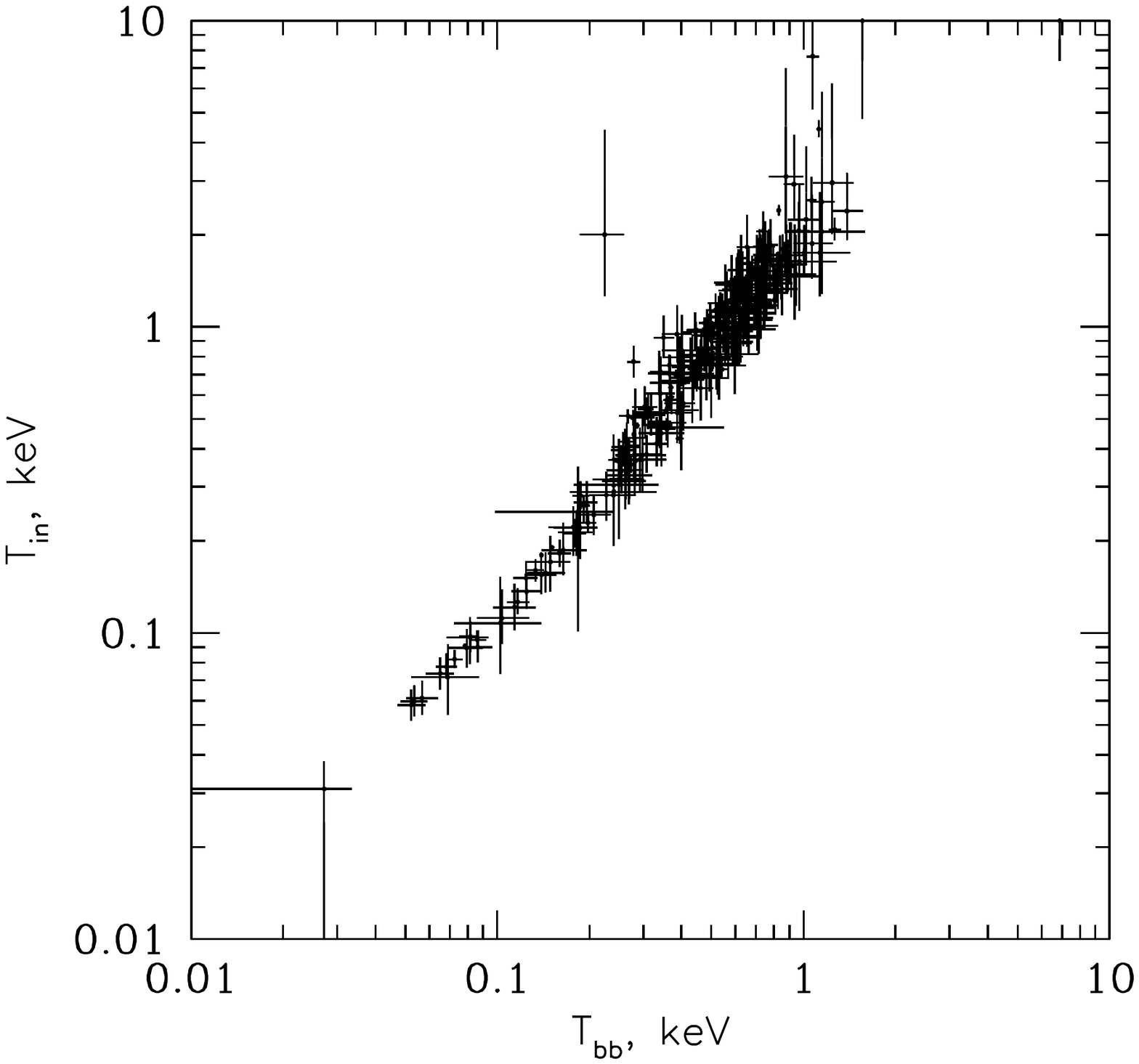}
\caption{{\sl Top panel:} Best-fit PL slope vs. best-fit BB
  temperature for 330 {\sl Chandra} sources with at least 100 counts
  in the spectrum. {\sl Bottom panel:} Best-fit DISKBB temperature
  vs. best-fit BB temperature for the same sources.
}
\label{fig:gamma_tbb_tin}
\end{figure}

\begin{figure}
\centering
\includegraphics[width=\columnwidth,viewport=30 200 560 710]{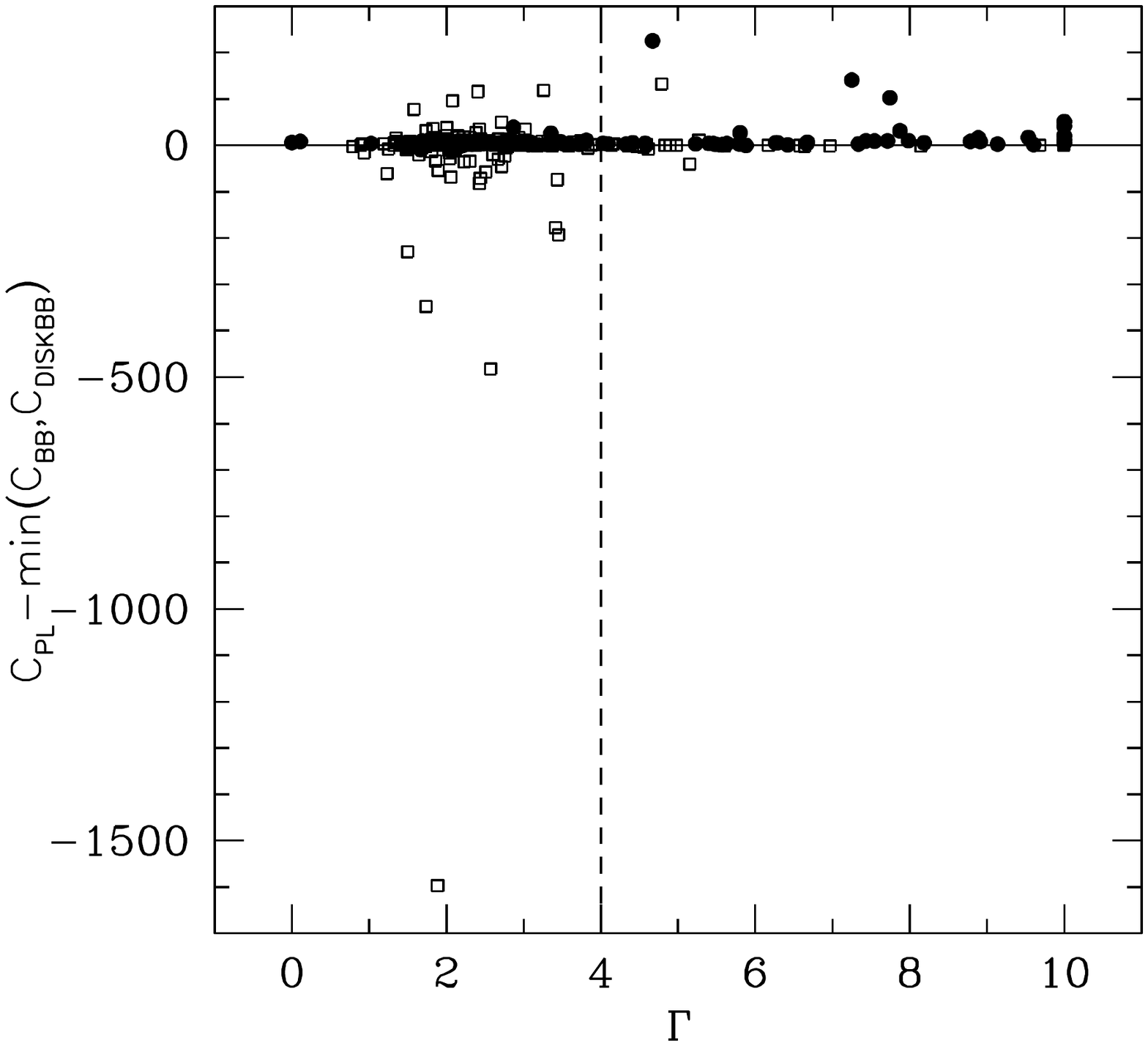}
\includegraphics[width=\columnwidth,viewport=30 200 560 710]{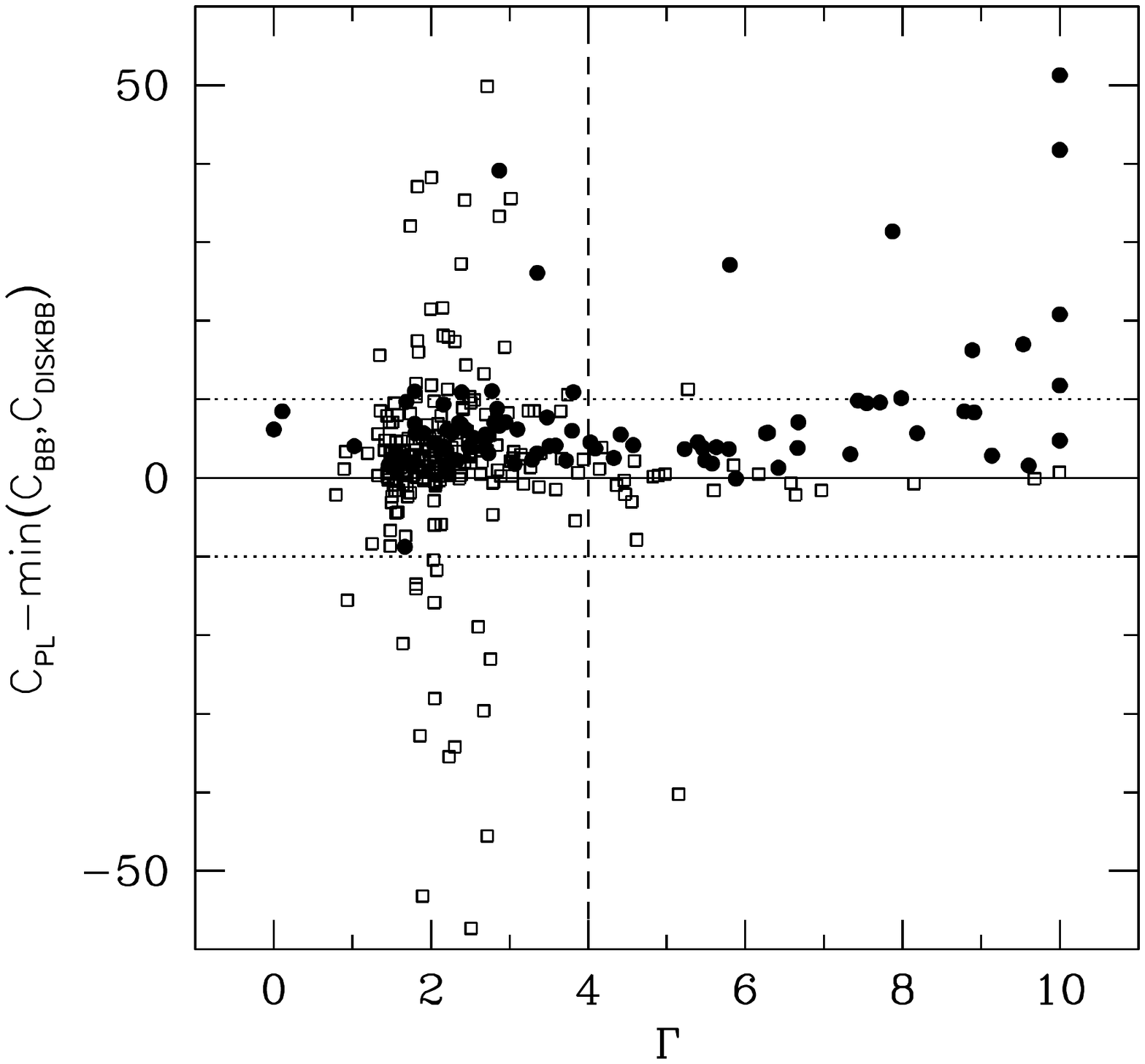}
\caption{{\sl Top panel:} Comparison of the fit quality
  ($C$-statistics) for the PL, BB and DISKBB models, for the same
  sources as in Fig.~\ref{fig:gamma_tbb_tin}. Filled circles and open
  squares denote sources whose spectra are better fit by BB than by
  DISKBB and vice versa, respectively. {\sl Bottom panel:} Zoom-in on
  the region where most sources are located. The vertical dashed line
  in both panels indicates the boundary of the $\Gamma>4$ region where
  we eventually select the best spectral model between BB and DISKBB
  only. The two horizontal dotted lines in the bottom panel indicate
  the  difference $\Delta C=\pm 10$, which may be regarded as
  statistically significant for favouring one of the models over the
  others.
}
\label{fig:cstat}
\end{figure}

As expected, there is strong correlation between the best-fit values
of $\Gamma$, $\Tbb$ and $\Tin$ for our sources, as shown in
Fig.~\ref{fig:gamma_tbb_tin}. The crucial question is which of the
considered spectral models should we prefer for a given source? 
To address this issue, we show in Fig.~\ref{fig:cstat} the difference
in the {\it cstat} values, $C$, for the PL, BB and DISKBB models as a
function of the best-fitting power-law index. 

For $\sim 80$\% of
the softest spectra with $\Gamma> 4$, PL provides a worse 
fit compared to one of the thermal models, with 39 and 11 out of 63
spectra being best fit by BB and DISKBB, respectively. This result
is anticipated, since such steep power-law spectra are observed
very rarely from astrophysical X-ray sources, with both synchrotron and
inverse Compton mechanisms usually giving rise to flatter
spectra. 

\begin{figure}
\centering
\includegraphics[width=\columnwidth,viewport=30 160 560 720]{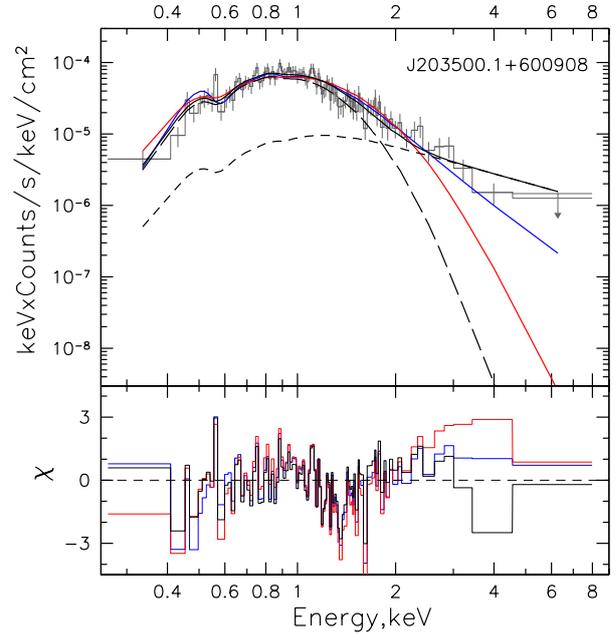}
\caption{{\sl Chandra} spectrum for a soft (with an effective power-law
  slope $\Gamma>4$) source for which a two-component DISKBB+PL model
  (solid black curve, with the long-dashed and short-dashed curves
  showing the DISKBB and PL components, respectively) provides a
  significant improvement in fit quality compared to a single component
  DISKBB model (red curve). Nearly the same fit quality can be
  achieved by an unphysically steep ($\Gamma=5.2$) PL model (blue
  curve), at the expense of much higher absorption column
  density and intrinsic luminosity. The data 
  points (rebinned for visualisation purposes and shown in grey) are
  the number of source counts per bin multiplied by the bin's central
  energy and divided by the product of the exposure time, the bin's width and
  effective area at its central energy. The error bars
  correspond to the 1$\sigma$ uncertainty. The lower panel shows
  residuals of the same models in units of 1$\sigma$ uncertainties 
  with the same colour-coding as in the panel above. 
  }
\label{fig:twocomp_spectrum}
\end{figure}

Nevertheless, 13 of the 63 sources with $\Gamma>4$ are better
described by PL than by BB or DISKBB. However, the improvement in the
fit has low statistical significance ($\Cpl-\min(\Cbb,\Cdiskbb)<10$),
according to standard criteria such as the Akaike information
criterion (AIC)\footnote{In our case, ${\rm AIC}=2k+C$, where $k=3$ 
  for all three considered spectral models, so a difference $\Delta
  C=10$ between two models implies that one of them is
  $\exp(-10/2)\sim 1$\% as probable as the other.}, except for one
source -- CXOGSG~J203500.1+600908\footnote{Hereafter, we use source
  names from \cite{wanetal16}.}. In this case, a steep PL model (with
$\Gamma=5.2$) provides a clear improvement: $\Cpl-\Cdiskbb=40.3$. We
explored the spectrum of this source (see
Fig.~\ref{fig:twocomp_spectrum}) and found that the addition of a
relatively hard ($\Gamma=2.5$) PL component to the supposedly dominant
soft thermal (DISKBB) one leads to a fit of similar to the
single-component PL model quality. Moreover, if PL is used as a
secondary rather than single component, its contribution to the total
unabsorbed luminosity proves to be small ($\sim 20$\%), while the
temperature of the dominant thermal component does not change much
compared to modeling by DISKBB alone ($\Tin=0.21$ vs. 0.31~keV), with
the total unabsorbed luminosity $\Lunabs$ increasing by a factor of
$\sim 2$. In contrast, the PL fit implies unrealistically high values
of $\NH$ and $\Lunabs$. 

For the above reasons, we imposed an additional condition on
$\Gamma>4$ sources: even if PL is the statistically best model, we
select the better of BB and DISKBB as our preferred model. 

For softer spectra with $\Gamma<4$, there is no clear preference
between the models (see Fig.~\ref{fig:cstat}): 66 (25\%), 48 (18\%)
and 153 (57\%) out of 267 spectra are best fit by PL, BB and DISKBB,
respectively. Since all these models are reasonable, we did not
introduce any additional criteria for $\Gamma<4$ sources and picked
their best spectral models based on $C$-statistics alone. However, for
relatively soft sources ($3\lesssim\Gamma\lesssim4$) there is a risk
that choosing the wrong model from two or three models with similar
$C$ values may lead to a large error in the absorption column and
hence unabsorbed luminosity of the object. We made a special test to
estimate the impact of this systematic uncertainty on our results (see 
Section~\ref{s:sample_test}). 

\subsection{Luminous sample}
\label{s:luminous_sample}

Based on the results of the spectral analysis, we compiled a sample of
sources (hereafter, the 'luminous sample'), 219 in total, with
estimated unabsorbed luminosities (assuming the sources indeed belong
to the corresponding galaxies) $\Lunabs>10^{38}$~erg~s$^{-1}$
(0.25--8~keV). This sample provides the basis for the subsequent
analysis (in Sections~\ref{s:final_selection} and \ref{s:lumfunc}). We
have not extended our study to lower luminosities partially because the
problem of separating HMXBs from other types of sources such as LMXBs
and AGN, is greatly exacerbated at $\Lunabs<10^{38}$~erg~s$^{-1}$.

\section{Survey coverage as a function of source luminosity and
  spectrum}
\label{s:sfr_lum}

We do not expect our sample of X-ray sources to directly represent the
intrinsic distribution of luminosities and proportion of hard and
soft sources, because it is based on a photon limited survey. Even in
the absence of line-of-sight absorption, detection of an X-ray source
with a given luminosity in a given galaxy must depend on the source's
spectral properties and the {\sl Chandra}/ACIS spectral response. This
may bias the resulting sample of sources in favour of soft ones. On the
other hand, the presence of interstellar gas in the Milky Way and
especially in the host galaxy in the direction of the source may
significantly lower its chances of being detected, and this negative
bias is expected to mainly affect soft sources. The situation is
exacerbated by the fact that HMXBs are usually located close to sites
of active star formation, which in turn are correlated with
concentrations of cold molecular gas.

As mentioned before, the unique property of the sample of nearby
galaxies studied here is that their high-quality {\sl Chandra} X-ray
maps can be superposed on similarly high-quality maps of the SFR and
atomic and molecular gas. This makes it possible, under reasonable
assumptions about the ISM distribution perpendicular to the plane of
the galaxy (see \S\ref{s:abs_model} below), to evaluate the X-ray 
source selection effects discussed above and eventually integrate the
SFR over those 3D regions of the galaxies where {\sl Chandra} can
detect a source with given intrinsic luminosity and spectral type. In
what follows, the so-derived SFR will be sometimes referred to as the
SFR probed by {\sl Chandra}. By dividing the number of actually
detected X-ray sources of a given spectral type in a given luminosity
bin by the corresponding SFR probed by {\sl Chandra}, we later
construct XLFs of different types of sources per unit SFR
(\S\ref{s:lumfunc}). 

We assumed that the spatial distribution of HMXBs in a galaxy
approximately follows that of its current SFR. As known from HMXB
surveys in the Milky Way \citep{bodetal12,lutetal13}, the two are
correlated on a spatial scale of few hundred~pc. This is the 
distance that a HMXB can travel (due to a kick received from the
supernova explosion of the progenitor of the relativistic compact
object) from its birth site over its lifetime of several 
million years. By the same argument and as also confirmed by Galactic
observations \citep{lutetal13}, HMXBs are expected to be distributed
approximately uniformly in the vertical direction within $\sim 100$~pc
of the central plane of their galaxy. It is worth noting in this
connection that the Galactic microquasar SS~433 is located $\sim
200$~pc away from the Galactic plane. 

The HI component of the host galaxy ISM, again by analogy with the 
Milky Way (e.g. \citealt{naksof16}), is expected to have a vertical
scale height of a few hundred pc (which may strongly depend on the
galactocentric distance). Therefore, the distribution of HMXBs along the 
normal to the galactic plane should be approximately random with
respect to the atomic interstellar gas. 

The situation is somewhat different with the colder, H$_2$, phase of
the ISM. For the Galaxy it is known that a substantial or even
dominant fraction of H$_2$ is concentrated in giant molecular clouds
(GMCs), but there is a broad distibution of cloud sizes and
masses. However, even GMCs may be regarded as small objects
($<100$~pc) compared to the other relevant spatial scales discussed
above. The molecular gas is expected to be concentrated near the
galactic plane, with a vertical scale $\sim 60$~pc (as in the
Milky Way, \citealt{heydam15}). All this suggests that i) one should {\it not}
expect HMXB positions in a given galaxy to correlate with the
positions of individual molecular clouds, although it is natural to
expect some correlation between HMXB locations and the large-scale
distribution of H$_2$ in the galaxy (because the latter is correlated
with the SFR); ii) to a first approximation, HMXBs are as likely
to be located between the H$_2$ disk of their galaxy and the observer
as behind the H$_2$ disk.

\subsection{SFR and ISM maps of the galaxies}
\label{s:galaxy_profiles}

We used a linear combination of {\sl GALEX} UV and {\sl Spitzer}/MIPS
24~$\mu$m intensity maps to produce maps of the SFR surface density,
$\SSFR$, for our galaxies. Our analysis closely followed
\cite{leretal08} and \cite{bigetal08}, and in fact the $\SSFR$ maps
for 20 of our 27 galaxies have already been presented by
\cite{leretal08}.

Specifically, we calculated $\Sigma_{\rm SFR}$ according to
equation~(1) in \citealt{bigetal08}: 
\begin{eqnarray}
\label{eq-sfr}
\SSFR\left[ {\rm M}_\odot~{\rm yr}^{-1}~{\rm kpc}^{-2}\right]
&=&3.2 \times 10^{-3} I_{24} \left[{\rm MJy~ster}^{-1}\right] \\
\nonumber &+& 8.1 \times 10^{-2} I_{\rm FUV} \left[{\rm MJy~ster}^{-1}\right]~,
\end{eqnarray}
where I$_{24}$ and I$_{\rm FUV}$ are the 24~$\mu$m and FUV
intensities, respectively, derived directly from \textit{Spitzer} and
\textit{GALEX} images. We additionally corrected the FUV intensities
for the Galactic extinction assuming $A_{\rm FUV}=4.63\times
E_{B-V}$\citep{youetal13} and adopting $E_{B-V}$ from \cite{schfin11}
(using IRSA Galactic Dust Reddening and Extinction
Service\footnote{http://irsa.ipac.caltech.edu/applications/DUST/}). Note
the significant difference between the $A_{\rm FUV}/E_{B-V}$ ratio
adopted here (4.64) and that used by \cite{bigetal08} (8.24, as in
\citealt{wydetal07}); see \cite{youetal13} for a relevant discussion.

For most of our galaxies, we used HERACLES CO $J=2\to 1$ line
intensity ($I_{\rm CO}$, K km s$ ^{-1}$) maps \citep{leretal09} to
produce maps of the H$_2$ surface density, $\SHM$. This was done using
equation~(4) in \cite{leretal09}, which reduces to
\begin{equation}
  \SHM[M_\odot~{\rm pc}^{-2}]=5.5~I_{\rm CO}[{\rm K~km~s}^{-1}],
  \label{eq:shm_co}
\end{equation}
assuming the CO $J = 1\to 0$ to H$_{2}$ conversion factor
$X_{\rm CO}=2\times 10^{20}$ cm$^{-2}$ (K km s$ ^{-1} $)$^{-1}$ and the CO
$J = 2\to 1$ to $J= 1\to 0$ ratio $R_{21}=0.8$ (see \citealt{leretal09}
and references therein for more details). Note that $\SHM$ makes
allowance for the contribution of helium.

Seven of our galaxies have not been covered by the HERACLES survey
(see Section~\ref{s:galaxies}). For these, we constructed synthetic
$\SHMS$ maps from the SFR maps using a well-established correlation
between $\SSFR$ and $\SHM$ \citep{bigetal08,leretal08}:
\begin{equation}
  \SSFR/\SHM=(5.25\pm 2.5)\times 10^{-10}~\textrm{yr}^{-1}.
  \label{eq:sfr_hm}
\end{equation}
We used the same method to estimate H$_2$ surface densities in the
outer regions of some galaxies if the HERACLES maps were
insufficiently large (usually, $\SHM$ is negligibly small in these
regions, though). As has been shown by \cite{bigetal08}, the
$\SSFR$--$\SHM$ correlation appears to work universally well in spiral
galaxies in a broad range of H$_2$ surface densities, $\sim
3$--50~$M_\odot$~pc$^{-2}$, i.e. $\sim (1.4$--$23)\times
10^{20}$~H$_2$~molecules~cm$^{-2}$, with the interpretation being that
H$_2$ forms stars at a constant efficiency. However, this correlation
may not be applicable both below and above this range. The former case
does not present any problem, since usually regions with
$\SHI+\SHM\lesssim 10$~$M_\odot$~pc$^{-2}$ are dominated by HI
\citep{bigetal08,leretal08}. The $\SHM\gtrsim 50$~$M_\odot$~pc$^{-2}$
limit pertains to starburst regions, which are usually the central
regions of some of our galaxies, and our use of the $\SSFR$--$\SHM$
correlation in this regime is limited because we have excluded the
$R<0.05R_{25}$ regions from the analysis. 

The resulting $\SSFR$, $\SHI$ and $\SHM$ (or $\SHMS$) maps for the
studied galaxies, with the positions of the X-ray sources from our
clean sample (defined below in Section~\ref{s:clean}) superposed on
them, are presented in Appendix~\ref{s:maps}.

We next constructed radial profiles (with a bin size of $0.2R_{25}$)
of $\SSFR$, $\SHI$ and $\SHM$ from the corresponding maps. They are 
are in good agreement with \citep{leretal08}, where most of our
galaxies have been analysed before. For the galaxies with available
HERACLES data, there is satisfactory agreement between the directly
measured H$_2$ surface densities and the SFR-based estimates (see
further discussion in Appendix~\ref{s:maps}), which justifies our
usage of synthetic $\SHMS$ maps as a substitute for missing CO data.

\subsection{Simulations of X-ray absorption in the ISM}
\label{s:abs_model}

Using the derived SFR and ISM radial profiles for the galaxies, we
performed simulations of {\sl Chandra} observations of X-ray sources
of various spectral types with different intrinsic luminosities and
locations within the galaxies.

For each galaxy, we chose $\sim 10$ random celestial positions within
each of the five $0.2R_{25}$-wide bins (elliptical rings in the sky
plane) within $R_{25}$ and simulated spectra that could be measured in a
particular existing {\sl Chandra}/ACIS observation from X-ray sources
located in these positions, depending on the sources' intrinsic
spectral shape and line-of-sight absorption. The 10 different
positions per radial bin are needed to take into account possible
differences in the {\sl Chandra} response from one position to
another. We used the same three spectral models, {\it wabs(powerlaw)},
{\it wabs(bb)} and {\it wabs(diskbb)}, as we used before in the
spectral analysis of the real {\sl Chandra} sources in our sample, and
ran simulations on a large grid of spectral parameter values:
$0<\Gamma<10$, $0.05<\Tbb<5$~keV, $0.05<\Tin<5$~keV and
$\NH=10^{20}$--$10^{24}$~cm$^{-2}$.

\begin{figure}
\centering
\includegraphics[width=\columnwidth,viewport=50 300 562 662]{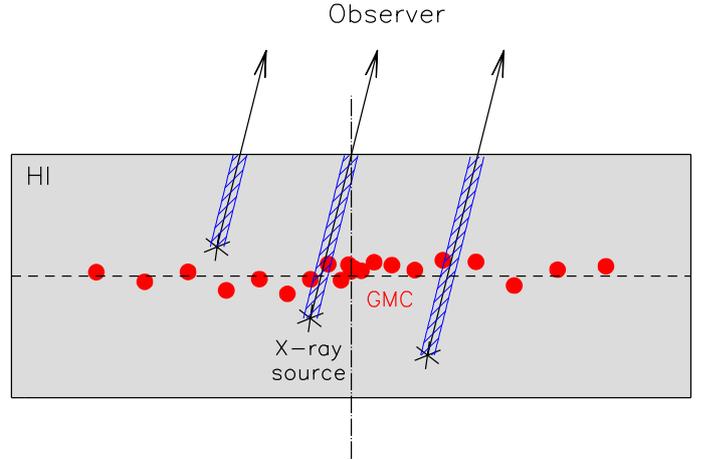}
\caption{A schematic of the absorption of X-rays from a source in the
  atomic and molecular ISM of its host galaxy.
}
\label{fig:sketch}
\end{figure}

We then assumed that an X-ray source with a given position on the sky 
can be located quasi-randomly along the line of sight within the 
ISM of its host galaxy. Namely, motivated by the discussion at the
beginning of Section~\ref{s:sfr_lum}, we modeled the HI component of
the galaxy as a thick homogeneous slab whereas we assumed the H$_2$
gas to be concentrated in GMCs located in the central plane of the
galaxy (see a schematic in Fig.~\ref{fig:sketch}).

In this model, an X-ray source falling into a given elliptical ring of
the galaxy, with the ring-averaged HI column density $\SHI$, has
a probability $dP=d\NHI/\SHI$ to have a column $\NHI+\NHIMW$ of atomic
gas between itself and the observer, with $0\le\NHI\le\SHI$. Here,
$\NHIMW$ is the Galactic HI column density in the direction of
the source, adopted from \cite{kaletal05} (we use the {\it nh} tool from
{\sc heasoft
  ftools}\footnote{http://heasarc.gsfc.nasa.gov/docs/software/ftools/ftools\_menu.html}).

The same X-ray source has a 50\% chance of being located in front of
the galactic plane and thus at least 50\% probability to have no
GMC between itself and the observer. If the source is located 
behind the galactic plane, its view can be obscured or not by a GMC,
as discussed below.

We assumed that a typical column density through a GMC is
$\SGMC=2\times 10^{21}$~H$_2$ molecules cm$^{-2}$, or, equivalently,
$\sim 42$~$M_\odot$~pc$^{-2}$ (taking into account helium). This
is the median mass surface density  of Galactic GMCs determined by
\cite{heyetal09} using the Boston University--FCRAO Galactic Ring
Survey of CO emission, but a factor of $\sim 4$ smaller than
170~$M_\odot$~pc$^{-2}$, the value from \cite{soletal87} (with a small
correction by \citealt{heyetal09}) which was for a long time regarded
as standard in astrophysics. \cite{heyetal09} explain a number of
reasons (including the use of $^{13}$CO~$J=1-0$ instead of
$^{12}$CO~$J=1-0$ and improved angular and spectral resolution) that
led to this significant revision. However, \cite{heyetal09}
point out that their new reference value for $\SGMC$ may be
underestimated by a factor of $\sim 2$.

We implemented the following algorithm. If the radial-bin averaged
$\SHM$ is smaller than $\SGMC$, we posit that a source behind the
galactic plane has a $\fGMC=\SHM/\SGMC$ chance of having a GMC between
itself and the observer. In this case, a fraction $\fGMC$ of such
sources will have an H$_2$ column $\NHM=\SGMC$ toward the observer,
while the remaning $(1-\fGMC)$ will have $\NHM=0$. If $\SHM>\SGMC$, we
adopt that $\fGMC=1$ and $\NHM=\SHM$, i.e. use the bin-averaged $\SHM$
instead of $\SGMC$ as the H$_2$ column density in front of the
source. The last assumption may reflect a situation when several
GMCs form a larger molecular complex (in a starburst region) and/or
there are individual GMCs with column densities higher that our
fiducial value of $2\times 10^{21}$~cm$^{-2}$. There is clearly
significant uncertainty here, and we performed some tests (see below) 
to estimate its effects on our results.

Our final assumption is that the total number of X-ray sources of a
given kind (i.e. intrinsic luminosity and spectral type) in a given
radial bin is proportional to the total SFR ($\SSFR$) in that
bin. This assumption is reasonable, since HMXBs generally follow
ongoing star formation activity in galaxies, as mentioned before.  

Using the above assumptions and prescriptions, we evaluated for each
galaxy in our sample the SFR probed by {\sl Chandra} as a
function of X-ray source intrinsic luminosity and spectral type. Just
as for real sources in our sample, a simulated source was
considered to be detected if it provided at least 100 counts in any of
the available {\sl Chandra}/ACIS observations for the given position
within the galaxy. We integrated the SFR over
$0.05R_{25}<R<R_{25}$. We assumed that the innermost $0.05R_{25}$
region contributes 1/16th of the SFR encompassed within $0.2R_{25}$
and subtracted this amount from this radial bin. This
is unlikely to lead to a significant error.

\begin{figure}
\centering
\includegraphics[width=0.8\columnwidth,viewport=30 170 560 710]{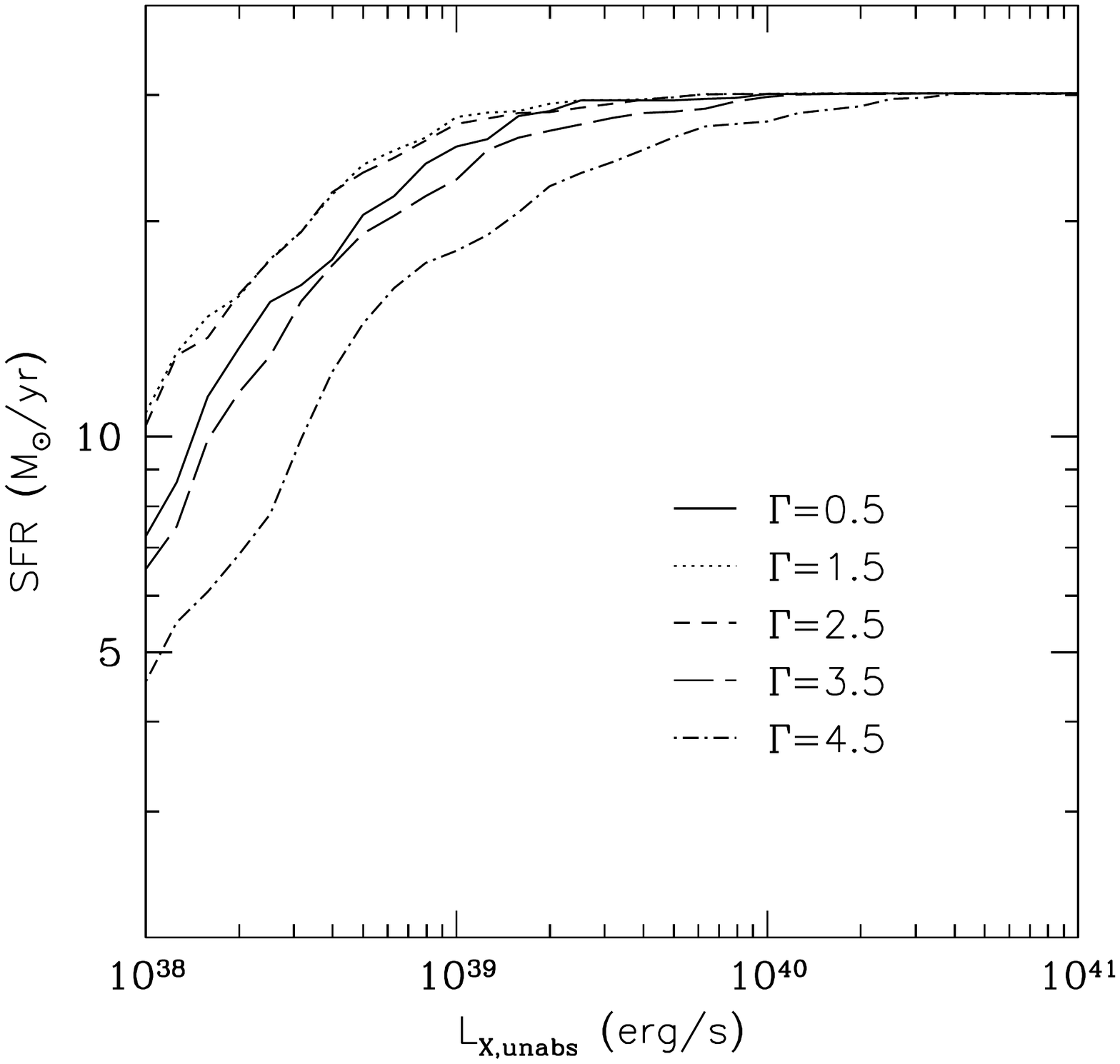}
\includegraphics[width=0.8\columnwidth,viewport=30 170 560 710]{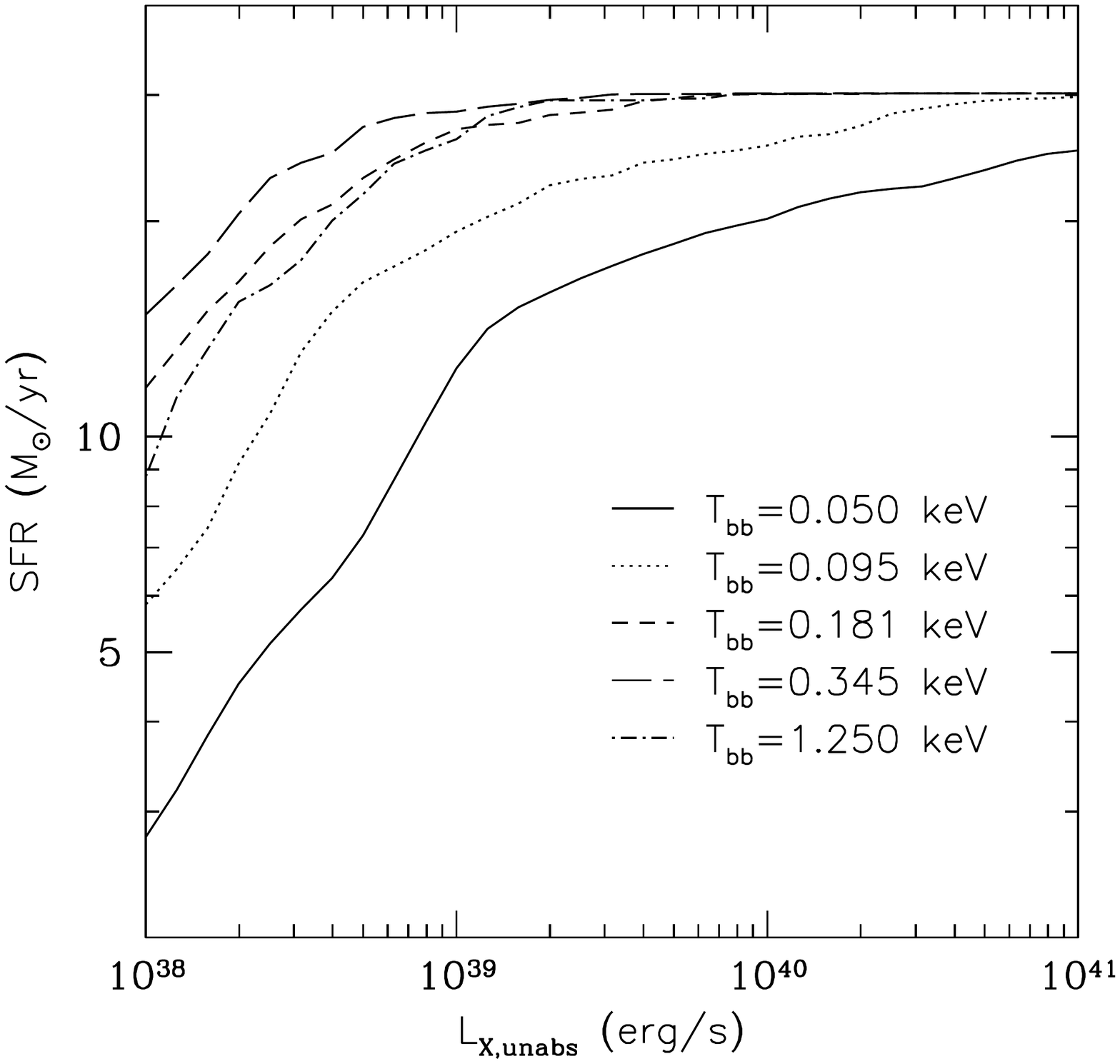}
\includegraphics[width=0.8\columnwidth,viewport=30 170 560 710]{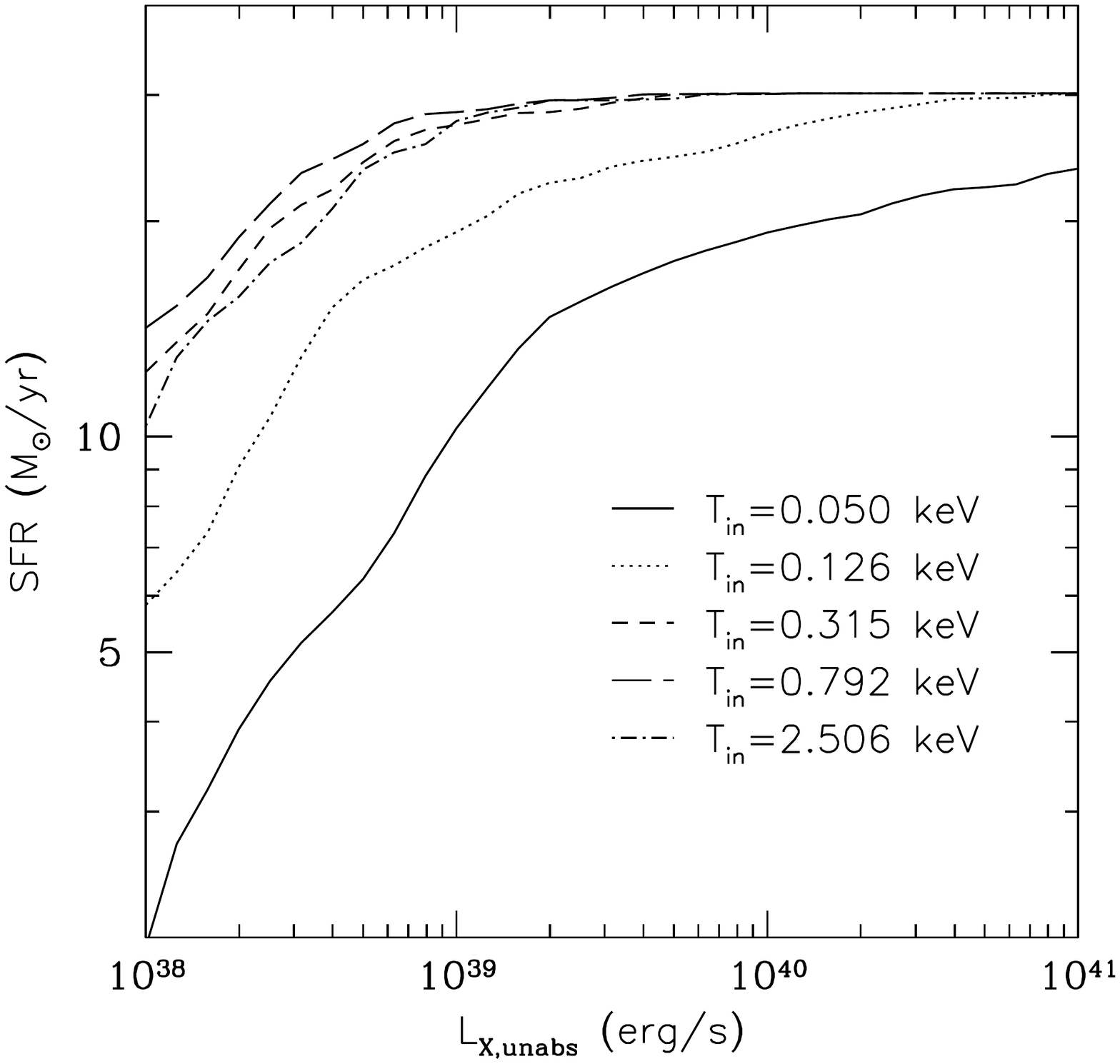}
\caption{{\sl Top panel:} Integrated SFR over the space within the
  27 studied galaxies where {\sl Chandra} can detect a source with a
  power-law intrinsic spectrum, as a function of the source's PL slope
  and intrinsic luminosity in the 0.25--8~keV energy band. {\sl Middle
    panel:} The same, but for sources with BB spectra of different
  temperatures. {\sl Bottom panel:} The same, but for sources with
  DISKBB spectra of different temperatures.
}
\label{fig:lum_sfr}
\end{figure}

Figure~\ref{fig:lum_sfr} shows the resulting SFR ($\Lunabs$)
dependencies for the PL, BB and DISKBB models, summed over all 27
galaxies. 

As could be expected, at very high luminosities $\Lunabs\gtrsim
10^{40}$~erg~s$^{-1}$ {\sl Chandra} probes the total SFR of the 27
galaxies of $\sim 30$~$M_\odot$ for all spectral types of sources,
except for the softest ones ($\Tbb\lesssim 0.1$~keV or $\Tin\lesssim
0.1$~keV). This is because very high ISM column densities are needed
to suppress the observed X-ray flux of a luminous source to less than
100 counts, given that all of the considered galaxies are located
within 15~Mpc. However, the difference between various spectral types
of sources becomes noticeable below $\Lunabs\sim 10^{40}$~erg~s$^{-1}$ 
and grows with decreasing luminosity. As also aniticipated, for a
given $\Lunabs$, the integrated SFR first increases going from hard to
soft sources (e.g., for the PL model this occurs  for $\Gamma\lesssim
2$) due to the increasing number of photons per unit flux, but the 
opposite trend sets in for yet softer spectra because of
the strong suppression of soft X-rays in the ISM.

\subsection{Additional tests}
\label{s:abs_tests}

\begin{figure}
\centering
\includegraphics[width=0.8\columnwidth,viewport=30 170 560 710]{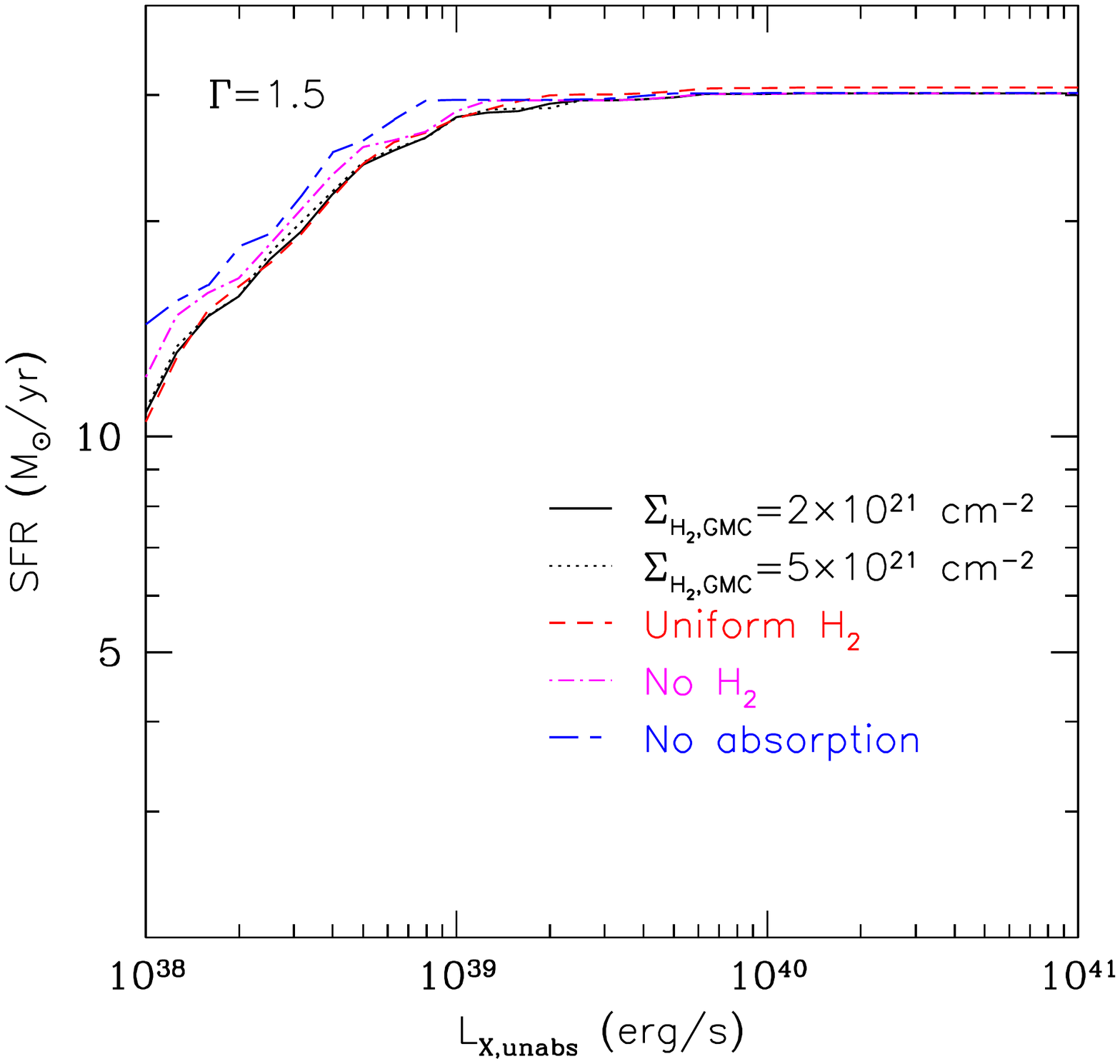}
\includegraphics[width=0.8\columnwidth,viewport=30 170 560 710]{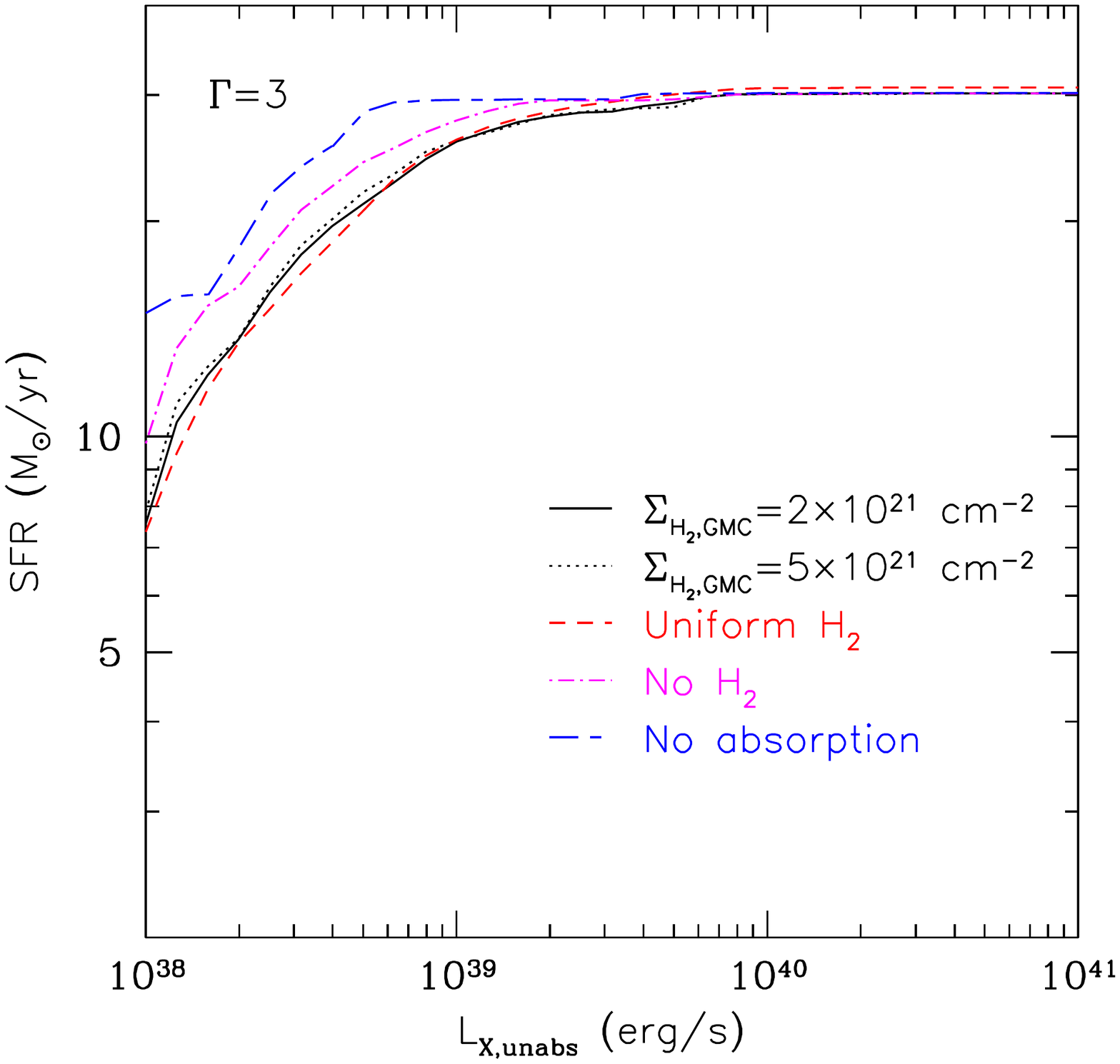}
\includegraphics[width=0.8\columnwidth,viewport=30 170 560 710]{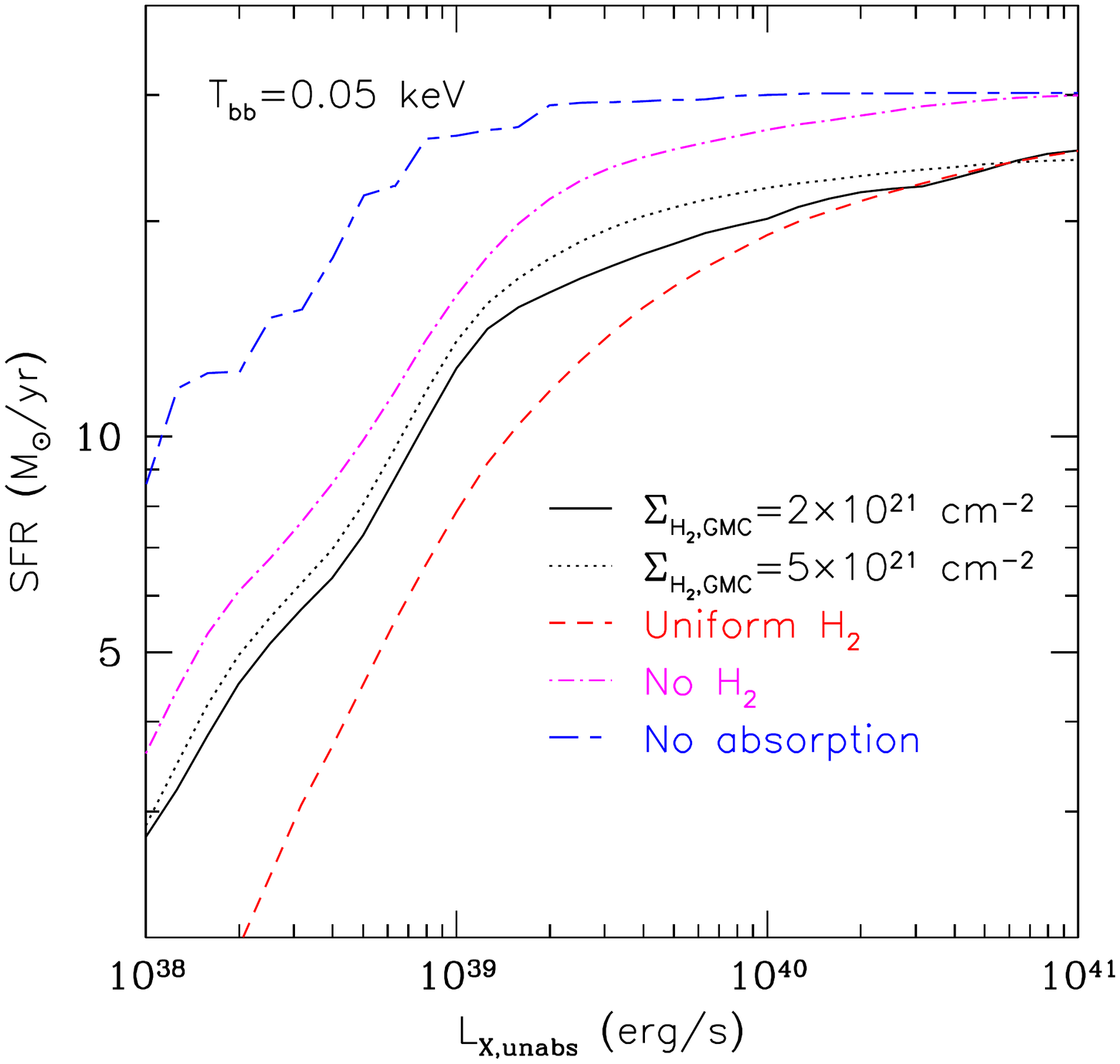}
\caption{{\sl Top panel:} Similar to Fig.~\ref{fig:lum_sfr}, for
  sources with hard ($\Gamma=1.5$) PL spectra, under various
  assumptions about the ISM (see text). {\sl Middle panel:} The same,
  but for sources with soft ($\Gamma=3$) PL spectra. {\sl Bottom
    panel:} The same, but for sources with supersoft ($\Tbb=0.05$~keV)
  BB spectra.
}
\label{fig:lum_sfr_test}
\end{figure}

The largest systematic uncertainty in our simulations is associated
with the distribution of H$_2$ in the galaxies. It is clear that our
assumption that all molecular gas seen in the CO maps is located in
GMCs of fixed column density simplifies the real situation. We
therefore tested the sensitivity of our SFR ($\Lunabs$) curves to this
assumption. 

First, we tried a model where there is no absorbing gas in the
direction of the X-ray sources. In this imaginary scenario we, as
expected, see (in Fig.~\ref{fig:lum_sfr_test}) a large enhancement of
the total SFR probed by {\it Chandra} for sources with soft and
supersoft spectra (PL with $\Gamma=3$ and BB with $\Tbb=0.05$~keV,
respectively) and a noticeable increase ($\sim 20$\% at
$\Lunabs<10^{39}$~erg~s$^{-1}$) for hard sources ($\Gamma=1.5$). As a
second test, we retained the atomic gas but removed all molecular gas
from the modeled galaxies. In this (also unrealistic) scenario, the
impact of absorption on the cumulative SFR is still substantial for
soft and supersoft sources. 

We next tried using different values of $\SGMC$ in our 'standard'
model and found that the SFR does not change significantly if $\SGMC$
is increased from $2\times 10^{21}$ to $5\times 10^{21}$~cm$^{-2}$ (see
Fig.~\ref{fig:lum_sfr_test}). 

Finally, we implemented a more radical change of the model, namely
distributed all the molecular gas seen in the CO map of a given
elliptical ring homogeneously over the volume of the galaxy subtended
by this ring, i.e. essentially allowed H$_2$ to be mixed with HI. This
model leads to a much stronger negative bias for soft sources (see
Fig.~\ref{fig:lum_sfr_test}) compared to our baseline model. This has
a clear explanation: in our standard model, there is always at least a
50\% chance (due to the assumed location of all GMCs in the galactic
plane) for sources of any spectral type to have no molecular gas
obscuring their view. This probability is in fact often higher,
because even if a source is located on the other side of the galactic
plane there may still be holes in the molecular layer above
it. In the alternative scenario, all sources are immersed
in the HI+H$_2$ ISM and their X-ray fluxes inevitably suffer some
attenuation in the molecular gas in addition to absorption in the
atomic gas.

Although we find the above 'homogeneous H$_2$ model' unrealistic, it
indicates that our baseline absorption model is probably too
conservative, i.e. it may somewhat {\sl underestimate} the negative
bias associated with soft sources if the molecular gas has a more
fragmentary structure than assumed in our baseline model.

\section{Removal of contaminants}
\label{s:final_selection}

We cross-correlated our luminous sample with standard astronomical
databases in an attempt to find out the nature of our
sources. Although some published catalogues ascribe detailed
classifications, e.g. 'HMXB' or 'LMXB', to many of our sources, we
considered such classifications unreliable, because they are usually
based solely on the location of the source within its presumed host
galaxy (e.g. a source within the bulge of a galaxy would be designated
a LMXB). Similarly, many of our most luminous ($\Lunabs\gtrsim
10^{39}$~erg~s$^{-1}$) sources are marked 'ULX' in various catalogues,
but we do not use this purely empirical classification either.

Specifically, we searched for objects located within 3~arcsec of the
{\sl Chandra} positions of our sources. Since all these sources are
bright (more than 100 counts), their X-ray positions are known to
better than $1''$, except for  3 sources (observed at large offset
angles) for which the uncertainties are between $1.0''$ and $1.4''$
\citep{wanetal16}.

\subsection{Foreground stars}
\label{s:stars}

We looked for the possible presence of foreground stars in the
luminous sample and found 3 likely associations (see
Appendix~\ref{s:stars_agn}).

\subsection{Background AGN}
\label{s:agn}

We next looked for the presence of AGN in the luminous sample and
found 8 likely associations, 7 of which may be considered reliable (see
Appendix~\ref{s:stars_agn}). However, the total number of CXB sources
contaminating the sample may be higher. 

To estimate this number, we adopted the AGN $\log N$--$\log S$ curve in
the 0.5--10~keV band from (\citealt{geoetal08}, their equation~(2) and
table~2; see also \citealt{koletal13} for a comparison with other
published $\log N$--$\log S$ curves) and made a small correction from
0.5--10~keV to our operative energy band of 0.25--8~keV. We assumed all
CXB sources to have power-law spectra with $\Gamma=1.5$, which is
approximately equal to the slope of the CXB spectrum and is
effectively an average between the $\Gamma\sim 1.8$ spectra of
intrinsically unobscured (type 1) AGN and the harder spectra of type-2
objects (see, e.g., \citealt{sazetal08}). We then took into account
that the X-ray flux from an AGN located behind any of our studied
galaxies will suffer some absorption in its ISM (and in the Galaxy),
by performing simulations similar to our modeling of SFR 
($\Lunabs$) above. 
 
We found that there are expected to be 19.7 background AGN in our luminous
sample. This estimate is quite robust due to the hardness of CXB
sources. In fact, we also tried to completely switch off line-of-sight
absorption in our model, which led to a negligible enhancement of the
expected number: 20.0. Therefore, the luminous sample likely contains
$\sim 20\pm 5$ background AGN, 8 of which we have already
identified. Hence, $\sim 12\pm 5$ CXB sources likely remain hidden in
the sample. 

\subsection{Measured vs. expected absorption columns, further AGN candidates}
\label{s:nhobs_nhmodel}

\begin{figure}
\centering
\includegraphics[width=\columnwidth,viewport=30 170 560 710]{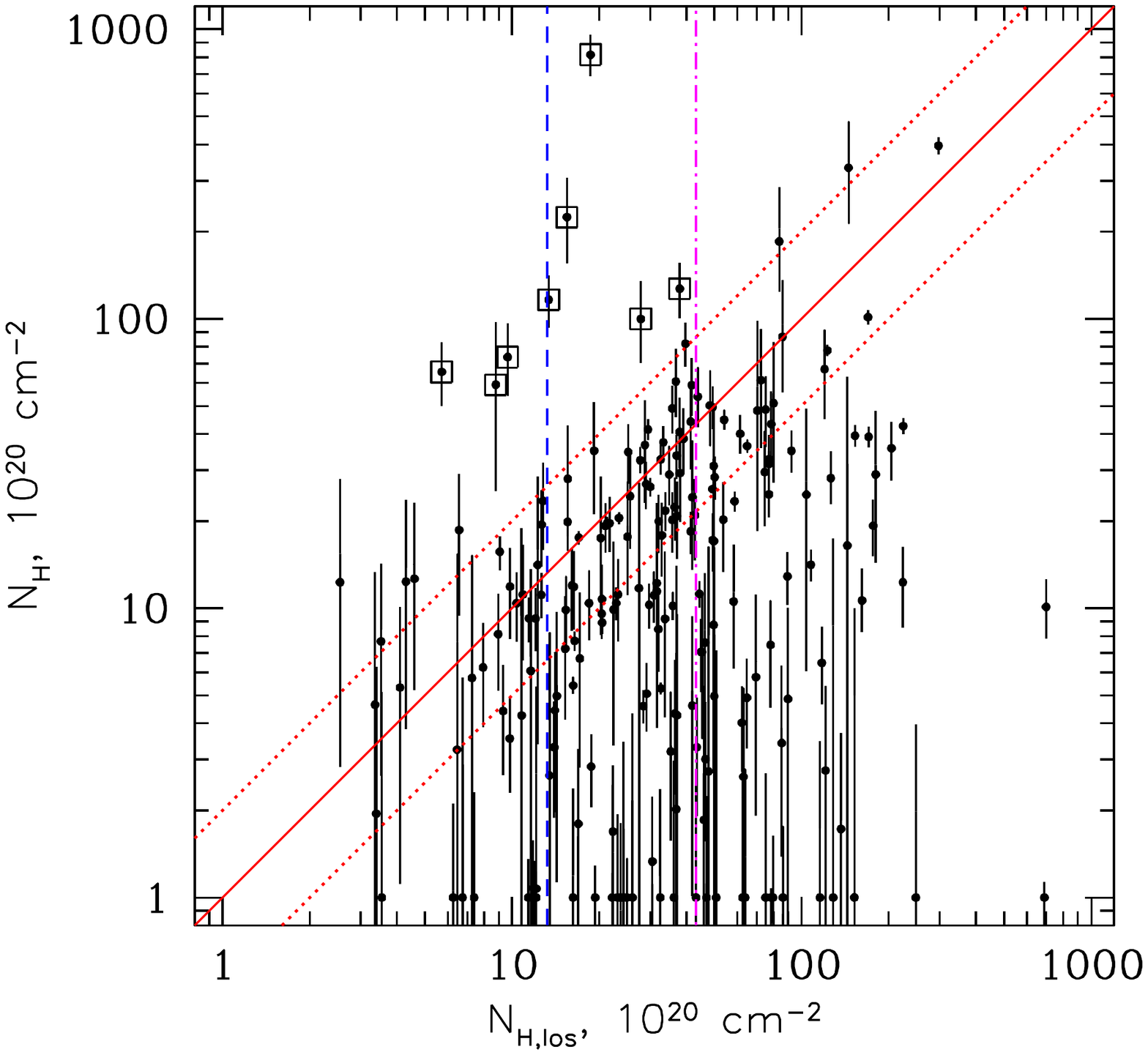}
\caption{Absorption columns, $\NH$, measured in the X-ray spectra of
  sources from the luminous sample (excluding 3 likely foreground
  stars and 8 likely background AGN) vs. the corresponding total
  line-of-sight column densities, $\NHlos$, through the Galaxy and the
  host galaxy. The red solid and dotted lines show the $\NH=\NHlos$,
  $\NH=0.5\NHlos$ and $\NH=2\NHlos$ dependences. The magenta
  dash-dotted and blue dashed vertical lines indicate the median
  line-of-sight H+H$_2$ and H column densities, respectively. The
  statistical uncertainties for $\NHlos$ (estimated based on the noise
  level in the corresponding HI and H$_2$ maps) are negligible. The
  squares denote 8 outliers (located above the $\NH=2\NHlos$ line,
  taking into account the $\NH$ uncertainties), some of which may be
  background AGN. 
}
\label{fig:nhlos_nh}
\end{figure}

It is interesting to compare the $\NH$ columns inferred from our
spectral analysis of the {\sl Chandra} sources with the total column
density of gas in their direction: $\NHlos=\NHIMW+\SHI+2\SHM$ (where
$\SHI$ and $\SHM$ are measured in H~atoms~cm$^{-2}$ and
H$_2$~molecules~cm$^{-2}$, respectively). The latter information can be 
taken from the same HI and H$_2$ (or synthetic H$_2$) maps
that we used before for computing the SFR ($\Lunabs$) dependencies for
the galaxies.  

Figure~\ref{fig:nhlos_nh} shows the result of comparison of $\NH$ with
$\NHlos$ for the luminous sample, excluding the 11 objects associated
with either stars or AGN mentioned above, i.e. for 208 sources in
total. Disregarding the (significant) statistical  uncertainties
associated with the $\NH$ measurements, 105 (55\%) and 163 (78\%)
sources lie below the $\NH=0.5\NHlos$ and $\NH=\NHlos$ lines,
respectively. Taking the uncertainties into account, 181 (87\%) sources are
consistent with having $\NH\le\NHlos$.

The above numbers appear to be in fairly good agreement with the
expectations of our ISM absorption model (see
Section~\ref{s:abs_model}). Indeed, the total line-of-sight gas column
density for most of our sources proves to be dominated by H$_2$ in the
host galaxy: the mean and median values of $\NHlos$ are 5.5 and
$3.4\times 10^{21}$~cm$^{-2}$, respectively, whereas the corresponding
values for $\NHIMW+\SHI$ are 1.6 and $1.3\times
10^{21}$~cm$^{-2}$. Hence, our typical source has $\NHM\sim
(1$--$2)\times 10^{21}$~cm$^{-2}$ in its direction, which corresponds
to $\sim 0.5$--1~GMC per line of sight (for our adopted $\SGMC=2\times 
10^{21}$~cm$^{-2}$). Taking into account that the source may be
located either in front or behind the molecular disk of its galaxy, we 
may expect $\sim 25$--50\% of our sources to be obscured by molecular
gas and hence have $\NH\sim \NHlos$.

The satisfactory agreement of the measured X-ray absorption columns
with the expectations of our ISM absorption model implies that the
absorption evident in the X-ray spectra of most of our sources is
indeed due to the ISM in their host galaxies, rather than due to gas
intrinsic to the sources. This result fits well into the assumed
picture that most sources with $\Lunabs>10^{38}$~erg~s$^{-1}$ are
binary systems where accretion onto the compact object proceeds via
Roche-lobe overflow of the (massive) companion, rather than through
its wind. If the opposite were true, we would expect to see a lot of
additional absorption in the wind. Such absorption, with $\NH$ up to
$\sim 10^{24}$~cm$^{-2}$, is indeed observed in many Galactic HMXBs,
in particular those discovered with the {\sl INTEGRAL} observatory
\citep{waletal15}, but all such systems have
$\Lunabs<10^{38}$~erg~s$^{-1}$. 

However, we do see several clear outliers in the $\NH$ vs. $\NHlos$
diagram. Specifically, there are 8 sources (denoted by squares in
Fig.~\ref{fig:nhlos_nh} and listed in Appendix~\ref{s:stars_agn}) that
are located above the $\NH=2\NHlos$ line, taking into account the
$\NH$ uncertainties. If they belonged to the explored galaxies (3
objects in NGC~5194 and 5 objects in NGC~5457), the bulk of the
absorption observed in their spectra would have to be intrinsic to the
sources rather than arise in the ISM (even allowing for metallicity
variations). Although this possibility cannot be excluded, we may also
consider an alternative explanation that some or all of these objects
are background AGN with intrinsic absorption. In fact, the spectra of
6 of them have slopes consistent with $\Gamma=1.8$ if fitted by an
absorbed PL model, which is typical for AGN (the remaining two sources
have somewhat softer spectra with the 1$\sigma$ lower limits on
$\Gamma$ being 2.3 and 2.5). Since we previously concluded there must
remain $\sim 12\pm 5$ unidentified AGN in our luminous sample, of
which a substantial fraction should have significant X-ray absorption
(as is common for AGN), the 8 absorption-diagram outliers are more
likely to be these missing AGN compared to other sources in the
sample. We therefore excluded these 8 objects from further
consideration. This disputable decision has a very limited impact on
the results of our study since if affects just $\sim 4$\% of the
entire sample of sources.

\section{Clean sample}
\label{s:clean}

\begin{table*}

\caption{Clean sample of X-ray sources
  \label{tab:clean_lum39}
}

\begin{tabular}{|l|r|l|l|r|r|r|r|r|r|l|l|}
\hline
  \multicolumn{1}{|c|}{Source} &
  \multicolumn{1}{c|}{{\sl Chandra}} &
  \multicolumn{1}{c|}{Galaxy} &
  \multicolumn{1}{c|}{$R/R_{25}$} &
  \multicolumn{1}{c|}{$\NHlos$} &
  \multicolumn{1}{c|}{$\NH$} &
  \multicolumn{1}{c|}{$\Lunabs$} &
  \multicolumn{1}{c|}{$\LSunabs$} &
  \multicolumn{1}{c|}{$\Lobs$} &
  \multicolumn{1}{c|}{$\LSobs$} &
  \multicolumn{1}{c|}{Type} & 
  \multicolumn{1}{c|}{Note} \\
\multicolumn{1}{c}{CXOGSG} & \multicolumn{1}{c|}{obs.} & & &
\multicolumn{2}{c}{$10^{20}$~cm$^{-2}$} &
\multicolumn{4}{c}{$10^{38}$~erg~s$^{-1}$} &
& \\
\hline
  J131519.5$+$420302 & 2197 & NGC~5055 & 0.962 & 22.9 &
  $10.4_{9.1}^{11.8}$ & $271.53_{258.96}^{286.04}$ &
  $195.26_{180.69}^{212.14}$ & 171.44 & 96.23 & S & \\[1.5mm]
  J022727.5$+$333442 & 7104 & NGC~925 & 0.508 & 49.3 &
  $17.2_{13.4}^{21.2}$& $194.02_{181.87}^{206.90}$ &
  $80.96_{73.87}^{88.81}$ & 162.19 & 51.61 & H & \\[1.5mm] 
  J125055.6$+$410719 & 808 & NGC~4736 & 0.150 & 85.9 &
  $86.6_{55.6}^{136.1}$& $140.63_{11.72}^{7664.73}$ &
  $146.11_{11.83}^{2142.29}$ & 0.15 & 0.15 & SS & \\[1.5mm]
  J073721.8$+$653317 & 4629 & NGC~2403 & 0.546 & 36.8 &
  $60.7_{45.8}^{78.7}$& $134.18_{31.13}^{759.94}$ &
  $134.21_{31.13}^{759.94}$ & 0.22 & 0.22 & SS & \\[1.5mm]
  J110545.6$+$000016 & 9552 & NGC~3521 & 0.617 & 54.0 &
  $44.7_{41.4}^{48.2}$& $129.34_{118.92}^{141.47}$ &
  $101.96_{91.02}^{114.62}$ & 43.89 & 18.08 & S & \\[1.5mm]
  J111815.1$-$324840 & 9278 & NGC~3621 & 0.120 & 77.1 &
  $32.7_{30.7}^{34.8}$& $128.46_{118.02}^{140.47}$ &
  $109.94_{98.94}^{122.54}$ & 38.40 & 20.73 & S & \\[1.5mm]
  J112020.9$+$125846 & 9548 & NGC~3627 & 0.584 & 11.8 &
  $1.1_{1.0}^{1.6}$& $104.70_{102.81}^{106.65}$ &
  $37.91_{37.29}^{38.65}$ & 103.00 & 36.30 & H & \\[1.5mm]
  J140229.9$+$542118 & 4732 & NGC~5457 & 0.537 & 15.5 &
  $19.9_{17.7}^{22.0}$& $93.68_{82.35}^{107.11}$ &
  $89.38_{77.85}^{103.03}$ & 20.38 & 16.24 & SS & \\[1.5mm]
  J203500.7$+$601130 & 1043 & NGC~6946 & 0.471 & 49.2 &
  $25.8_{24.9}^{26.7}$& $86.63_{83.52}^{90.01}$ &
  $70.03_{66.62}^{73.69}$ & 33.93 & 17.89 & S & 1 \\[1.5mm]
  J081929.0$+$704219 & 15771 & HO~II & 0.831 & 28.4 &
  $4.6_{4.0}^{5.2}$& $82.65_{81.42}^{83.90}$ & $40.47_{39.24}^{41.77}$
  & 72.41 & 30.44 & H & \\[1.5mm]
  J133007.5$+$471106 & 13813 & NGC~5194 & 0.873 & 17.0 &
  $17.5_{16.6}^{18.5}$& $65.37_{64.04}^{66.74}$ &
  $40.44_{38.92}^{42.02}$ & 41.77 & 17.34 & S & \\[1.5mm]
  J133001.0$+$471343 & 13813 & NGC~5195 & 0.744 & 20.4 &
  $9.6_{8.7}^{10.4}$& $56.95_{56.24}^{57.66}$ &
  $22.36_{21.70}^{23.07}$ & 48.93 & 14.66 & H & \\[1.5mm] 
  J133705.1$-$295207 & 12994 & NGC~5236 & 0.129 & 32.7 &
  $5.3_{5.0}^{5.5}$ & $49.35_{49.00}^{49.72}$ &
  $27.26_{26.87}^{27.66}$ & 41.23 & 19.26 & H &\\[1.5mm] 
  J095532.9$+$690033 & 735 & NGC~3031 & 0.410 & 23.4 &
  $20.5_{19.5}^{21.5}$& $45.30_{44.74}^{45.87}$ & $9.50_{9.24}^{9.78}$
  & 40.91 & 5.67 & H & 1 \\[1.5mm] 
  J093206.1$+$213058 & 11260 & NGC~2903 & 0.493 & 30.0 &
  $26.2_{24.3}^{28.2}$& $42.38_{40.35}^{44.66}$ &
  $29.33_{27.03}^{31.91}$ & 21.78 & 9.14 & S & \\[1.5mm] 
  J140303.9$+$542734 & 4731 & NGC~5457 & 0.580 & 14.0 &
  $3.3_{1.9}^{4.8}$& $42.14_{41.16}^{43.16}$ & $15.30_{14.80}^{15.82}$
  & 40.31 & 13.56 & H & 1\\[1.5mm]
  J235751.0$-$323726 & 3954 & NGC~7793 & 0.554 & 20.5 &
  $8.9_{8.1}^{9.7}$& $36.54_{36.03}^{37.06}$ & $10.76_{10.44}^{11.09}$
  & 33.24 & 7.65 & H & \\[1.5mm]
  J095524.7$+$690113 & 735 & NGC~3031 & 0.406 & 64.8 &
  $36.3_{34.7}^{38.0}$& $32.64_{29.18}^{36.69}$ &
  $31.25_{27.73}^{35.37}$ & 4.39 & 3.09 & SS & 2\\[1.5mm] 
  J093209.6$+$213106 & 11260 & NGC~2903 & 0.259 & 152.9 &
  $39.4_{36.0}^{43.0}$& $32.41_{28.46}^{37.30}$ &
  $27.94_{23.80}^{33.02}$ & 8.65 & 4.43 & S & \\[1.5mm] 
  J101954.7$+$453248 & 9551 & NGC~3198 & 0.089 & 36.4 &
  $22.3_{17.1}^{28.2}$& $32.27_{21.73}^{52.27}$ &
  $31.21_{20.45}^{51.43}$ & 5.54 & 4.52 & SS & \\[1.5mm] 
  J132959.0$+$471318 & 13813 & NGC~5195 & 0.579 & 122.6 &
  $77.7_{74.3}^{81.3}$ & $32.02_{31.31}^{32.77}$ &
  $14.44_{13.88}^{15.03}$ & 20.08 & 4.24 & H & \\[1.5mm]
  J140332.3$+$542102 & 934 & NGC~5457 & 0.247 & 16.3 &
  $5.4_{5.1}^{5.7}$& $31.51_{30.53}^{32.53}$ & $31.47_{30.49}^{32.49}$
  & 17.94 & 17.90 & SS & \\[1.5mm]
  J132950.6$+$471155 & 13813 & NGC~5195 & 0.124 & 296.9 &
  $396.0_{369.9}^{423.0}$& $30.65_{29.27}^{32.17}$ &
  $8.68_{7.84}^{9.61}$ & 16.06 & 0.30 & H & \\[1.5mm]  
  J122810.9$+$440648 & 10875 & NGC~4449 & 0.881 & 224.3 &
  $42.6_{39.8}^{45.4}$ & $26.45_{22.55}^{31.30}$ &
  $25.30_{21.33}^{30.21}$ & 3.21 & 2.14 & SS & 1\\[1.5mm]
  J131602.2$+$420153 & 2197 & NGC~5055 & 0.452 & 32.6 &
  $32.6_{28.9}^{36.6}$ & $24.66_{22.82}^{26.78}$ &
  $20.00_{18.04}^{22.31}$ & 12.39 & 8.06 & S & \\[1.5mm]
  J140414.2$+$542604 & 4736 & NGC~5457 & 0.867 & 11.4 &
  $1.0_{1.0}^{1.4}$ & $24.16_{23.72}^{24.61}$ &
  $17.55_{17.18}^{17.93}$ & 23.26 & 16.67 & S & \\[1.5mm]
  J013651.1$+$154547 & 16000 & NGC~628 & 0.528 & 16.9 &
  $1.8_{1.0}^{3.3}$ & $20.68_{19.92}^{21.48}$ & $7.41_{7.10}^{7.74}$ &
  20.16 & 6.91 & H & \\[1.5mm]
  J132953.3$+$471042 & 15553 & NGC~5194 & 0.261 & 117.4 &
  $6.5_{4.7}^{8.6}$ & $18.06_{17.25}^{ 18.97}$ & $9.77_{8.87}^{10.87}$
  & 14.80 & 6.56 & H & \\[1.5mm] 
  J132951.8$+$471137 & 13814 & NGC~5194 & 0.057 & 204.3 &
  $35.7_{27.6}^{44.1}$ & $18.01_{10.52}^{32.02}$ &
  $17.85_{10.52}^{32.03}$ & 1.53 & 1.53 & SS & 1\\[1.5mm] 
  J112018.3$+$125900 & 9548 & NGC~3627 & 0.329 & 89.3 &
  $12.9_{10.3}^{15.6}$ & $17.74_{16.95}^{18.55}$ &
  $7.87_{7.38}^{8.40}$ & 15.12 & 5.43 & H & \\[1.5mm] 
  J133719.8$-$295348 & 12994 & NGC~5236 & 0.630 & 9.1 &
  $15.7_{13.5}^{17.7}$ & $17.14_{16.11}^{18.25}$ &
  $13.05_{11.94}^{14.26}$ & 8.96 & 4.97 & S & \\[1.5mm] 
  J223706.6$+$342619 & 2198 & NGC~7331 & 0.767 & 41.8 &
  $58.9_{46.1}^{73.3}$ & $17.12_{15.01}^{19.64}$ &
  $10.03_{8.12}^{12.61}$ & 9.77 & 3.34 & H & \\[1.5mm] 
  J073625.5$+$653539 & 2014 & NGC~2403 & 0.557 & 21.3 &
  $19.7_{18.7}^{20.7}$ & $16.85_{16.56}^{17.15}$ &
  $7.81_{7.62}^{8.01}$ & 13.50 & 4.71 & H & \\[1.5mm]
  J140228.2$+$541626 & 5322 & NGC~5457 & 0.660 & 41.4 &
  $18.4_{15.4}^{21.6}$ & $15.53_{14.93}^{16.15}$ &
  $7.95_{7.49}^{8.45}$ & 12.21 & 4.84 & H & \\[1.5mm] 
  J101959.1$+$453403 & 9551 & NGC~3198 & 0.401 & 23.1 &
  $1.0_{1.0}^{1.9}$ & $14.93_{14.08}^{15.84}$ & $7.38_{7.03}^{7.76}$ &
  14.60 & 7.07 & H & \\[1.5mm]  
  J110549.1$-$000257 & 9552 & NGC~3521 & 0.230 & 145.1 &
  $332.2_{213.5}^{477.9}$ & $14.44_{12.13}^{17.65}$ &
  $2.15_{1.32}^{3.72}$ & 9.42 & 0.17 & H & \\[1.5mm]  
  J203500.1$+$600908 & 1043 & NGC~6946 & 0.183 & 58.6 &
  $23.4_{21.5}^{25.3}$ & $13.24_{12.23}^{14.38}$ &
  $12.85_{11.79}^{14.01}$ & 4.81 & 4.44 & SS & \\[1.5mm] 
  J095542.1$+$690336 & 735 & NGC~3031 & 0.117 & 16.4 &
  $7.7_{7.1}^{8.3}$ & $11.34_{10.44}^{12.34}$ &
  $11.34_{10.44}^{12.34}$ & 3.43 & 3.43 & SS & \\[1.5mm] 
  J133702.4$-$295126 & 12994 & NGC~5236 & 0.080 & 120.0 &
  $66.9_{45.0}^{90.9}$ & $10.85_{2.45}^{59.32}$ &
  $11.09_{2.44}^{59.70}$ & 0.10 & 0.10 & SS & 1 \\[1.5mm] 
  J110548.4$-$000250 & 9552 & NGC~3521 & 0.265 & 128.5 &
  $1.0_{1.0}^{17.4}$ & $10.64_{9.55}^{11.85}$ & $0.24_{0.20}^{0.28}$ &
  10.64 & 0.23 & H & \\[1.5mm]
  J132953.7$+$471435 & 13814 & NGC~5195 & 0.754 & 19.4 &
  $1.0_{1.0}^{1.3}$ & $10.36_{10.13}^{10.61}$ & $3.80_{3.71}^{3.89}$ &
  10.20 & 3.65 & H & \\[1.5mm]
  J132946.1$+$471042 & 13812 & NGC~5194 & 0.485 & 30.9 &
  $11.1_{8.9}^{13.3}$ & $10.01_{9.25}^{10.90}$ & $7.57_{6.80}^{8.67}$
  & 5.78 & 3.49 & S & \\[1.5mm] 
\hline\end{tabular}

\flushleft
\textbf{References:} (1) Possible SNR association; (2) young SN. 

Note. This table is given in its entirety in the electronic version of
the paper (see Supporting information). The abbreviated version of the
table shown here contains the most luminous sources with
$\Lunabs>10^{39}$~erg~s$^{-1}$.
  
\end{table*}

After exclusion of 3 foreground stars, 8 background AGN and another 8
sources suspected to be AGN based on their high X-ray absorption columns,
we obtained a 'clean sample' of 200 sources
(Table~\ref{tab:clean_lum39}), the vast majority of which presumably
belong to the 27 nearby galaxies under consideration.  

It is unlikely that all sources in the clean sample are HMXBs (and
ULXs). First, it definitely contains a significant number of
LMXBs. Unfortunately, it is practically impossible, even in nearby
galaxies, to separate HMXBs from LMXBs on a source by source
basis. We therefore made an attempt to estimate the contribution of
LMXBs in a statistical way, as described below (Section~\ref{s:lmxb}).

In addition, our clean sample may contain supernovae (SNe) recently
exploded in the studied galaxies. We indeed found 4 SNe positionally
coincident with our X-ray sources (see
Table~\ref{tab:clean_lum39}). These are all core-collapse ones, and in
fact were the targets of the {\sl Chandra} observations used in our
analysis. Three of these observations (for SN~2002hh, SN~2004dj and
SN~2004et) were performed within two months after discovery,
whereas the observation of the famous SN~1993J was taken in 2000,
i.e. 7 years after the explosion (see \citealt{dwagru12} for the X-ray
light curves of these SNe). It is well known that young core-collapse
SNe can remain luminous ($\Lx>10^{38}$~erg~s$^{-1}$) X-ray sources for
months, years and sometimes decades after the explosion
(e.g. \citealt{immlew03,dwagru12}). For this reason and because
the 4 identified SNe constitute just $\sim 2$\% of our clean sample,
we did not exclude them from further analysis despite them being the
{\sl Chandra} targets. It is actually interesting to retain these
objects in the sample because young core-collapse SNe are another
manifestation of star-formation activity, complementary to HMXBs. As
for the X-ray spectra of the 4 identified SNe, two of them (SN~2004dj
and SN~1993J) are best fit by an absorbed PL among our simple
one-component models, with $\Gamma=2.0$ and $3.5$, respectively, one
(SN~2002hh) by a BB model with $\Tbb=1.24$~keV and one (SN~2004et) by
a DISKBB model with $\Tin=1.67$~keV. We also tried to desribe these
spectra in terms of optically thin thermal emission, {\it wabs(apec)},
and obtained worse fits. 

Finally, we found 15 possible associations of our sources with
supernova remnants (SNRs, see Table~\ref{tab:clean_lum39}). All of
these SNRs had been identified as such based on their optical
\citep{blaetal12,vucetal15} and/or radio \citep{choetal09a,choetal09b} 
  emission. However, the SNR is unlikely to be the actual source of
  the X-ray emission observed by {\sl Chandra} in most 
of these cases. First, these SNRs are expected to be at least few
hundred years old, while, to our knowledge, there are no
well-documented cases of an SNR maintaining an X-ray luminosity in
excess of $10^{38}$~erg~s$^{-1}$ for such a long time. On the other
hand, some core-collapse SNRs are known to be physically associated
with a bright HMXB, the best-known example perhaps being the pair of
the Galactic microquasar SS~433 (believed to be a 'misaligned' ULX)
and the W50 nebula (SNR~G039.7-02.0). A similar situation may well be 
realised in at least some of our 15 possible SNR--X-ray source
associations. 

\subsection{Spectral properties of the sample}
\label{s:sample_properties}

According to the results of our spectral analysis (see
Section~\ref{s:spectra}), the clean sample consists of the following
spectral types: there are 49 PL spectra with $0.79<\Gamma<3.84$,
40 BB spectra with $0.05<\Tbb<6.88$~keV (excluding one, exceptionally 
hard spectrum, the range is $0.05<\Tbb<1.24$~keV), and 
111 DISKBB spectra with $0.11<\Tin<3.10$~keV.

For the following treatment it is convenient to split the sources
into three categories: 'hard', 'soft' and 'supersoft'. We define these
according to the ratio of the unabsorbed flux in the 0.25--2~keV band,
$\Fsoft$, to that in the total (0.25--8~keV) band, $\Ftot$:
\begin{eqnarray}
{\rm Hard:}& \frac{\Fsoft}{\Ftot}\le 0.6,\\
{\rm Soft:}& 0.6<\frac{\Fsoft}{\Ftot}\le 0.95,\\
{\rm Supersoft:}& \frac{\Fsoft}{\Ftot}>0.95.
\label{eq:types}
\end{eqnarray}
This is a purely empirical division, with no direct relation to
similar terminology frequently used in the literature to describe
different types of X-ray sources. However, our hard class
approximately corresponds to X-ray pulsars and black-hole X-ray
binaries in their hard state, our soft class to black-hole X-ray
binaries in their high/soft and very high states and our supersoft
class to ULSs.

In practice, the hard category corresponds to $\Gamma\le 2$ for the PL
model, $\Tbb\ge 0.496$~keV for BB and $\Tin\ge 0.859$~keV for DISKBB. For
the soft class, the corresponding ranges are $2<\Gamma\le 3.37$,
$0.260\le\Tbb<0.496$~keV and $0.354\le\Tin<0.859$~keV; and for the
supersoft class they are $\Gamma>3.37$, $\Tbb<0.260$~keV and 
$\Tin<0.354$~keV.

\begin{figure}
\centering
\includegraphics[width=0.82\columnwidth,viewport=30 200 560 710]{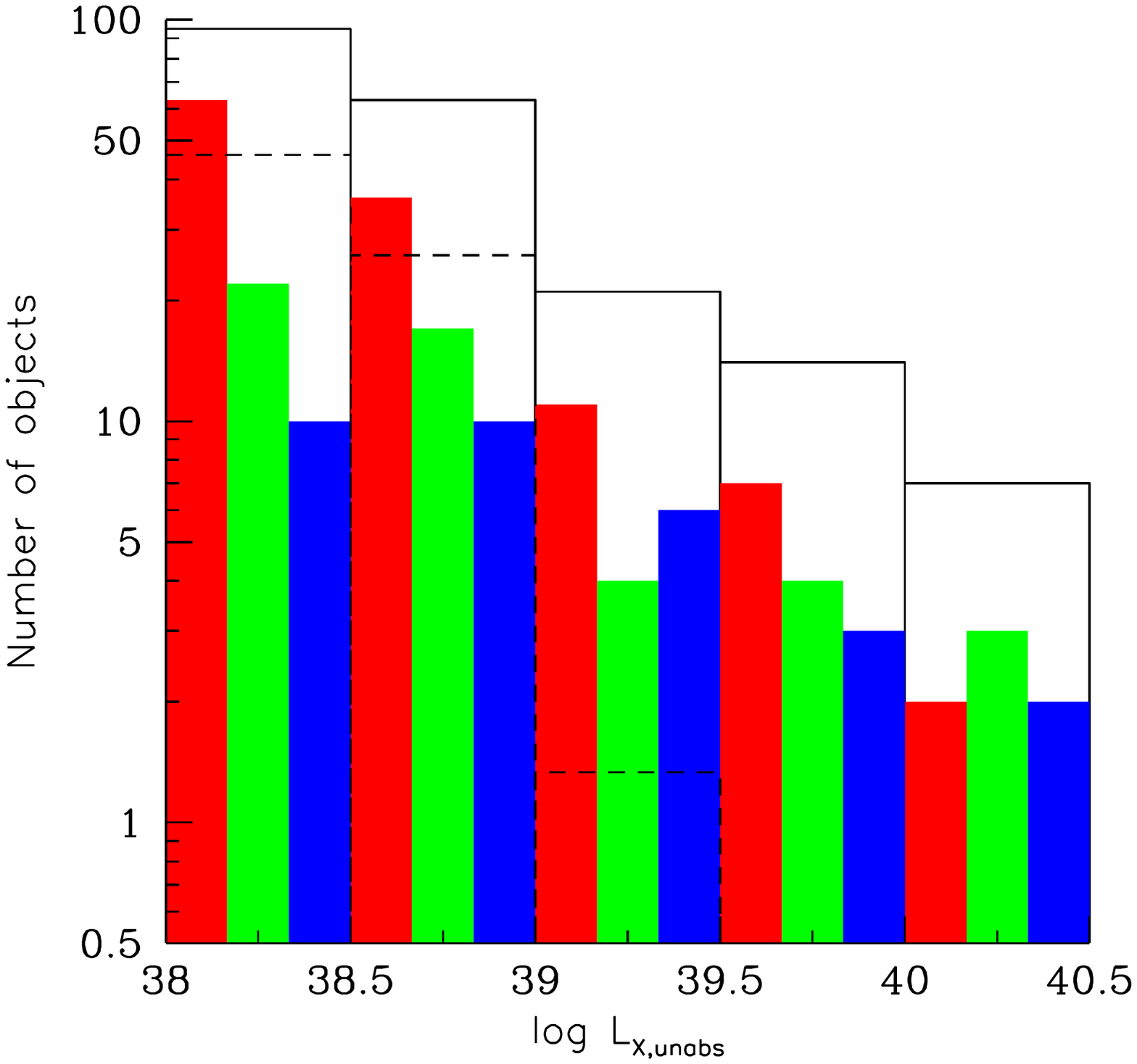}
\includegraphics[width=0.82\columnwidth,viewport=30 200 560 710]{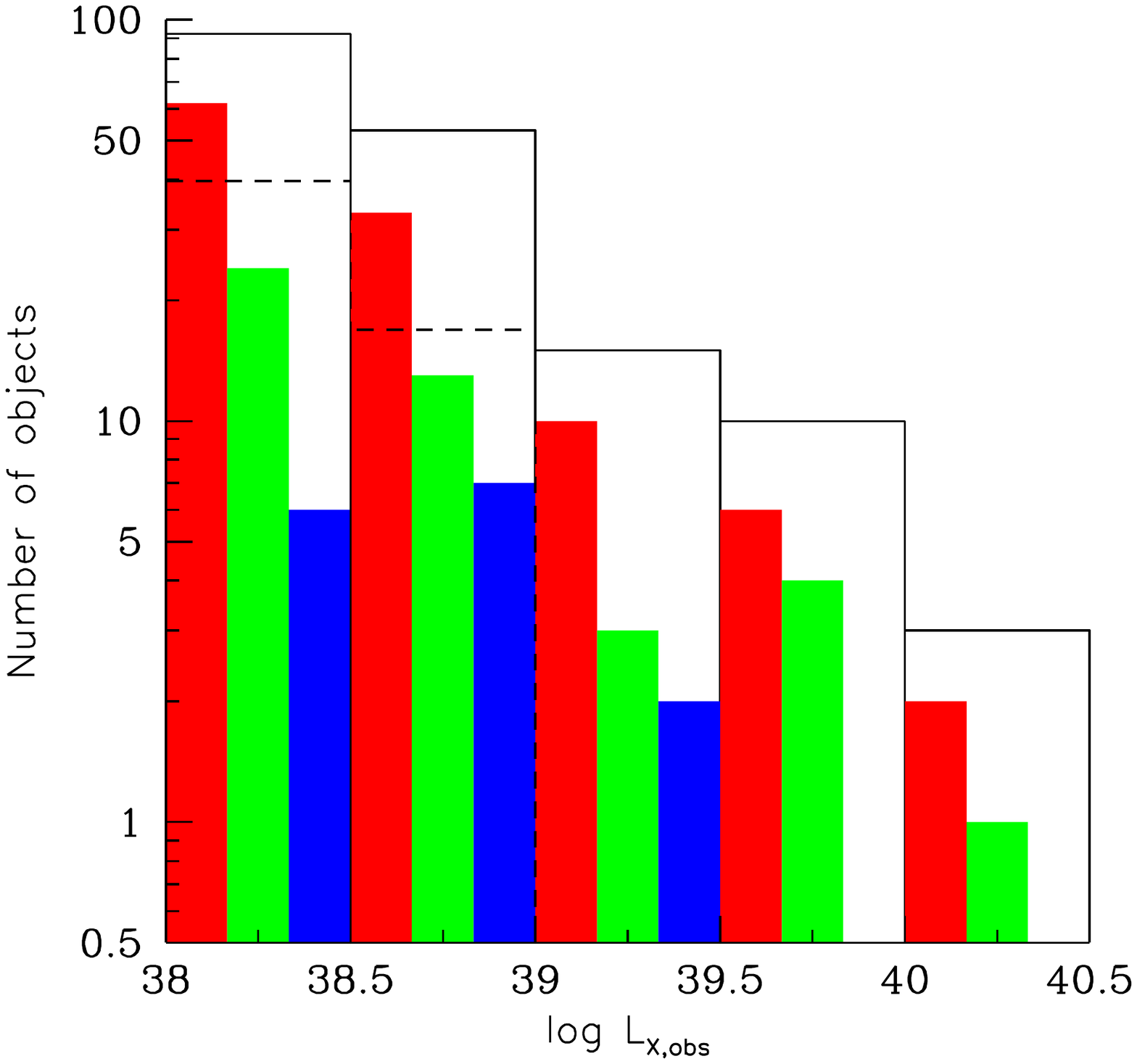}
\includegraphics[width=0.82\columnwidth,viewport=30 200 560 710]{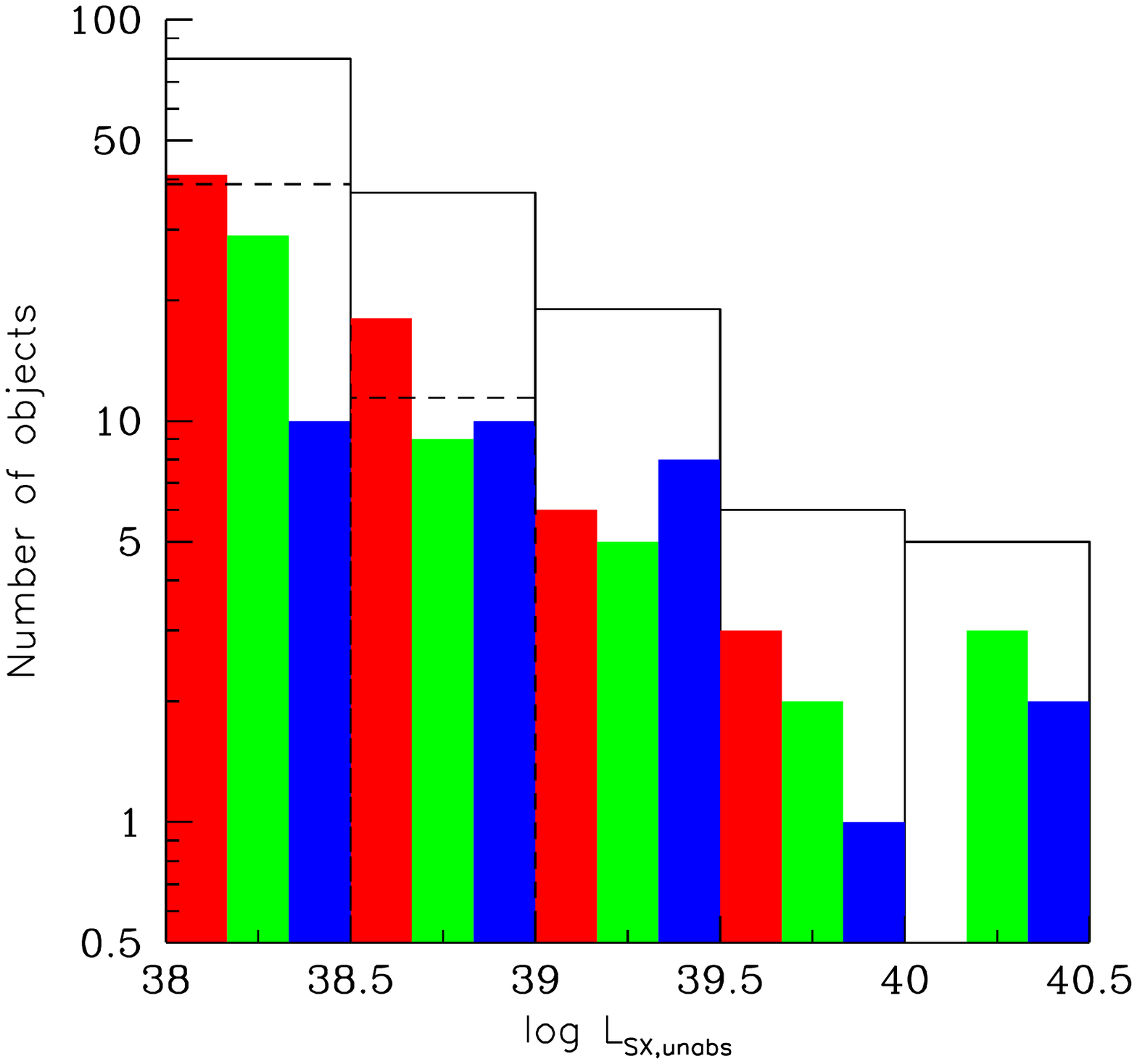}
\caption{{\sl Top panel:} Histrogram of intrinsic luminosities in the
  0.25--8~keV energy band for the clean sample (solid line) and its
  hard, soft and supersoft constituents (red, green and blue
  columns, drawn from left to right). The dashed line shows the
  estimated contribution of LMXBs. {\sl Middle panel:} The same for
  the observed luminosities in the 0.25--8~keV energy band. {\sl
    Bottom panel:} The same for the intrinsic luminosities in the soft
  band (0.25--2~keV). Note that some sources have $\Lobs$ and/or
  $\LSunabs<10^{38}$~erg~s$^{-1}$ and do not appear in the middle and
  bottom panels.
}
\label{fig:lum_hist}
\end{figure}

There are in total 119 hard, 50 soft and 31 supersoft sources in the
clean sample. Figure~\ref{fig:lum_hist} (upper panel) shows the
distribution of unabsorbed 0.25--8~keV luminosities, $\Lunabs$, for
these classes of sources as well as for the whole sample. We see that
the relative fraction of hard sources is highest ($\sim 2/3$) in the
lowest luminosity bin, $10^{38}<\Lunabs<10^{38.5}$~erg~s$^{-1}$,
diminishes to $\sim 1/2$ in the next 3 bins and drops to $\sim 1/3$ in
the highest luminosity bin ($10^{40}<\Lunabs<10^{40.5}$~erg~s$^{-1}$),
although there are only 7 sources in total in this last interval. We
also see that both soft and supersoft sources are important over the
entire studied luminosity range.  

\begin{figure}
\centering
\includegraphics[width=0.9\columnwidth,viewport=30 160 560 720]{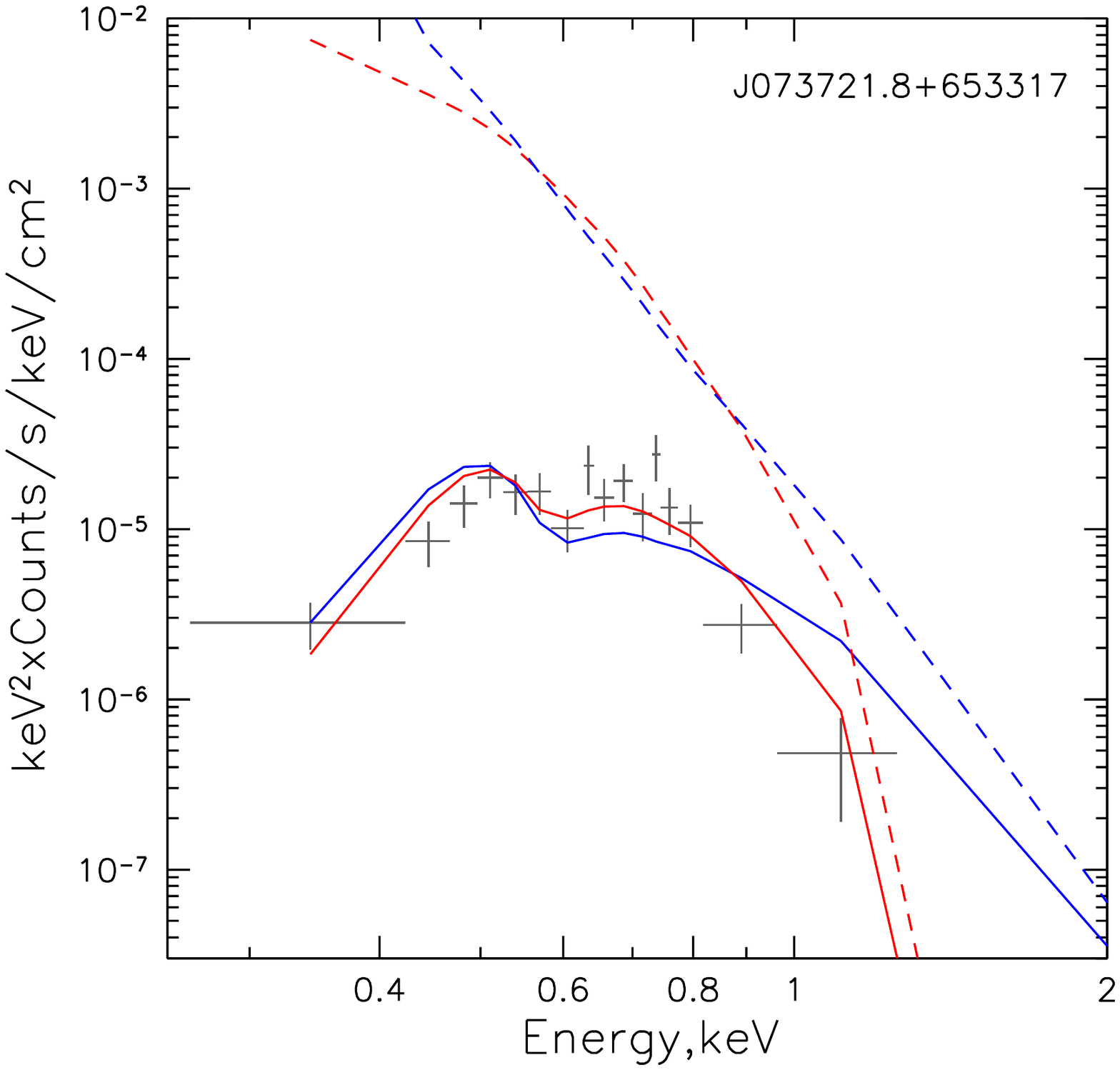}
\includegraphics[width=0.9\columnwidth,viewport=30 160 560 720]{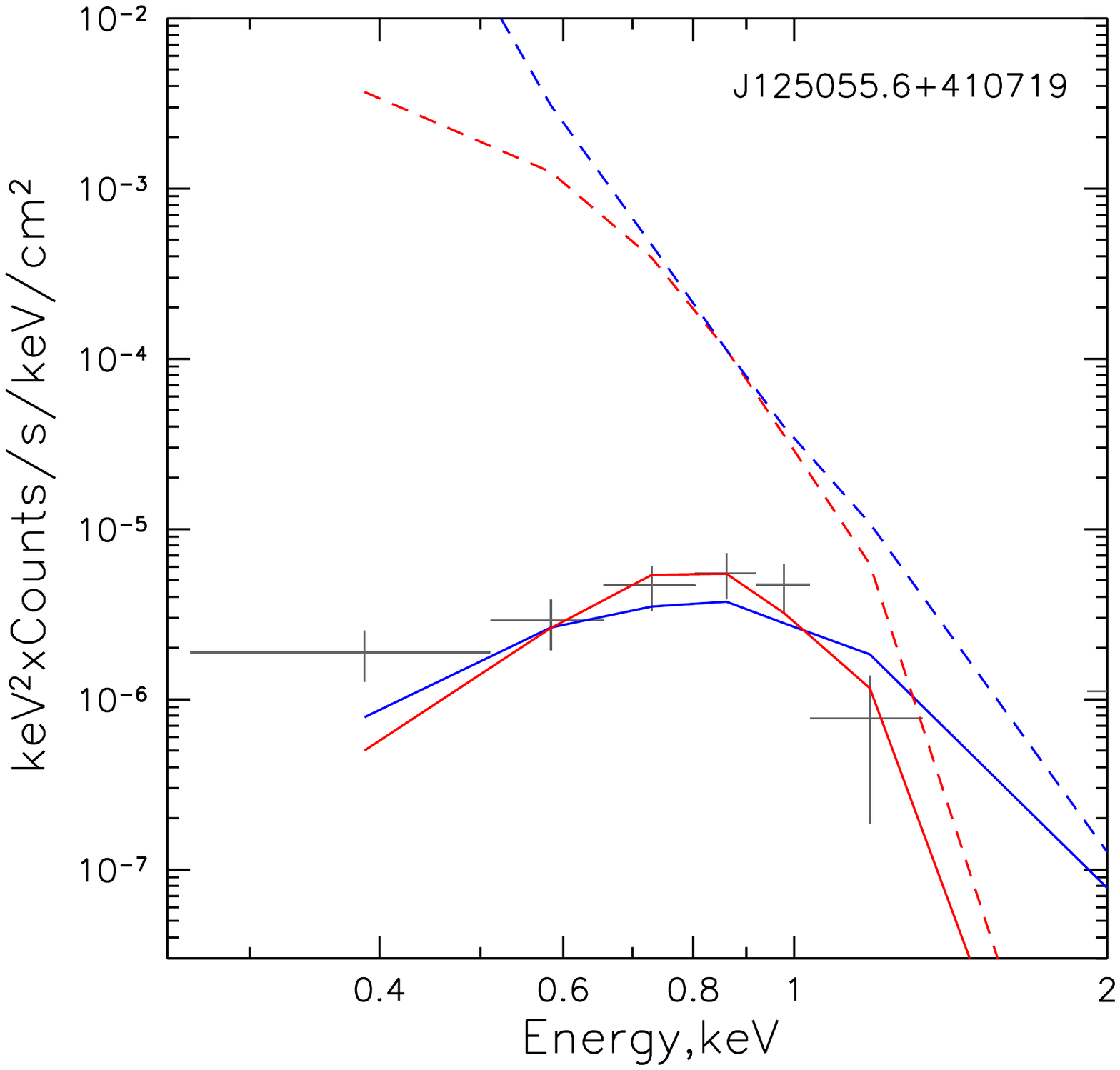}
\caption{{\sl Chandra} spectra for two supersoft sources for which the
  absorption correction factors turn out to be extremely high. The
  data points (rebinned for visualisation purposes and shown in grey)
  are the number of source counts per bin multiplied by the square of the
  bin's central energy and divided by the product of the exposure
  time, the bin's width and effective area at its central energy. The
  error-bars correspond to the 1$ \sigma$ uncertainty. The solid blue
  and red curves show absorbed PL and BB fits (the latter are
  statistically favoured), respectively, while the dashed curves are
  the same models with no absorption applied.
}
\label{fig:extreme_spectra}
\end{figure}

As for the supersoft sources, we caution that there is very large
uncertainty associated with the two objects of this type present in the
$10^{40}$--$10^{40.5}$~erg~s$^{-1}$ bin. Both have extremely soft
spectra (see Fig.~\ref{fig:extreme_spectra}), which are best fit by a 
heavily absorbed ($\NH\sim 5\times 10^{21}$--$10^{22}$~cm$^{-2}$) BB
model with $\Tbb\sim 0.06$--0.07~keV. As a result they have huge
intrinsic/observed luminosity (0.25--8~keV) ratios of $\sim 10^{3}$
(see Table~\ref{tab:clean_lum39}), and these absorption corrections
are uncertain by 1--2 orders of magnitude. Both sources are located in
the direction of gas rich regions of their presumed host galaxies,
with $\NHlos\sim 9\times 10^{21}$ and $\sim 4\times
10^{21}$~cm$^{-2}$, so the $\NH$ values inferred from our X-ray
spectral analysis are not unexpected, although for one of the sources
the lower limit on $\NH$ somewhat exceeds $\NHlos$. Moreover, fitting
by other models, namely absorbed PL and absorbed DISKBB, leads to even
higher $\Lunabs$. Note that the lower luminosity bins in the $\Lunabs$
histogram are not strongly affected by this kind of systematic
uncertainty, since these intervals contain a significant number of
supersoft sources with relatively small absorption corrections (see
Table~\ref{tab:clean_lum39}). 

For comparison, the middle panel of Fig.~\ref{fig:lum_hist} shows the
corresponding distribution of {\it observed} (i.e. uncorrected for
line-of-sight absorption) luminosities, $\Lobs$, for the same
sources. As could be expected, the relative number of soft and
especially supersoft sources is smaller in this case. Finally, the
bottom panel of Fig.~\ref{fig:lum_hist} shows the distribution of
soft-band (0.25--2~keV) unabsorbed luminosities, $\LSunabs$. In this
case, we see the opposite and also expected trend: the relative
numbers of soft and supersoft sources are higher.

\subsection{Contribution of LMXBs} 
\label{s:lmxb}

Our goal is to obtain the XLF of HMXBs. However, the histograms
shown in Fig.~\ref{fig:lum_hist} must inevitably contain a
contribution of LMXBs. The XLF of LMXBs is known fairly
well from studies of the bulge of the Galaxy \citep{revetal08},
galactic bulges and early-type galaxies \citep{gilfanov04},
i.e. environments nearly free of HMXBs (apart from projection
effects). We can use this information to estimate the contibution of
LMXBs to our sample of sources.

We adopted the analytic form of the LMXB XLF from
(\citealt{gilfanov04}, namely their equations~(8) and (9) and the
parameters for 'All galaxies' from table~3). In contrast to HMXBs, the
number of LMXBs is proportional to the stellar mass of a galaxy rather
than its SFR. We thus multiplied the XLF per stellar mass from
\cite{gilfanov04} by the stellar masses of our galaxies, given in
Table~\ref{tab:galaxies}. Our clean sample by
construction contains objects located within the $0.05R_{25}$--$R_{25}$ 
regions of the galaxies. Because these regions encompass nearly all of
the stellar mass of the galaxies, we did not apply any correction to 
the normalisation, bearing in mind much larger uncertainties
associated with the LMXB XLF itself (see relevant discussions in
\citealt{gilfanov04} and \citealt{minetal12}). 

The majority of Galactic LMXBs have hard spectra, with $\Gamma\sim
1.5$--2 if fitted by a power law (e.g. \citealt{revetal08}). However,
we are interested here in LMXBs with $\Lx>10^{38}$~erg~s$^{-1}$, which
are very rare in the Galaxy. The majority of such luminous LMXBs are
black-hole transients (mostly X-ray novae), which are known to
experience transitions between hard and soft spectral states
(e.g. \citealt{donetal07}). Therefore, our clean sample may well
contain LMXBs in both hard and soft states. Furthemore, among our
lowest luminosity sources ($\Lunabs\gtrsim 10^{38}$~erg~s$^{-1}$) there 
may be present classical supersoft sources, associated with 
accreting white dwarfs, but such objects are not expected to have
$\Lunabs\gtrsim 2\times 10^{38}$~erg~s$^{-1}$ (e.g. \citealt{soretal16}).

Since it is hardly possible to predict the relative numbers of hard,
soft and supersoft sources among the LMXBs contaminating our HMXB
sample, and in view of the substantial uncertainties associated with
the LMXB XLF, we assumed that $\Gamma=2$ for all LMXBs. We further made
a flux correction from 0.5--8~keV (the effective energy band in
\citealt{gilfanov04}, although the lower boundary of this range is
somewhat uncertain in that study) to 0.25--8~keV and simulated the 
appearence of LMXBs in {\sl Chandra} observations, as we did
before in deriving the SFR ($\Lunabs)$ curves and estimating the AGN
contribution to the sample. Specifically, we assumed that LMXBs are
distributed uniformly over the face-on image of a given galaxy within
$R_{\rm max}=0.6 R_{25}$ (to roughly take into account that LMXBs are
more concentrated toward the centre of the galaxy compared to HMXBs,
although the results prove to be almost insensitive to the choice of
$R_{\rm max}$) and randomly with respect to the ISM in the direction
perpendicular to the plane of the galaxy. We thus assumed that LMXBs
are subject to the same kind of X-ray absorption as HMXBs. We also 
assumed that the \cite{gilfanov04} LMXB XLF is an intrinsic rather
than observed one. This is a reasonable first-order approximation,
since most of the \cite{gilfanov04} galaxies are elliptical ones,
presumably containing very little cold gas, although his sample also
contains a few bulges of spiral galaxies (including NGC~3031 and
NGC~5457, which are also present in our sample) and the measured
fluxes of some of the \cite{gilfanov04} X-ray sources may well be
affected by absorption in the H$_2$  disks of these galaxies.

The estimated contribution of LMXBs to the $\Lunabs$ distribution for
the clean sample is shown by the dashed line in
Fig.~\ref{fig:lum_hist} (upper panel). It proves to be substantial in
the first two bins ($10^{38}$--$10^{38.5}$ and
$10^{38.5}$--$10^{39}$~erg~s$^{-1}$) but vanishes at higher
luminosities. We also estimated the LMXB contributions to the
distributions of observed 0.25--8~keV luminosities
(Fig.~\ref{fig:lum_hist}, middle panel) and intrinsic 0.25--2~keV
luminosities (Fig.~\ref{fig:lum_hist}, bottom panel).

\subsection{Robustness of the sample}
\label{s:sample_test}

As mentioned before, selection of a suitable spectral model for a
given source in the presence of line-of-sight absorption is an
unreliable procedure, which may lead to significant systematic
uncertainties in estimating unobsorbed source fluxes.

To evaluate this uncertainty, we constucted an alternative sample of
sources as follows. We retained the previous algorithm for very soft
sources with $\Gamma>4$ (i.e. selected the better of the BB and DISKBB
models based on $C$-statistics, see Section~\ref{s:spectral_models}
above), while for the harder sources ($\Gamma<4$) we introduced the 
logarithmic mean between the unabsorbed luminosities for the PL and BB
models, $\Lmean=10^{0.5*[\log\Lunabs({\rm PL})+\log\Lunabs({\rm BB})]}$, as a
replacement for the $\Lunabs$ corresponding to the statistically best
spectral model (PL, BB or DISKBB). Note that usually, for a given
source, $\Lunabs({\rm PL})>\Lunabs({\rm DISKBB})>\Lunabs({\rm BB})$. 
We thus obtained a sample of sources with
$\Lmean>10^{38}$~erg~s$^{-1}$, from which we excluded likely
associations with stars and AGN (including dropouts on the $\NH$
vs. $\NHlos$ diagram), as we did before. 

The properties of the alternative sample proved to be not very 
different from the clean sample. The total number of sources in the
new sample is 214 vs. 200. The number of sources with luminosity
higher than $10^{39}$ and $10^{40}$~erg~s$^{-1}$ is 46 and 5 vs. 42
and 7. The number of hard, soft and supersoft sources is 128, 55 and
31 vs. 119, 50 and 31.

This test gives us some confidence that the XLF derived below
using the clean sample is not strongly affected by the systematic
uncertainty in determining both the line-of-sight absorption
column and intrinsic source spectral shape from X-ray spectra with
limited photon statistics.

\subsection{Impact of variability}
\label{s:variable}

As was explained in Section~\ref{s:raw_sample}, our spectral analysis
is based on the {\sl Chandra}/ACIS observation providing the most
total counts for a given source. Since some of the studied galaxies
have been observed by {\sl Chandra} more than once, our reported
measurements of the fluxes of the sources from the same galaxy may
represent different observations, and the sum of these fluxes may be
biased with respect to the instantaneous total flux of the HMXBs in that
galaxy.  
 
\begin{figure}
\centering
\includegraphics[width=\columnwidth,viewport=30 200 560 710]{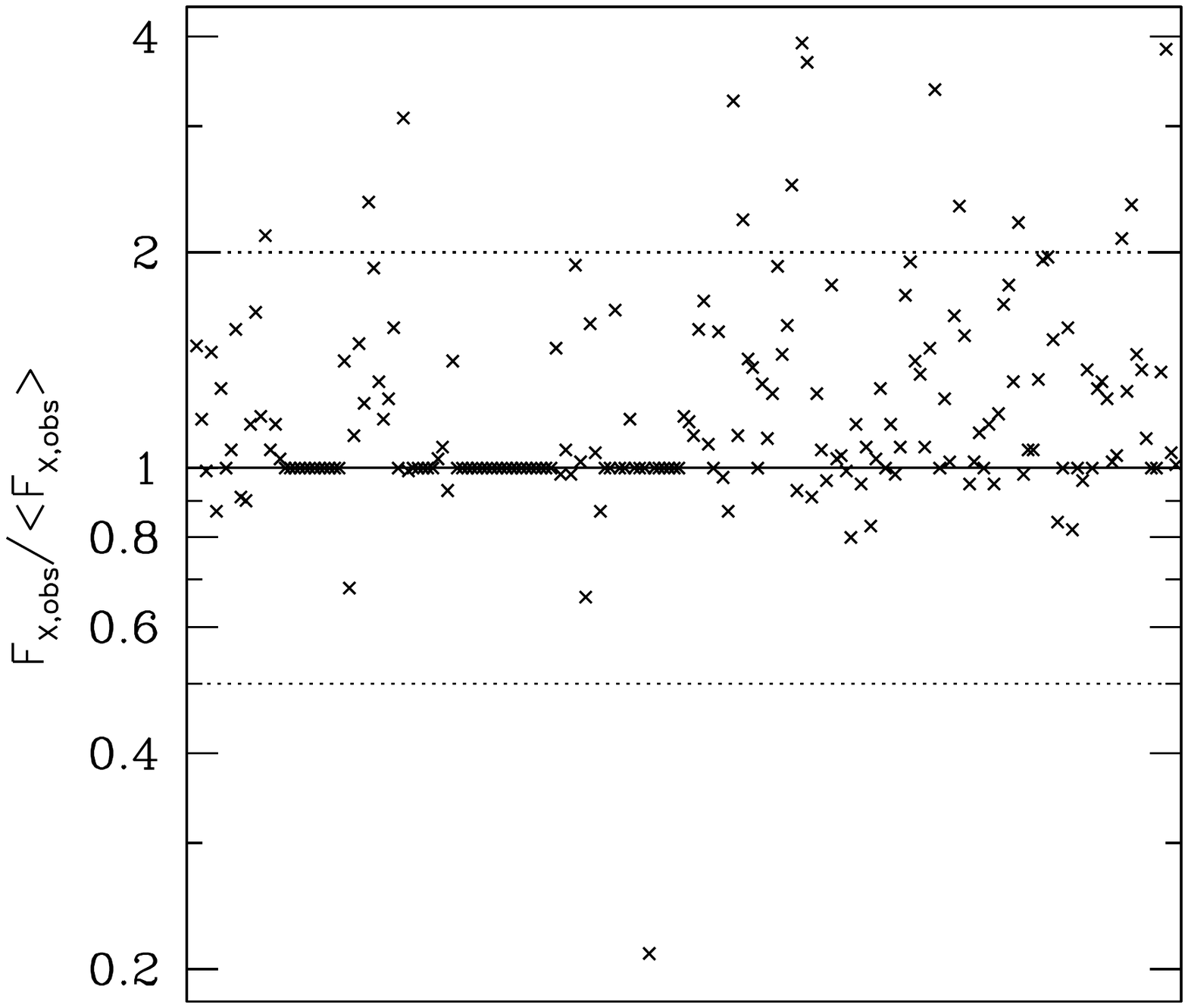}
\caption{Ratio of the X-ray flux measured in the observation actually
  used in our analysis over that averaged between all available {\sl
    Chandra} observations for a given source, for the clean sample.
}
\label{fig:variability}
\end{figure}

We estimated this effect as follows. For each of the 200 sources
in the clean sample, we compared the mean of its observed 0.25--8~keV
fluxes (derived from fitting by an absorbed PL model), $\Fobsmean$, in
all available observations with the observed flux, $\Fobs$, measured
in the observation that was actually used in our
analysis. Figure~\ref{fig:variability} shows the resulting
distribution of $\Fobs/\Fobsmean$ ratios. We see that
$0.2<\Fobs/\Fobsmean<4$ for all of the sources. The mean value of this
ratio is 1.26. Of the total sample, 62 sources have only one archival
{\sl Chandra} observation and so have $\Fobs/\Fobsmean=1$, while for
the remaining 138 ones the mean and median values of $\Fobs/\Fobsmean$
are 1.41 and 1.18, respectively.

We conclude that our selection procedure (motivated by the goal of 
achieving the best quality of spectral modelling) led to an average
flux increase of $\sim 25$\% for the studied sources compared to an
ideal situation where we could use the same observation for all
sources in a given galaxy. The relative smallness of the effect
suggests that we often selected the longest of the available
observations for a given source rather than one where its flux 
(i.e. count rate) was highest.

Since the HMXB XLF derived below is fairly flat (has a slope $\sim
0.6$), the ratio $\Fobs/\Fobsmean\sim 1.25$ implies that we probably
overestimated the XLF normalisation by $\sim 20$\%. We have thus
corrected the resulting XLFs by this factor (i.e. divided them by 1.2).
 
\section{X-ray luminosity function} 
\label{s:lumfunc}

\begin{figure}
\centering
\includegraphics[width=0.9\columnwidth,viewport=30 180 560 710]{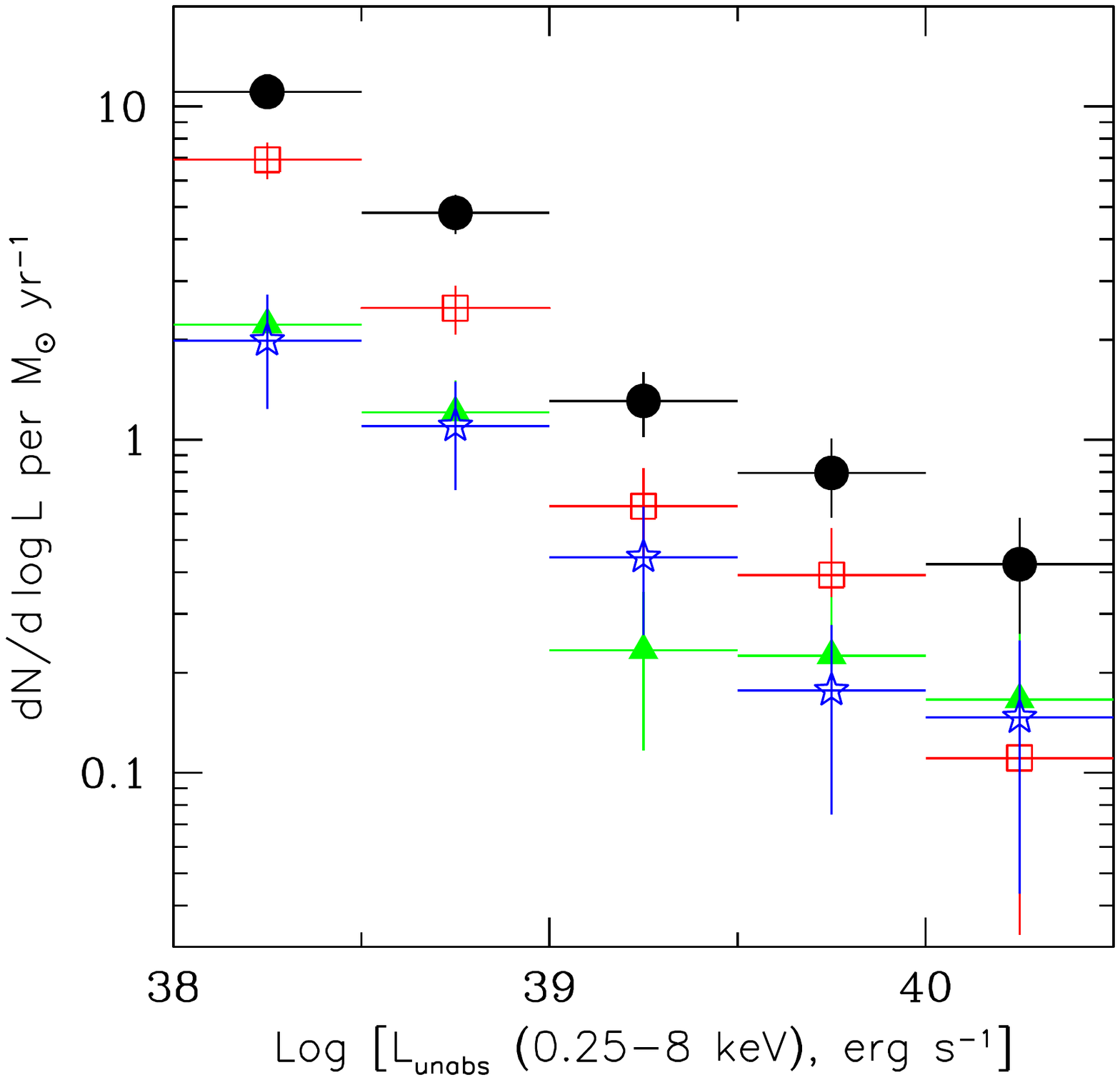}
\includegraphics[width=0.9\columnwidth,viewport=30 180 560 710]{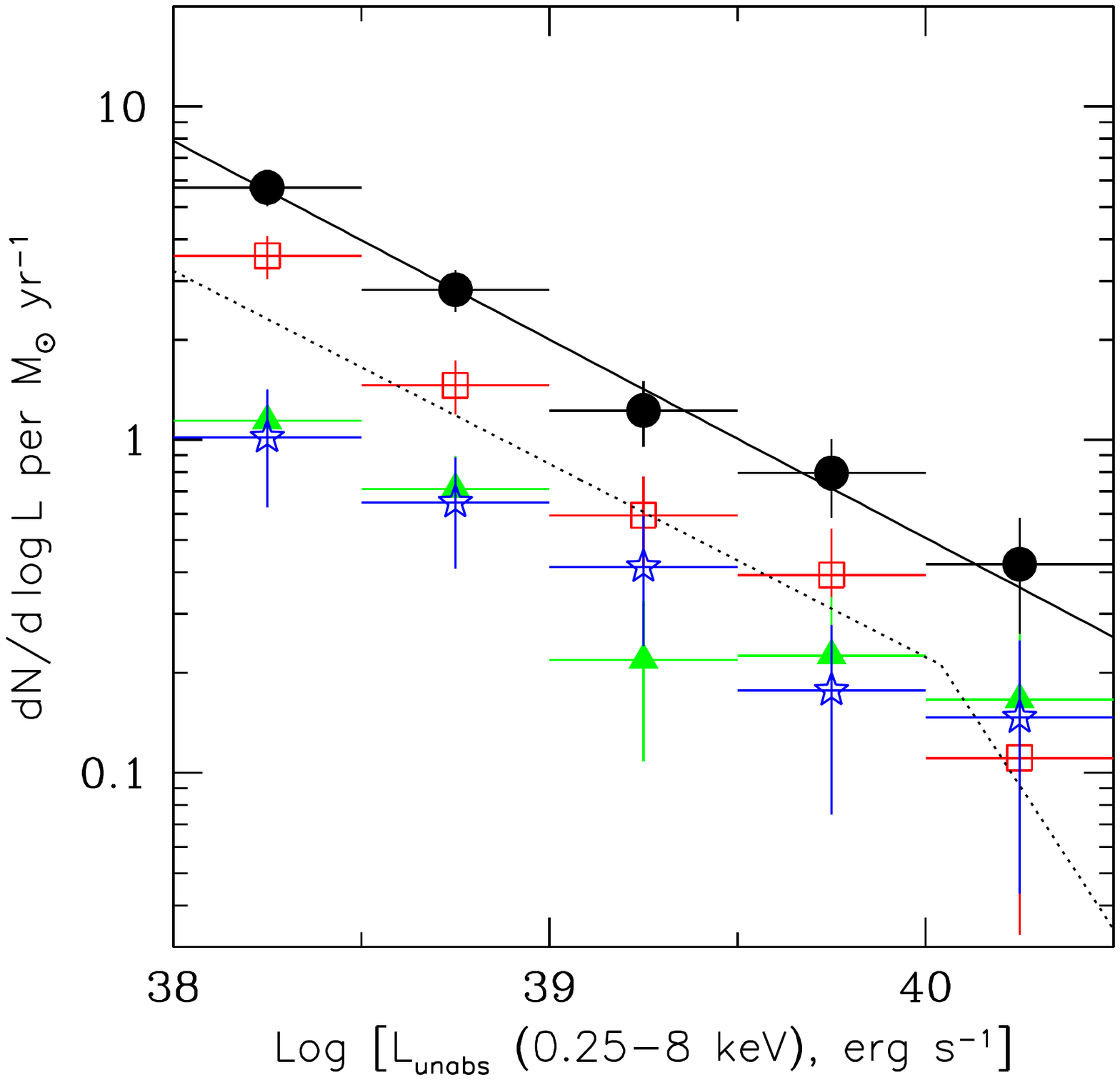}
\caption{{\sl Top panel:} Intrinsic HMXB luminosity function in the
  0.25--8~keV energy band corrected for variability bias but
  uncorrected for the contribution of LMXBs. Filled circles (with
  error bars) show the total XLF, while red squares, green triangles
  and blue stars show the XLFs of hard, soft and supersoft sources,
  respectively. {\sl Bottom panel:} The same, but with the expected
  LMXB contribution subtracted. The solid line shows the best-fit
  power-law model with the parameters given in
  Table~\ref{tab:lumfunc}. The dotted line is the fitting formula for
  the {\sl observed} HMXB XLF from \citep{minetal12}.
}
\label{fig:lumfunc}
\end{figure}

Using the clean sample of sources obtained in Section~\ref{s:clean}
and the SFR ($\Lunabs$) dependencies for different spectral types
computed in Section~\ref{s:sfr_lum}, we constructed the {\sl
  intrinsic} XLF of HMXBs at $\Lunabs>10^{38}$~erg~s$^{-1}$ per unit
SFR. Specifically, we calculated the XLF in binned form via the formula 
\beq
\frac{dN}{d\log\Lunabs}=\sum_i\frac{1}{{\rm SFR~}(\Lunabsi,\Speci)},
\label{eq:lumfunc}
\eeq
where the summation is performed over sources $i$ falling into a
given $\Lunabs$ bin, $\Lunabsi$ and $\Speci$ are the source's
luminosity and spectral type (for example a PL with $\Gamma=2.5$),
respectively, and SFR ($\Lunabsi,\Speci$) is read off from the
corresponding SFR ($\Lunabs,\Spec$) curves (computed on a fine mesh of
parameter values ($\Gamma$, $\Tbb$ and $\Tin$). Using
equation~(\ref{eq:lumfunc}), we also calculated the XLFs of hard, soft
and supersoft sources (as defined before in
equation~(\ref{eq:types})), by performing the summation over sources
of the spectral type of interest. Note that the algorithm used here is
analogous to the $1/V_{\rm max}$ method commonly used in astronomy,
with the SFR replacing $V_{\rm max}$. Figure~\ref{fig:lumfunc} (upper
panel) shows the resulting intrinsic XLF (0.25--8~keV) and its
composition in hard, soft and supersoft sources.

We next subtracted from the derived XLFs the expected contribution of
LMXBs. To this end, we assumed that the proportion of hard, soft and
supersoft sources among the LMXBs hidden in the clean sample is the
same as for the HMXBs in a given luminosity bin. We use this simple
approach because of the substantial uncertainty associated with the
LMXB XLF and its composition (see Section~\ref{s:lmxb}). Thorough
investigation of this problem is beyond the scope of this paper. As a
result, our estimates of the number density of HMXBs in the lowest two 
luminosity bins of $10^{38}$--$10^{38.5}$~erg~s$^{-1}$ and
$10^{38.5}$--$10^{39}$~erg~s$^{-1}$ (where the estimated fraction of
LMXBs among our sources is $\sim 40$--50\% and $\sim 30$--40\%,
respectively, see Fig.~\ref{fig:lum_hist}) and the corresponding
contributions of hard, soft and supersoft sources should be regarded
with caution. At higher luminosities, LMXB contamination is
negligible. 

The lower panel of Fig.~\ref{fig:lumfunc} shows the intrinsic XLF,
as well as its hard, soft and supersoft components, upon subtraction
of the expected LMXB contribution and correction for the variability
bias evaluated in Section~\ref{s:variable}. The uncertainties in
Fig.~\ref{fig:lumfunc} combine those associated with application of
equation~(\ref{eq:lumfunc}) ($\sqrt{\sum_i {\rm
    SFR}^{-2}(\Lunabsi,\Speci)}$) and the Poisson uncertainties
associated with the subtraction of LMXBs.

We derived intrinsic XLF is well fit by a simple power law with a
slope of $0.60\pm 0.07$ and normalisation given in
Table~\ref{tab:lumfunc}). Furthermore, the intrinsic XLFs of hard,
soft and supersoft sources are also well fit by power laws with the
same (within $\sim 1.5\sigma$) slope, but with larger associated
uncertainties. We can thus use the normalising constants given at
$\Lunabs=10^{39}$~erg~s$^{-1}$ in Table~\ref{tab:lumfunc} to estimate
the relative contributions of hard, soft and supersoft sources to the
combined XLF, which prove to be $\sim 50$\%, $\sim 25$\% and $\sim
25$\%, respectively. Hence, soft and supersoft sources constitute
together about half of all sources over the
$10^{38}$--$10^{40.5}$~erg~s$^{-1}$ luminosity range. 

\begin{table}
\begin{center}
  \caption{Power-law fits of different XLFs (corrected for variability
    bias and LMXB contamination):
    $dN/d\log L=A(L/10^{39}~{\rm erg~s}^{-1})^{-\gamma}$
\label{tab:lumfunc}
}

\begin{tabular}{l|r|c|c}
\hline
\multicolumn{1}{c}{Sample} &
\multicolumn{1}{c}{$N_{\rm src}$$^{\rm a}$ } &
\multicolumn{1}{c}{$\gamma$$^{\rm b}$} & 
\multicolumn{1}{c}{$A$$^{\rm b}$ } \\ 

& & & $(M_\odot$~yr$^{-1})^{-1}$\\

\hline
\multicolumn{4}{c}{Intrinsic XLF, 0.25--8~keV} \\[1.5mm]
All & 200 & $0.60\pm0.07$ & $2.00\pm0.18$ \\[1.5mm]
Hard & 119 & $0.73\pm0.09$ & $0.99\pm0.13$ \\[1.5mm]
Soft & 50 & $0.53\pm0.14$ & $0.46\pm0.08$ \\[1.5mm]
Supersoft & 31 & $0.46\pm0.14$ & $0.48\pm0.10$ \\
\multicolumn{4}{c}{Observed XLF, 0.25--8~keV} \\[1.5mm]
All & 173 & $0.73\pm0.07$ & $1.50\pm0.15$ \\[1.5mm]
Hard & 113 & $0.77\pm0.10$ & $0.98\pm0.13$ \\[1.5mm]
Soft & 45 & $0.70\pm0.16$ & $0.36\pm0.08$ \\[1.5mm]
Supersoft & 15 & $0.39\pm0.28$ & $0.18\pm0.07$ \\
\multicolumn{4}{c}{Intrinsic XLF, 0.25--2~keV} \\[1.5mm]
All & 147 & $0.63\pm0.08$ & $1.36\pm0.15$ \\[1.5mm]
Hard & 68 & $0.68\pm0.14$ & $0.51\pm0.10$ \\[1.5mm]
Soft & 48 & $0.68\pm0.19$ & $0.37\pm0.08$ \\[1.5mm]
Supersoft & 31 & $0.67\pm0.15$ & $0.38\pm0.09$ \\
\hline

\end{tabular}

\end{center}

$^{\rm a}$ Number of objects with $L>10^{38}$~erg~s$^{-1}$ in the sample.  

$^{\rm b}$ The best-fit value and $1\sigma$ uncertainty.

\end{table}

\subsection{Observed HMXB XLF} 
\label{s:obs_lumfunc}

For comparison, we also computed the {\sl observed} HMXB XLF. To this end,
we slightly modified equation~(\ref{eq:lumfunc}) as follows:  
\beq
\frac{dN}{d\log\Lobs}=\sum_i\frac{1}{{\rm SFR}(\Lunabsi,\Speci)},
\label{eq:obs_lumfunc}
\eeq
where $\Lobs$ is the observed luminosity in the 0.25--8~keV energy
band. The difference from the intrinsic XLF is that the summation is
now done over sources whose observed luminosities $\Lobsi$ fall into a
given $\Lobs$ bin. The rest of the calculation is unchanged, i.e. the
SFR entering the denominator on the right-hand side of
equation~(\ref{eq:obs_lumfunc}) is derived as a function of the
source's intrinsic luminosity, as before.

\begin{figure}
\centering
\includegraphics[width=\columnwidth,viewport=30 200 560 710]{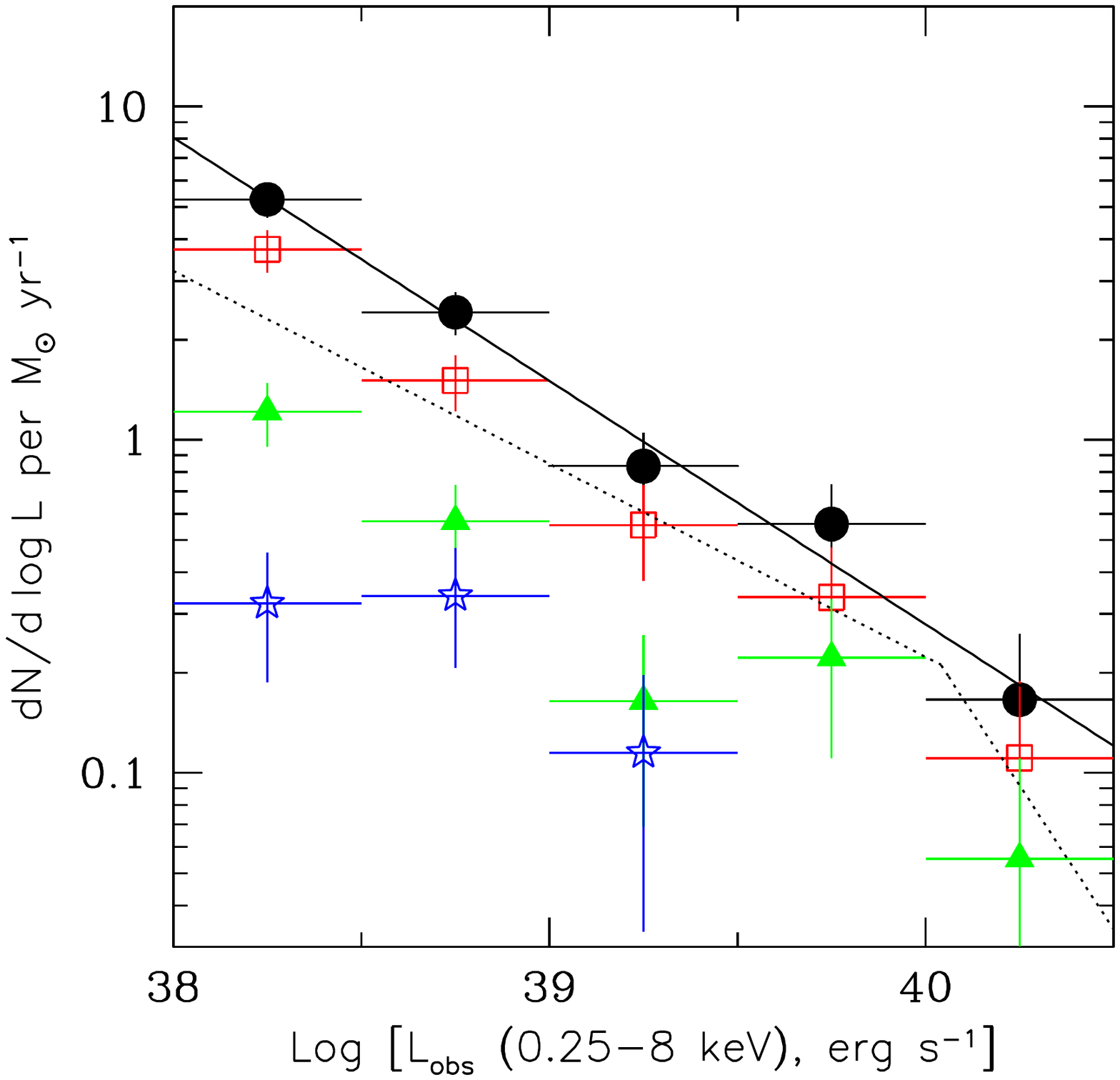}
\caption{Observed HMXB luminosity function in the 0.25--8~keV energy
  band with the LMXB contribution subtracted and corrected for
  variability bias, and its best fit by a power-law model (solid
  line). The dotted line is the fitting formula for the {\sl observed}
  HMXB XLF from \citep{minetal12}. The symbols as in
  Fig.~\ref{fig:lumfunc}.
}
\label{fig:lumfunc_obs}
\end{figure}

The resulting observed XLF, with the LMXB contribution subtracted and
corrected for variability bias, is shown in
Fig.~\ref{fig:lumfunc_obs}. The observed XLF and its hard, soft and
supersoft components are all well fit by power laws, with the slopes
consistent (within $\sim 1.5\sigma$) with the slope of the intrinsic
XLF (see Table~\ref{tab:lumfunc}). Importantly, the intrinsic number
density of HMXBs proves to be higher than their observed density by a
factor of $\sim 1.5$ ($\sim 2$) at $10^{39}$ ($10^{40}$)~erg~s$^{-1}$,
mostly due the contribution of soft and supersoft sources, and this
result is robust since LMXBs do not contaminate the HMXB sample at
these high luminosities.

\subsection{Comparison with previous studies}
\label{s:comparison}

It is interesting to compare the XLFs derived here with the
observed (by construction) HMXB XLF from \citep{minetal12}. Compared
to our study, their XLF was obtained using a larger sample
($\sim 1000$) of X-ray sources detected over a larger volume of the
local Universe, $D\lesssim 40$~Mpc, but without detailed information
on their spectra. Figures~\ref{fig:lumfunc} and \ref{fig:lumfunc_obs}
show the fitting dependence from \citep{minetal12} (their
equation~(18) and text below it), which is a broken power law with
slopes (in $dN/d\log \Lx$ representation) of $0.58\pm0.02$ and
$1.7^{+1.6}_{-0.5}$ below and above $\sim 1.1\times
10^{40}$~erg~s$^{-1}$, respectively.

\cite{minetal12} note that the break is only marginally detected in
their XLF and also provide a single power-law fit, which can be recast
as follows: $dN/d\log L=(0.86\pm0.06)(L/10^{39}\,{\rm
  erg~s}^{-1})^{-0.60\pm0.02}$. For comparison, our intrinsic XLF can
be described as (see Table~\ref{tab:lumfunc}): $dN/d\log
L=(2.00\pm0.18)(L/10^{39}\,{\rm erg~s}^{-1})^{-0.60\pm0.07}$, i.e. it
has the same slope, but its amplitude is $2.3\pm 0.2$ times larger,
which is a highly statistically significant ($\sim 6\sigma$)
difference. Our observed HMXB can be described as: $dN/d\log
L=(1.50\pm0.15)(L/10^{39}\,{\rm erg~s}^{-1})^{-0.73\pm0.07}$. Its
slope is thus marginally different from that of the \cite{minetal12}
XLF ($0.73\pm 0.07$ vs. $0.60\pm0.02$), whereas the factor of $\sim
1.7$ difference in the amplitudes is significant ($\sim 4\sigma$). We
can also calculate $\chi^2$ for our XLFs with respect to the best-fit
model of \cite{minetal12} (neglecting parameter uncertainties for the
latter). This yields $\chi^2=64$ and $\chi^2=47$ (for 5 luminosity
bins) for our intrinsic and observed XLF, respectively, which implies
that these XLFs are inconsistent with the \cite{minetal12} model at
$\sim 7\sigma$ and $\sim 6\sigma$, respectively.

Only a tiny part of this difference can be attributed to the somewhat
broader energy range used here (0.25--8~keV) for measuring fluxes and
luminosities, compared to that used by \citep{minetal12} (0.5--8~keV):
when doing counts-to-ergs conversion, \cite{minetal12} assumed all
sources to have absorbed PL spectra with $\Gamma=2$ and $\NH=3\times
10^{21}$, for which the difference in the observed 0.25--8~keV and
0.5--8~keV fluxes is less than 1\%. Furthermore, we have subtracted
the expected LMXB contribution from our HMXB XLF, whereas
\cite{minetal12} have not done so, arguing that the LMXB XLF derived
by \cite{gilfanov04} might overestimate the contribution of LMXBs to
the X-ray source population of actively star-forming galaxies (such as
are the galaxies in their sample) by a significant factor. If both
studies had taken a similar approach toward LMXBs, the difference in
the derived observed HMXB XLFs would have been even larger, albeit
only at $\Lobs\lesssim 10^{39}$~erg~s$^{-1}$ (at higher luminosities
the LMXB contribution is negligible).
 
Part of the explanation might be that the source sample used by
\cite{minetal12} is expected to be more biased against supersoft
sources compared to our clean sample, because it is based on source
detection in the 0.5--8~keV energy band, compared to 0.3--8~keV in the
\cite{wanetal16} catalogue used here. To estimate the resulting
effect, we calculated the expected 0.3--8 keV and 0.5--8~keV {\sl
  Chandra} count rates for sources in our mock sample (the one we used
for simulations in Section~\ref{s:abs_model}) and found that the
difference between these rates can indeed be significant for supersoft
sources (i.e. those with effective colour temperature $\Tbb\lesssim
0.25$~keV according to our definition). Namely, it is $\sim 5$--7\% for
$\Tbb=0.25$~keV and $\NH=10^{21}$~cm$^{-2}$ (the difference increases
with decreasing absorption), a factor of $\sim 2$ for $\Tbb=0.1$~keV and
$\NH=3\times10^{20}$~cm$^{-2}$ and a factor of $\sim 3$ for $\Tbb=0.05$~keV 
and $\NH=3\times 10^{21}$~cm$^{-2}$. Although it is difficult to evaluate the
resulting impact on the XLFs\footnote{Note that the [0.3--8]/[0.5--8]
count ratio also depends on the source position in the field of view
and the observation date, since the \textit{Chandra} soft band
sensitivity declined over the years.}, it is clear that the
\cite{minetal12} XLF may significantly underestimate the contribution
of supersoft sources. However, since the overall contribution of
supersoft sources to our observed XLF is $\sim 12$\% (see
Table~\ref{tab:lumfunc}) this effect is unlikely to lead to more
than a $\sim 10$\% difference between the XLFs. 

We believe that most of the difference between these observed XLFs is
due to the fact that the XLF calculation procedure used by
\cite{minetal12} is equivalent to assuming that i) all HMXBs have
intrinsic PL spectra with $\Gamma=2$ and ii) the observed fluxes of
all sources are subject to the same line-of-sight absorption
($\NH=3\times 10^{21}$). In contrast, we took into account the actual
diversity of both the intrinsic source spectra and their absorption
columns. The different assumptions about the spectral shape lead to
different counts-to-ergs conversion factors and hence different
inferred luminosities $\Lobs$. In addition, the quantity ${\rm
  SFR}(\Lunabsi,\Speci)$ (i.e. the SFR probed by {\sl Chandra}) in
equation~(\ref{eq:obs_lumfunc}) used for construction of the XLF in
our study, depends not only on the source luminosity but also on the
source spectrum; moreover, it takes into account the distribution of
the ISM within the galaxies (see \S\ref{s:abs_model} for details). In
contrast, \cite{minetal12} derived their observed XLF simply by
dividing the observed number of sources in a given $\Lobs$ bin by a
${\rm SFR} (\Lobs$) estimate obtained assuming some fiducial values of
the spectral slope ($\Gamma=2$) and absorption column ($\NH=3\times
10^{21}$).  

To summarise, the procedure of calculating the observed XLF used by
\cite{minetal12} is an approximation of the more accurate approach
used here. This approximation is certainly adequate for hard
sources but apparently leads to a significant systematic uncertainty
when dealing with the full diversity of HMXBs and their ISM
environments. Finally, we emphasise that the main goal of our study
was to obtain the intrinsic XLF, which proves to have a yet higher
amplitude compared to the observed one.

\subsection{Soft-band XLF}
\label{s:soft_xlf}

Finally, we computed the intrinsic HMXB XLF in the 0.25--2~keV band,
hereafter referred to as the soft X-ray luminosity function, SXLF. To
this end, we again modified equation~(\ref{eq:lumfunc}):
\beq
\frac{dN}{d\log\LSunabs}=\sum_i\frac{1}{{\rm SFR}(\Lunabsi,\Speci)},
\label{eq:soft_lumfunc}
\eeq
where $\LSunabs$ is the unabsorbed luminosity in the 0.25--2~keV
band.

\begin{figure}
\centering
\includegraphics[width=\columnwidth,viewport=30 200 560 710]{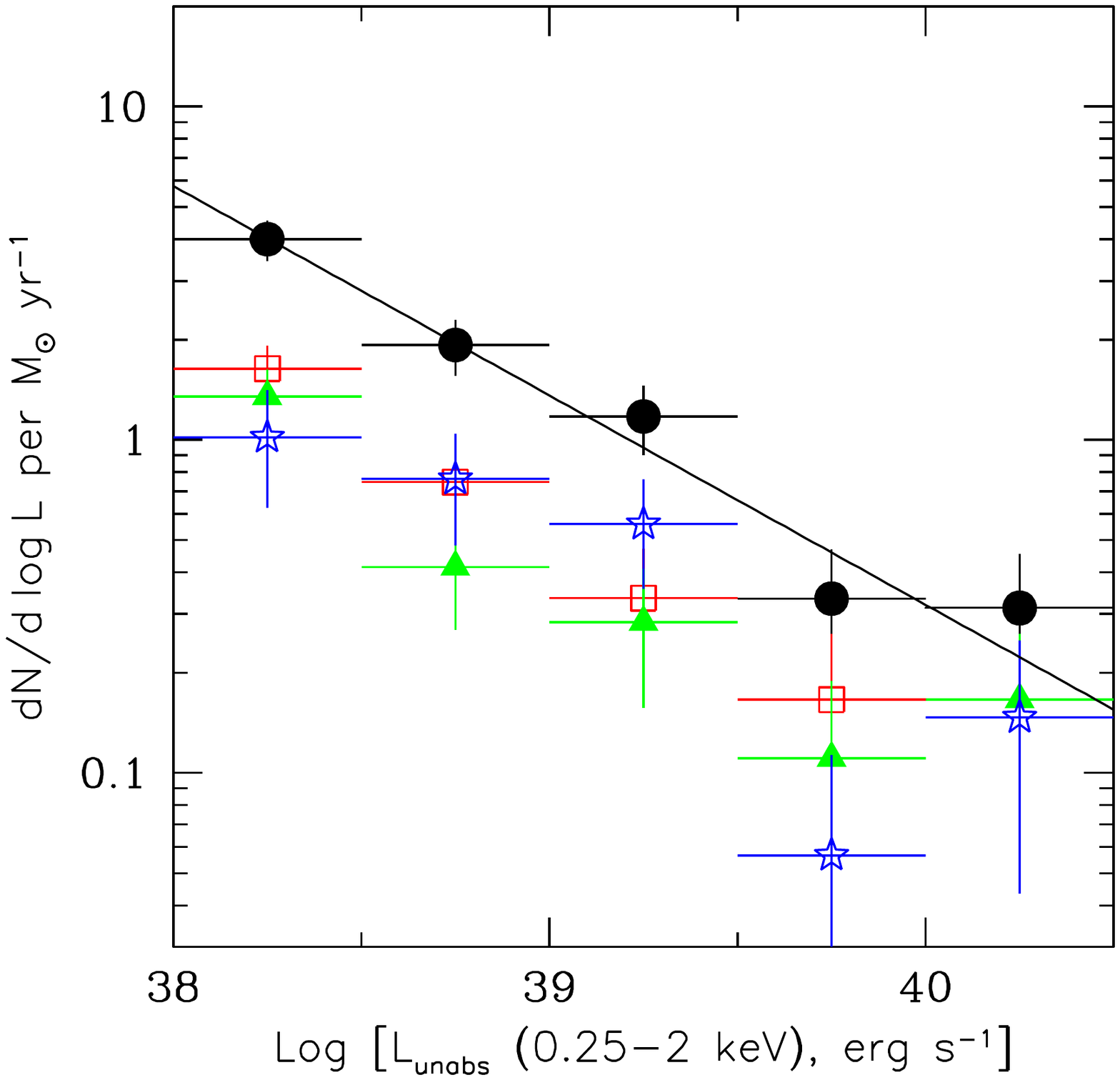}
\caption{Intrinsic HMXB luminosity function in the 0.25--2~keV energy
  band with the LMXB contribution subtracted and corrected for
  variability bias, and its best fit by a power-law model (solid
  line). The symbols as in Fig.~\ref{fig:lumfunc}.
}
\label{fig:slumfunc}
\end{figure}

The resulting SXLF is shown in Fig.~\ref{fig:slumfunc}, with the
estimated contribution of LMXBs (see Section~\ref{s:lmxb})
subtracted. The SXLF is well fit by a power law
with a slope consistent with that of the XLF in the full energy
band. The most important result, however, is that the SXLF proves to
be composed of nearly equal contributions of hard, soft and supersoft
sources (see Table~\ref{tab:lumfunc}). Note, however, that the
estimated fraction of supersoft sources in the bright end of the
SXLF, at $\LSunabs=10^{39.5}$--$10^{40.5}$~erg~s$^{-1}$ is fairly
uncertain as it is based on just 3 sources, of which the two most 
luminous ones have extremely uncertain $\LSunabs$ estimates due to the
large gas columns in their direction, as was discussed in
Section~\ref{s:sample_properties}.

\section{Conclusions}
\label{s:conc}

As was discussed in Section~\ref{s:intro}, previous studies of the
luminous HMXB population in the local Universe usually ignored the
impact of interstellar absorption on the observed statistical
properties of such sources. The purpose of this study was to evaluate 
these effects and uncover the intrinsic distribution of luminous HMXBs
in terms of their luminosity and spectral hardness. To this end, we
combined the results of i) our X-ray spectral analysis of 330 bright
{\sl Chandra} X-ray sources from the catalogue of \citep{wanetal16}
located in 27 nearby ($D<15$~Mpc) galaxies with ii) information
deduced from the detailed maps of SFR and ISM surface densities for
these galaxies. We then restricted our statistical study to objects
with inferred unabsorbed X-ray luminosities
$\Lunabs>10^{38}$~erg~s$^{-1}$ (0.25--8~keV) to minimise problems with
removal of the contributions of AGN and LMXBs (which are nevertheless
significant).

We found the distribution of absorption columns inferred from the
X-ray spectra of sources in our clean sample (200 objects) to be in
satisfactory agreemeent with the hypothesis that most of the observed
absorption is caused by the ISM in the host galaxies (and in the Milky
Way), with the vast majority of sources having $\NH\lesssim 5\times
10^{21}$~cm$^{-2}$. This gives credence to our determination of
$\Lunabs$ from the sources' observed luminosities ($\Lobs$). However,
our reported statistical uncertainties for the intrinsic luminosities
of supersoft sources probably underestimate the actual uncertainties,
since it is hardly possible to quantify the systematic uncertainty
arising from the substantial freedom in selecting the spectral model
for fitting.

The second key component of this study was determination of the total
SFR rate probed by {\sl Chandra} in the studied galaxies as a function
of source luminosity and spectral type. The resulting SFR ($\Lunabs$)
curves show strong dependence on the spectral type at $\Lunabs\lesssim
10^{40}$~erg~s$^{-1}$, which is non-monotonic: the SFR first grows
with increasing spectral softness (due to the increasing number of
photons produced per unit luminosity) and then falls (as a result of
absorption of a larger fraction of the emitted luminosity in the ISM).
 
By dividing the numbers of sources of different spectral types in
given luminosity bins by the corresponding SFR ($\Lunabs$) functions we
finally obtained the intrinsic XLF of (presumably) HMXBs
(Table~\ref{tab:lumfunc}). It is well fit by a power law
$dN/d\log\Lunabs\propto \Lunabs^{-\gamma}$ with $\gamma\approx 0.6$
over the sampled luminosity range of $10^{38}$--$10^{40.5}$. This
slope is in excellent agreement with that of the {\sl observed} HMXB
XLF constructed by \citet{minetal12} using a larger sample of sources
(albeit with fewer photon counts) in a larger volume of the local
Universe. However, the amplitude of the intrinsic XLF derived here is
a factor of $\sim 2.3$ higher.  

Therefore, our first conclusion is that the intrinsic number of
luminous HMXBs per unit SFR is a factor of $\sim 2$--2.5 larger than
appears directly from X-ray surveys. This enhancement is constant
within the uncertainties over the luminosity range from $10^{38}$ to
$10^{40}$~erg~s$^{-1}$, while there is large uncertainty associated
with the highest luminosity bin sampled here
($10^{40}$--$10^{40.5}$~erg~s$^{-1}$).

Secondly, the XLF slope of $\sim 0.6$ implies that at least half of
the total X-ray luminosity emitted by star-forming galaxies (excluding
diffuse gas emission) is produced by sources with $\Lunabs\gtrsim
10^{39}$~erg~s$^{-1}$, i.e. ULXs and ULSs.

Thirdly, the intrinsic HMXB XLF is composed of hard, soft and
supersoft sources (defined here as those with the 0.25--2~keV to
0.25--8~keV flux ratio of $<0.6$, 0.6--0.95 and $>0.95$), contributing
$\sim 50$\%, $\sim 25$\% and $\sim 25$\%, respectively. It is possible
(although we have not explicitly verified this) that our hard class
mainly consists of X-ray pulsars and black-hole binaries in their hard
spectral states and our soft class of black-hole HMXBs in their soft
states.

Most of our supersoft sources may be supercritically accreting
black-hole HMXBs viewed at a significant angle to the axis of the
thick disk, so that only wind-reprocessed emission from the central
source is observed (e.g. \citealt{urqsor16}). Classical supersoft
sources, associated with accreting white dwarfs, are unlikely to have
$\Lunabs\gtrsim 2\times 10^{38}$~erg~s$^{-1}$
(e.g. \citealt{soretal16}). If our supersoft sources are indeed
'misaligned' ULXs, then the inferred number ratio between hard, soft
and supersoft sources provides an important constraint on theoretical
models of supercritical accretors.

Finally, we derived the intrinsic XLF in the soft X-ray band
(0.25--2~keV, Table~\ref{tab:lumfunc}). As expected, the contribution
of supersoft sources is even more pronounced in the SXLF. Namely, the
numbers of hard, soft and supersoft sources are nearly equal over the
entire sampled luminosity range, although the uncertainties are too
large at $\LSunabs>10^{39.5}$~erg~s$^{-1}$ to make definitive conclusions
regarding the most luminous sources. The derived SXLF and its spectral 
composition is important (see our accompanying paper
\citealt{sazkha17}) for studying the photoionisation heating of the
early Universe by X-rays from the HMXBs in the first galaxies, since
only photons with $E\lesssim 2$~keV could efficiently deposit their
energy into the ambient intergalactic medium. The intrinsic (rather
than observed) SXLF constructed here may have direct application to
this situation, since soft X-ray photons were probably able to escape
nearly freely from the first galaxies due to the low metallicity of
the latter. By integrating the SXLF over $\LSunabs$ between $10^{38}$
and $10^{40.5}$~erg~s$^{-1}$, we find the cumulative emissivity of the
present-day HMXB population per unit star formation rate in the
0.25--2~keV energy band:
\beq
\epsilon_{\rm X}\equiv \int\frac{dN}{d\log\LSunabs}\Lx d\log\LSunabs \\
\sim 5\times 10^{39}~{\rm erg~s}^{-1}~(M_\odot~{\rm yr}^{-1})^{-1}. 
\label{eq:exnow}
\eeq

\section*{Acknowledgments}

We used {\sc topcat} \citep{taylor05} for table manipulation and
cross-correlation analysis. Image manipulation was performed partially
by {\sc montage}\footnote{http://montage.ipac.caltech.edu}, which is
funded by the National Science Foundation under Grant Number
ACI-1440620, and was previously funded by the National Aeronautics and
Space Administration's Earth Science Technology Office, Computation
Technologies Project, under Cooperative Agreement Number NCC5-626
between NASA and the California Institute of
Technology. \textit{Chandra} data retrieval, preparation and
processing were performed by standard tools of the {\sc ciao} 4.8
package \citep{fruetal06} with {\sc caldb} version 4.7.1. This
research made use of the SIMBAD database, operated at CDS, Strasbourg,
France, and the NASA/IPAC Extragalactic Database (NED), operated by
the Jet Propulsion Laboratory, California Institute of Technology,
under contract with the National Aeronautics and Space
Administration. The research was supported by the Russian Science
Foundation (grant 14-12-01315). We thank the referee for useful
suggestions. 

\begin{appendix}

\section{Stars and AGN among the X-ray sources}
\label{s:stars_agn}

We found 3 possible associations of our $\Lunabs\gtrsim
10^{38}$~erg~s$^{-1}$ sources with bright stars from the Tycho-2
catalogue (\citealt{hogetal00}, Table~\ref{tab:stars}). 

\begin{table}

\caption{Likely X-ray source--star associations
  \label{tab:stars}
}

\begin{tabular}{|l|l|r|r|}

\hline
  \multicolumn{1}{|c|}{CXOGSG} &
  \multicolumn{1}{c|}{Star} &
  \multicolumn{1}{c|}{$B_{\rm t}$} &
  \multicolumn{1}{c|}{$V_{\rm t}$} \\
\hline
  J131530.1+420313 & TYC~3024-814-1 & 9.91 & 9.362\\
  J140421.7+541921 & TYC~3852-1069-1$^{\rm a}$  & 12.663 & 12.466\\
  J203448.8+600554 & TYC~4246-779-1 & 12.184 & 11.624\\
  \hline
\end{tabular}

$^{\rm a}$ This star has a documented X-ray flare \citep{pyeetal15}. 

\end{table}

We also found 8 possible associations with AGN
(Table~\ref{tab:agn}). Seven of these may be considered reliable: for
6 objects, there is a spectroscopic redshift measurement by the Sloan
Digital Sky Survey (SDSS), and one has an SDSS-based photometric $z$
estimate. The 8th object, CXOGSG~J093205.3+213235, is also likely an
AGN, since it is relatively bright ($u=19.3$, $g=19.0$, $r=18.9$, 
$i=19.1$, $z=18.9$) and designated as an extended object ('galaxy') in
the SDSS source catalogue, being very similar in these properties to 
CXOGSG~J112019.6+130320 ($z_{\rm spec}=0.314$).

\begin{table}

\caption{Likely X-ray source--AGN associations
  \label{tab:agn}
}

\begin{tabular}{|l|l|l|l|}
\hline
  \multicolumn{1}{|c|}{CXOGSG} &
  \multicolumn{1}{c|}{SDSS} &
  \multicolumn{1}{c|}{$z$} &
  \multicolumn{1}{c|}{Ref.} \\
\hline
  J093205.3+213235 & J093205.36+213234.7 & \multicolumn{1}{c|}{?} & 1\\
  J095636.4+690028 & J095636.42+690028.4 & $z_{\rm spec}=1.975$ & 2\\
  J112019.6+130320 & J112019.62+130320.1 & $z_{\rm spec}=0.314$ & 2\\
  J131543.8+415910 & J131543.86+415910.6 & $z_{\rm spec}=1.368$ & 3\\
  J140232.5+542001 & J140232.52+542001.3 & $z_{\rm spec}=0.830$ & 3\\
  J140247.0+542655 & J140247.03+542654.9 & $z_{\rm spec}=2.746$ & 3\\
  J140346.1+541615 & J140346.15+541615.7 & $z_{\rm spec}=1.893$ & 3\\
  J140350.4+542413 & J140350.46+542413.5 & $z_{\rm phot}=2.4$   & 4\\
  \hline
\end{tabular}

\textbf{References:} (1) Association suggested based on SDSS
photometry \citep{abaetal09}; (2) \cite{schetal10}; (3) SDSS Data
Release 12, the Million Quasars Catalog
(http://heasarc.gsfc.nasa.gov/w3browse/all/milliquas.html,
\citealt{flesch15}; (4) \cite{ricetal09}.

\end{table}

Finally, we suspect another 8 sources to be AGN based on their high
$\NH/\NHlos$ ratios (see
Section~\ref{s:nhobs_nhmodel}). Table~\ref{tab:suspagn} provides
information on these objects.

\begin{table*}
  \caption{Possible X-ray absorbed AGN 
    \label{tab:suspagn}
  }
  \begin{tabular}{|l|r|l|r|r|c|r|c|c|}
\hline
  \multicolumn{1}{|c|}{Source} &
  \multicolumn{1}{c|}{{\sl Chandra}} &
  \multicolumn{1}{c|}{Galaxy} &
  \multicolumn{1}{c|}{$R/R_{25}$} &
  \multicolumn{1}{c|}{$\NHlos$} &
  \multicolumn{1}{c|}{Model} &
  \multicolumn{1}{c|}{$\NH$} &
  \multicolumn{1}{c|}{$\Gamma$ ($\Tbb$, $\Tin$)} &
  \multicolumn{1}{c|}{$\Funabs$} \\
\multicolumn{1}{c}{CXOGSG} & \multicolumn{1}{c}{obs.} & & & $10^{20}$~cm$^{-2}$ & & 
$10^{20}$~cm$^{-2}$ & (keV) & $10^{-13}$~erg~s$^{-1}$~cm$^{-2}$\\[1.5mm]
\hline
  J132942.2+471447 & 13813 & NGC 5194 & 0.97 & 5.7 & PL &
  $65.5_{50.0}^{82.7}$ & $2.12_{1.82}^{2.43}$ & $0.25_{0.20}^{0.34}$ \\[1.5mm]
  J132954.2+471300 & 13813 & NGC 5194 & 0.35 & 37.9 & PL &
  $126.9_{100.2}^{156.3}$ & $1.50_{1.25}^{1.75}$ &
  $0.54_{0.47}^{0.64}$ \\[1.5mm]
  J133007.8+471245 & 13814 & NGC 5194 & 0.95 & 27.8 & PL &
  $99.8_{70.2}^{135.0}$ & $2.79_{2.28}^{3.38}$ & $0.33_{0.18}^{0.82}$
  \\[1.5mm]
  J140248.1+541350 & 14341 & NGC 5457 & 0.66 & 18.7 & PL &
  $815.7_{687.9}^{958.1}$ & $2.04_{1.64}^{2.46}$ &
  $13.10_{8.22}^{25.39}$ \\[1.5mm]
  J140315.8+541748 & 4736 & NGC 5457 & 0.27 & 13.4 & PL &
  $116.3_{93.1}^{141.5}$ & $2.13_{1.83}^{2.46}$ & $1.14_{0.88}^{1.64}$
  \\[1.5mm]
  J140320.4+541632 & 934 & NGC 5457 & 0.39 & 15.5 & PL &
  $224.3_{155.5}^{307.4}$ & $3.38_{2.47}^{4.53}$ &
  $2.11_{0.52}^{18.88}$ \\[1.5mm]
  J140353.6+541559 & 6114 & NGC 5457 & 0.68 & 9.7 & DISKBB &
  $73.7_{54.0}^{96.1}$ & $1.11_{0.90}^{1.40}$ & $0.31_{0.27}^{0.37}$
  \\[1.5mm]
  J140405.9+541602 & 934 & NGC 5457 & 0.81 & 8.8 & BB &
  $59.1_{25.3}^{97.5}$ & $1.07_{0.93}^{1.25}$ & $0.25_{0.22}^{0.29}$
  \\[1.5mm]
  \hline\end{tabular}
\end{table*}

\section{Atlas of maps}
\label{s:maps}

\begin{figure*}
\centering
\includegraphics[width=0.75\textwidth]{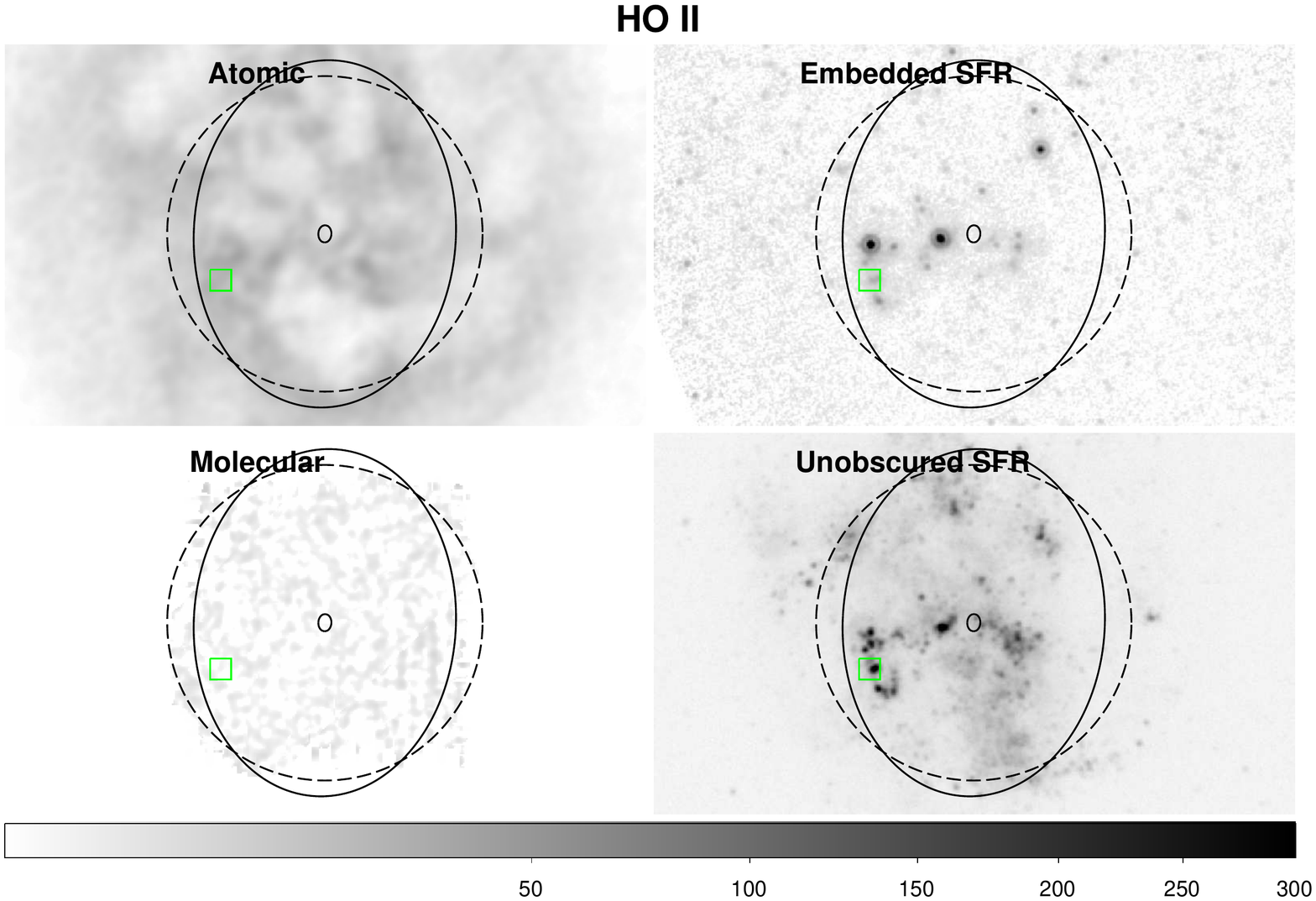}
\caption{
}
\label{fig:HOII_maps}
\end{figure*}

\begin{figure*}
\centering
\includegraphics[width=0.75\textwidth]{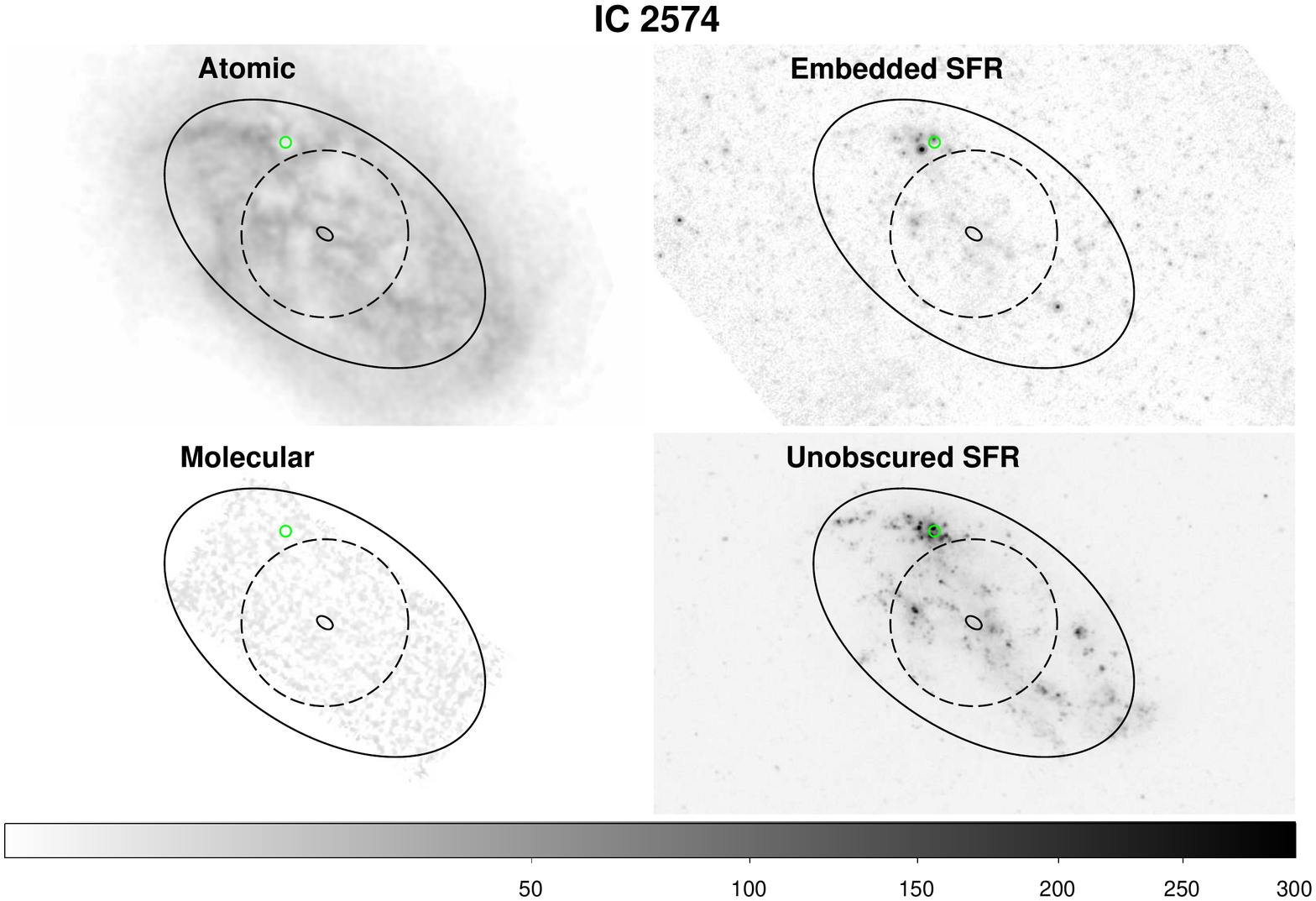}
\caption{
}
\label{fig:IC2574_maps}
\end{figure*}

\begin{figure*}
\centering
\includegraphics[width=0.75\textwidth]{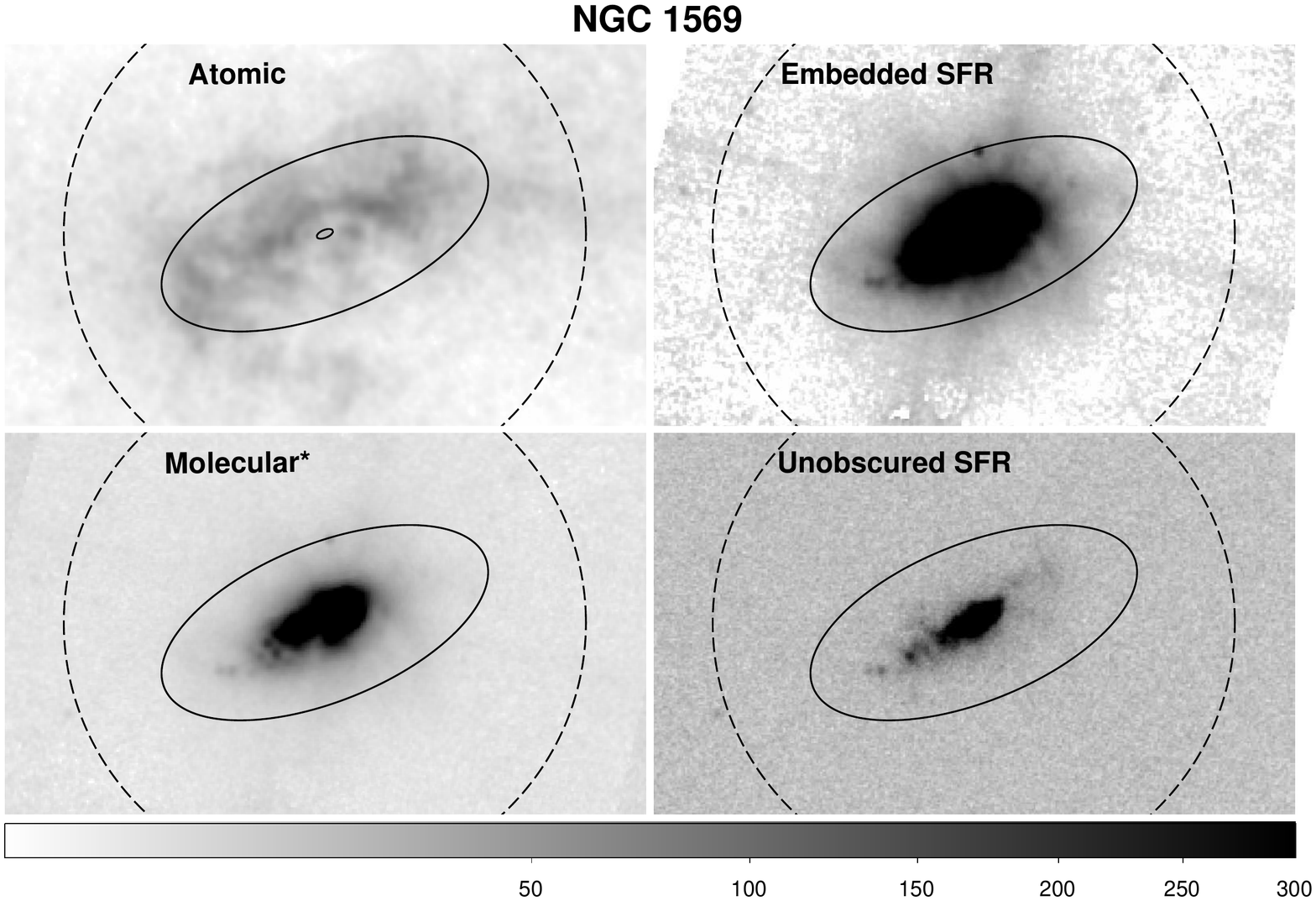}
\caption{
}
\label{fig:NGC1569_maps}
\end{figure*}

\begin{figure*}
\centering
\includegraphics[width=0.75\textwidth]{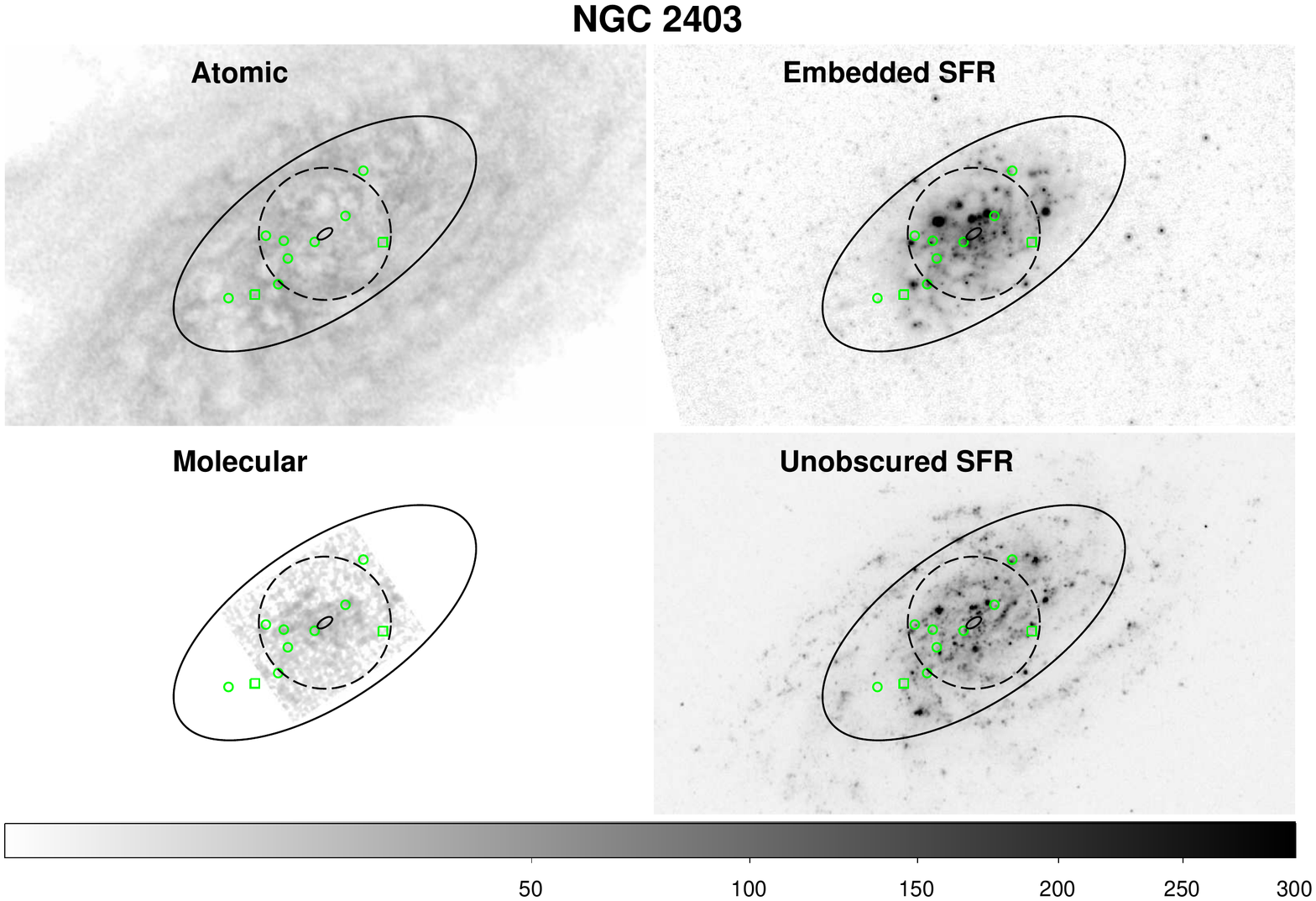}
\caption{
}
\label{fig:NGC2403_maps}
\end{figure*}

\begin{figure*}
\centering
\includegraphics[width=0.75\textwidth]{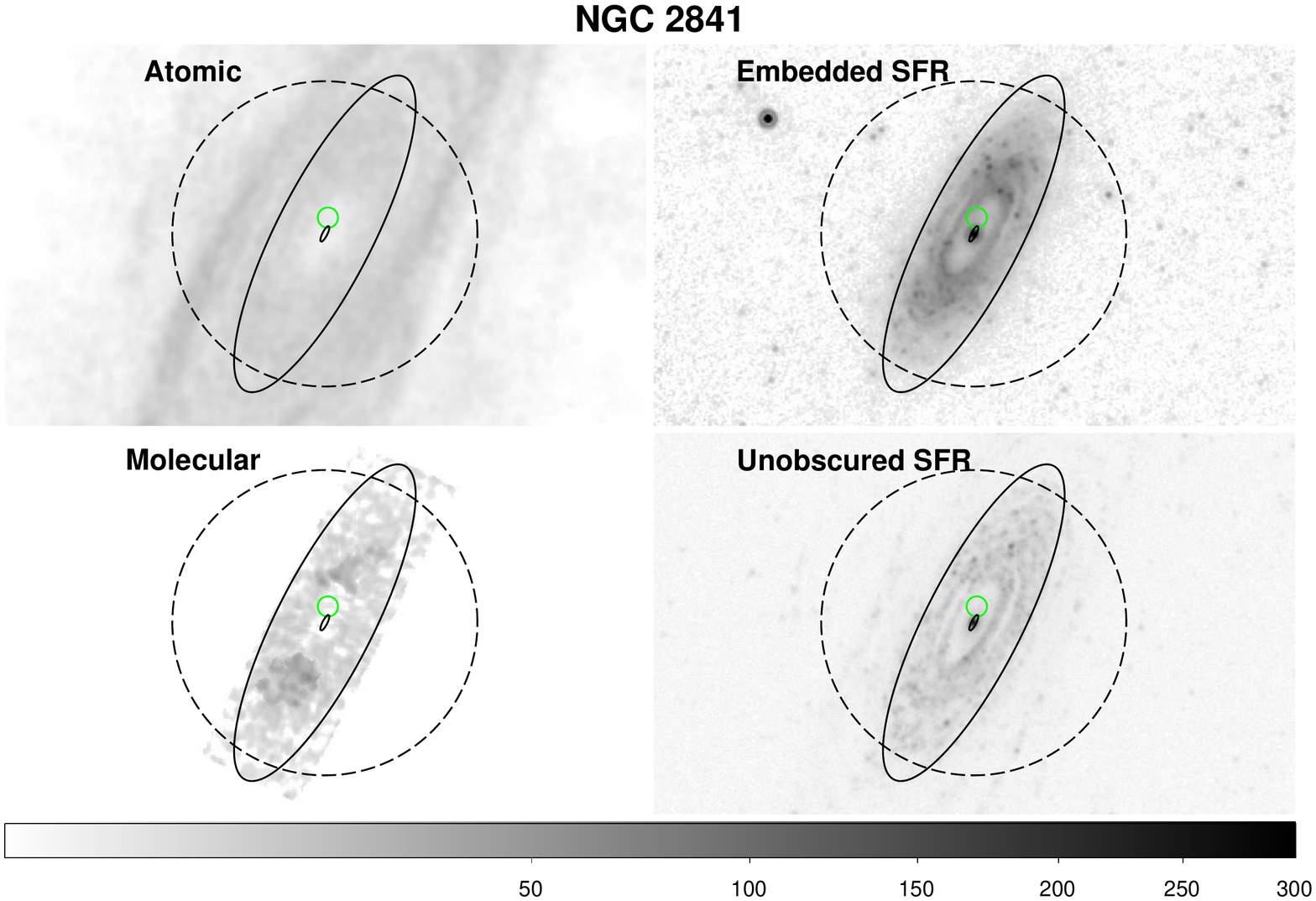}
\caption{
}
\label{fig:NGC2841_maps}
\end{figure*}

\begin{figure*}
\centering
\includegraphics[width=0.75\textwidth]{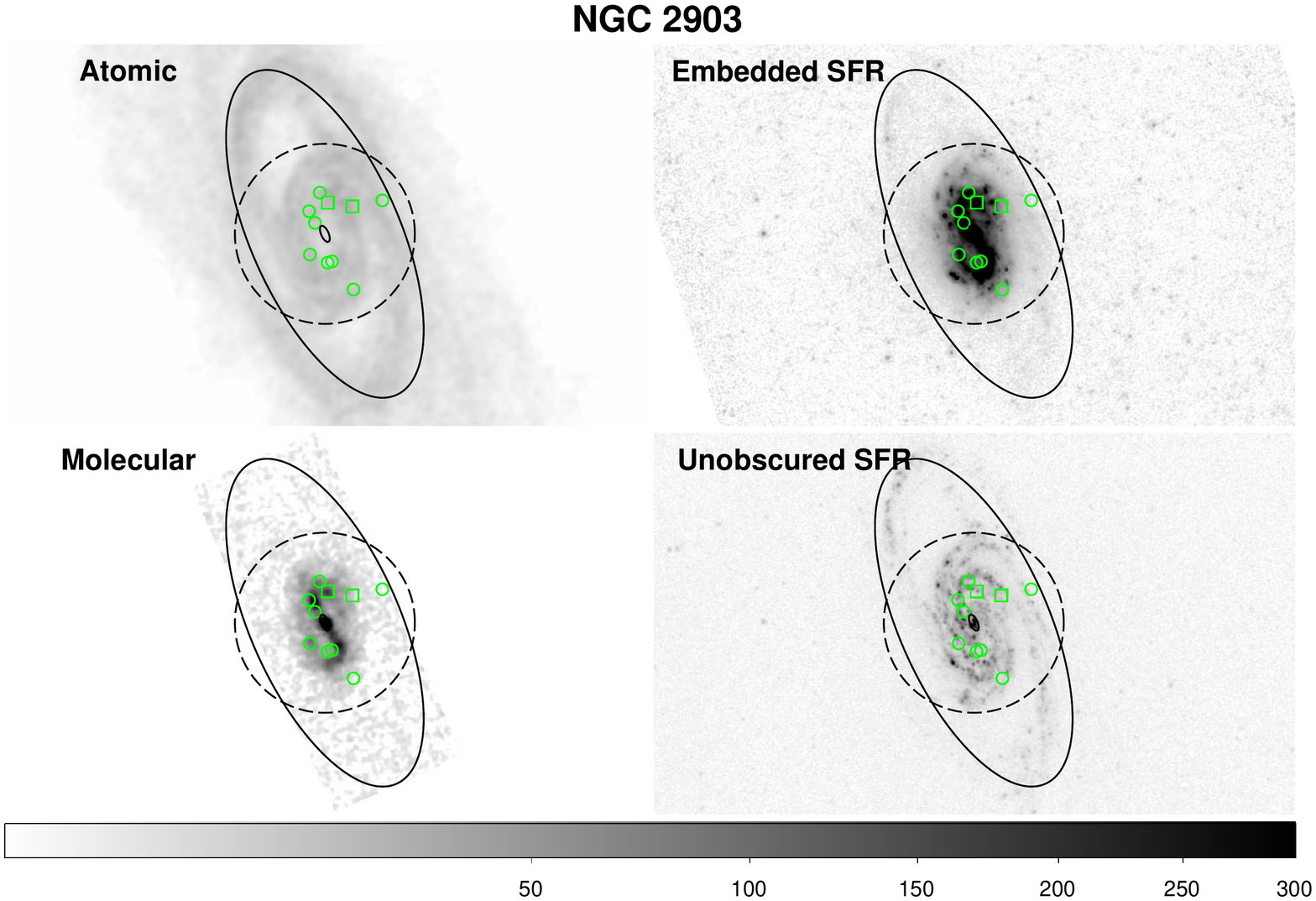}
\caption{
}
\label{fig:NGC2903_maps}
\end{figure*}

\begin{figure*}
\centering
\includegraphics[width=0.75\textwidth]{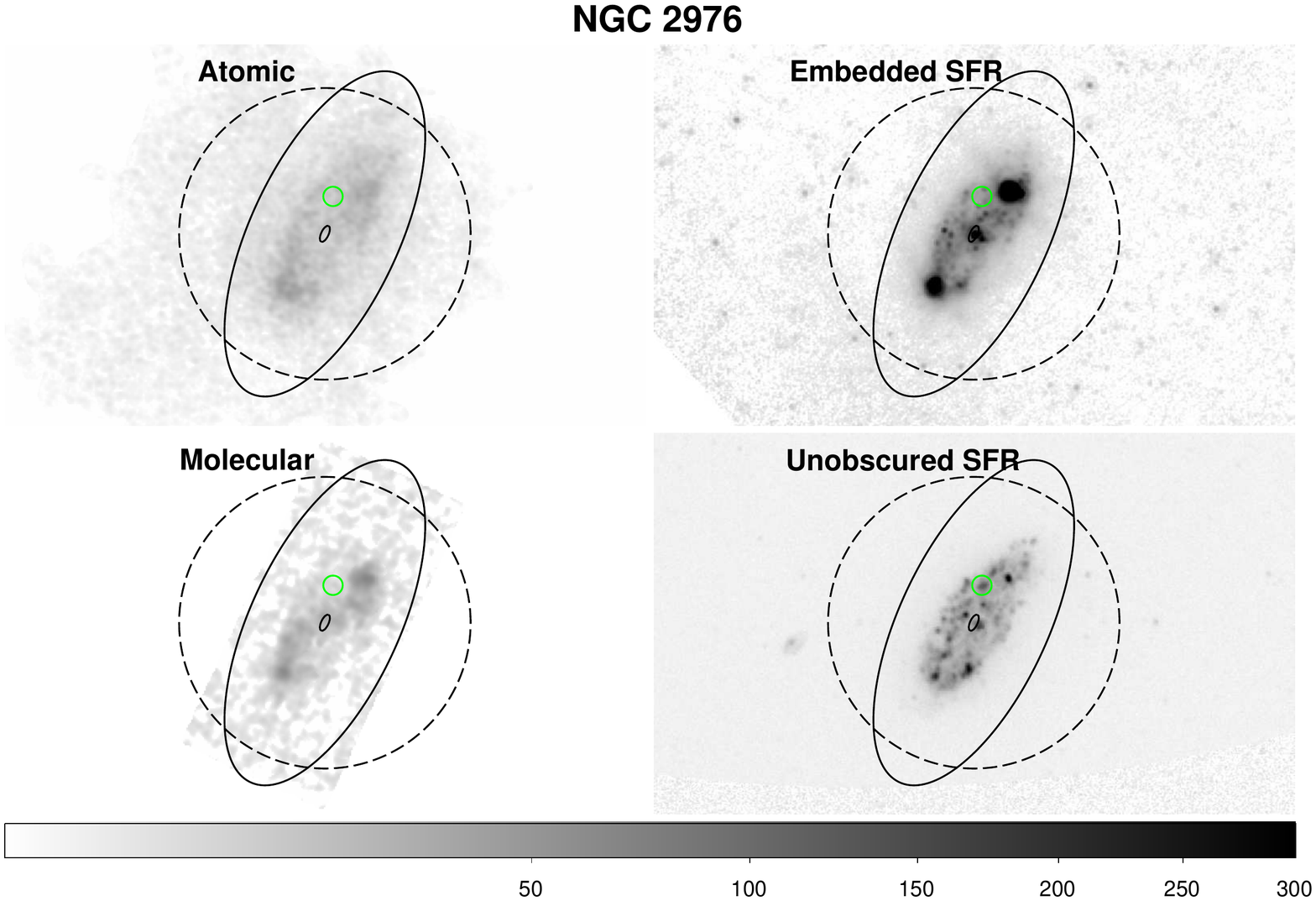}
\caption{
}
\label{fig:NGC2976_maps}
\end{figure*}

\begin{figure*}
\centering
\includegraphics[width=0.75\textwidth]{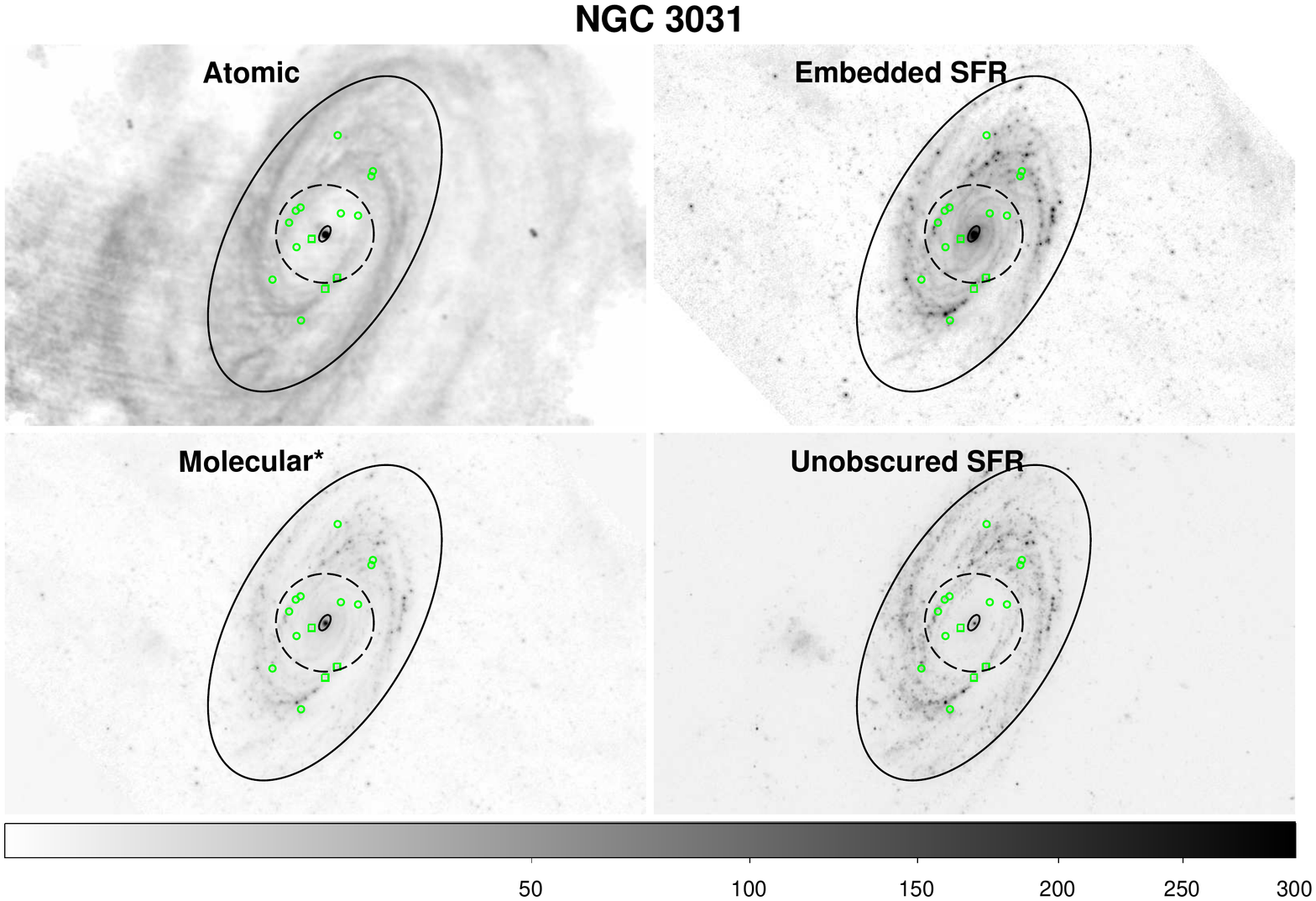}
\caption{
}
\label{fig:NGC3031_maps}
\end{figure*}

\begin{figure*}
\centering
\includegraphics[width=0.75\textwidth]{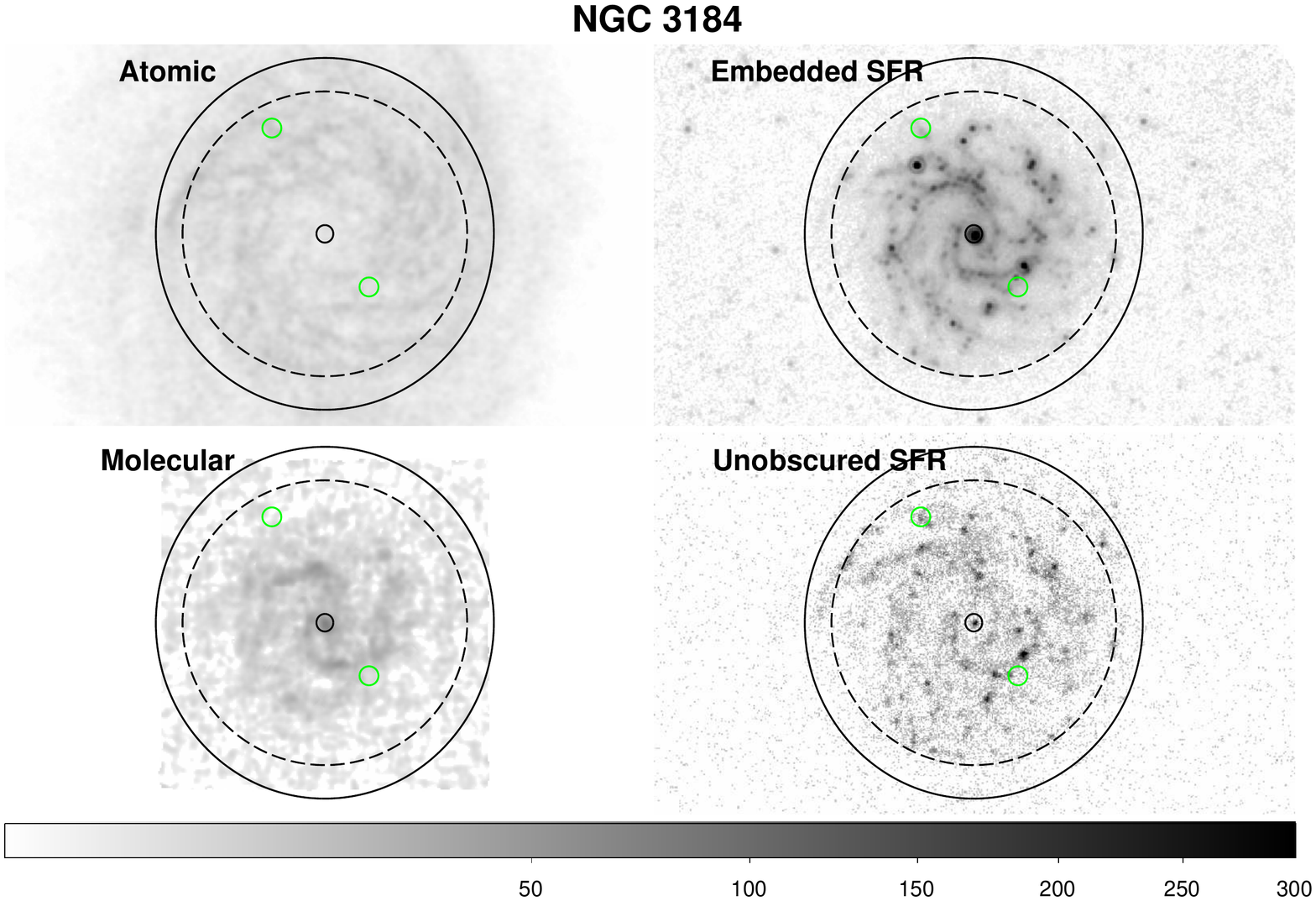}
\caption{
}
\label{fig:NGC3184_maps}
\end{figure*}

\begin{figure*}
\centering
\includegraphics[width=0.75\textwidth]{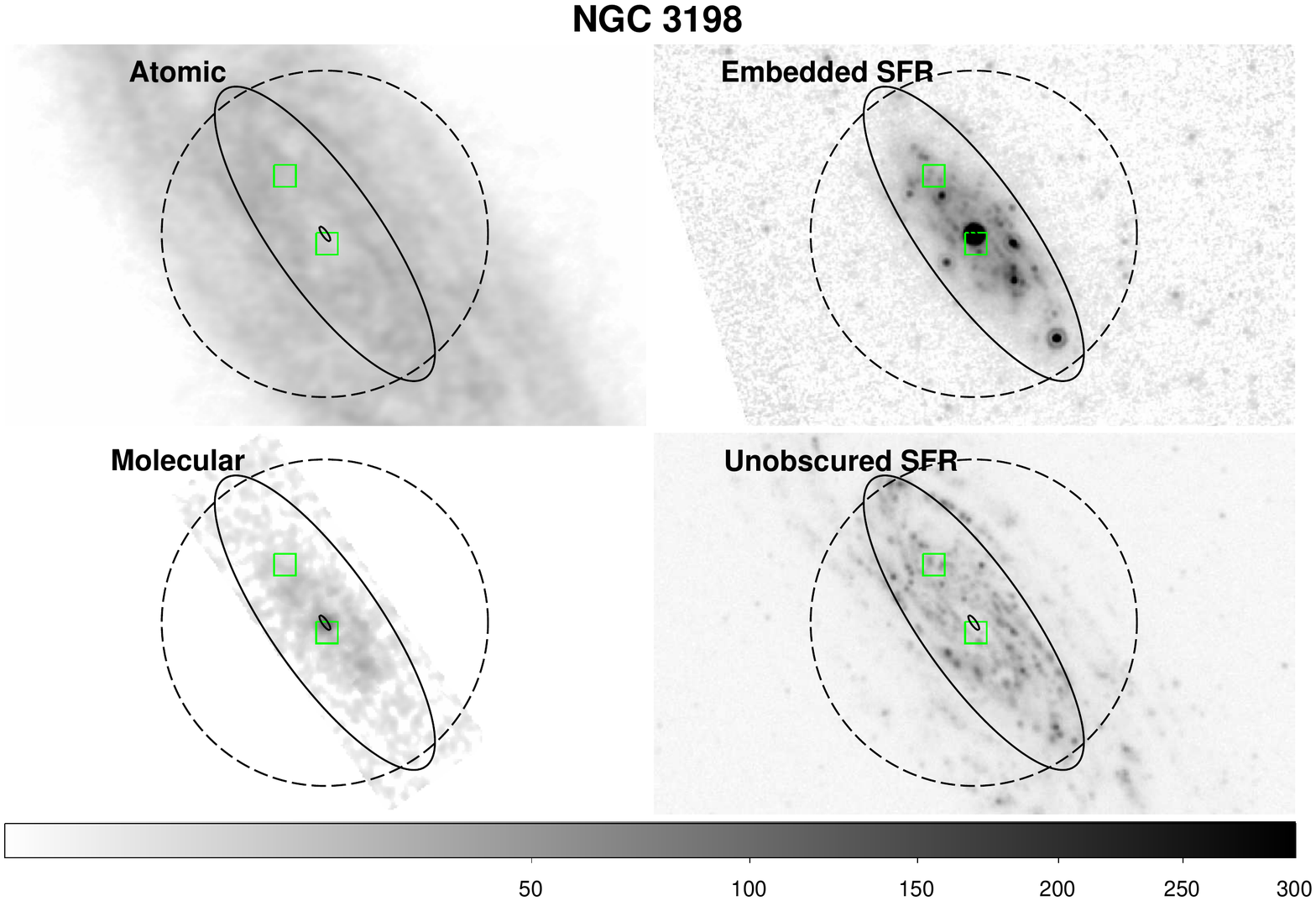}
\caption{
}
\label{fig:NGC3198_maps}
\end{figure*}

\begin{figure*}
\centering
\includegraphics[width=0.75\textwidth]{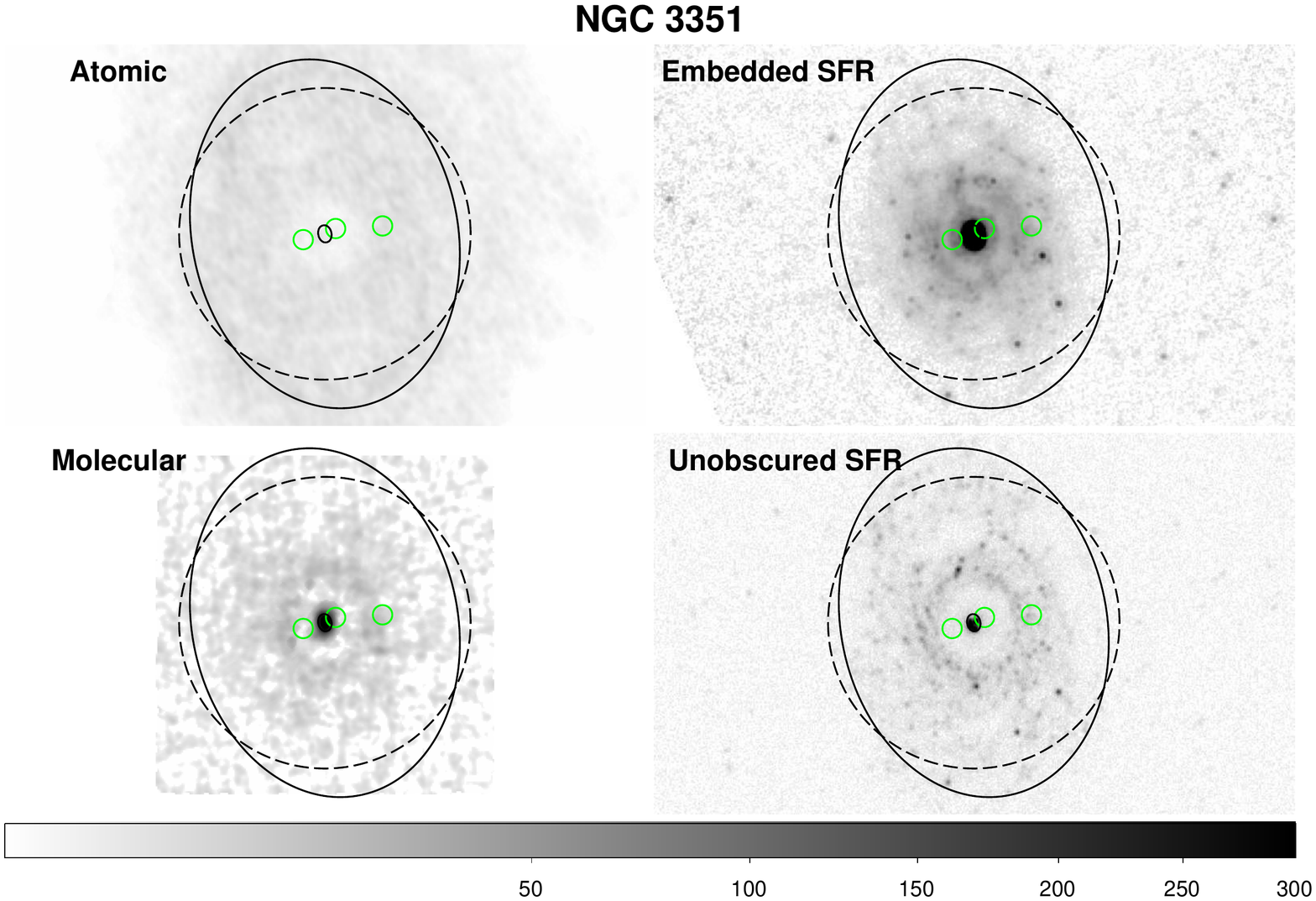}
\caption{
}
\label{fig:NGC3351_maps}
\end{figure*}

\begin{figure*}
\centering
\includegraphics[width=0.75\textwidth]{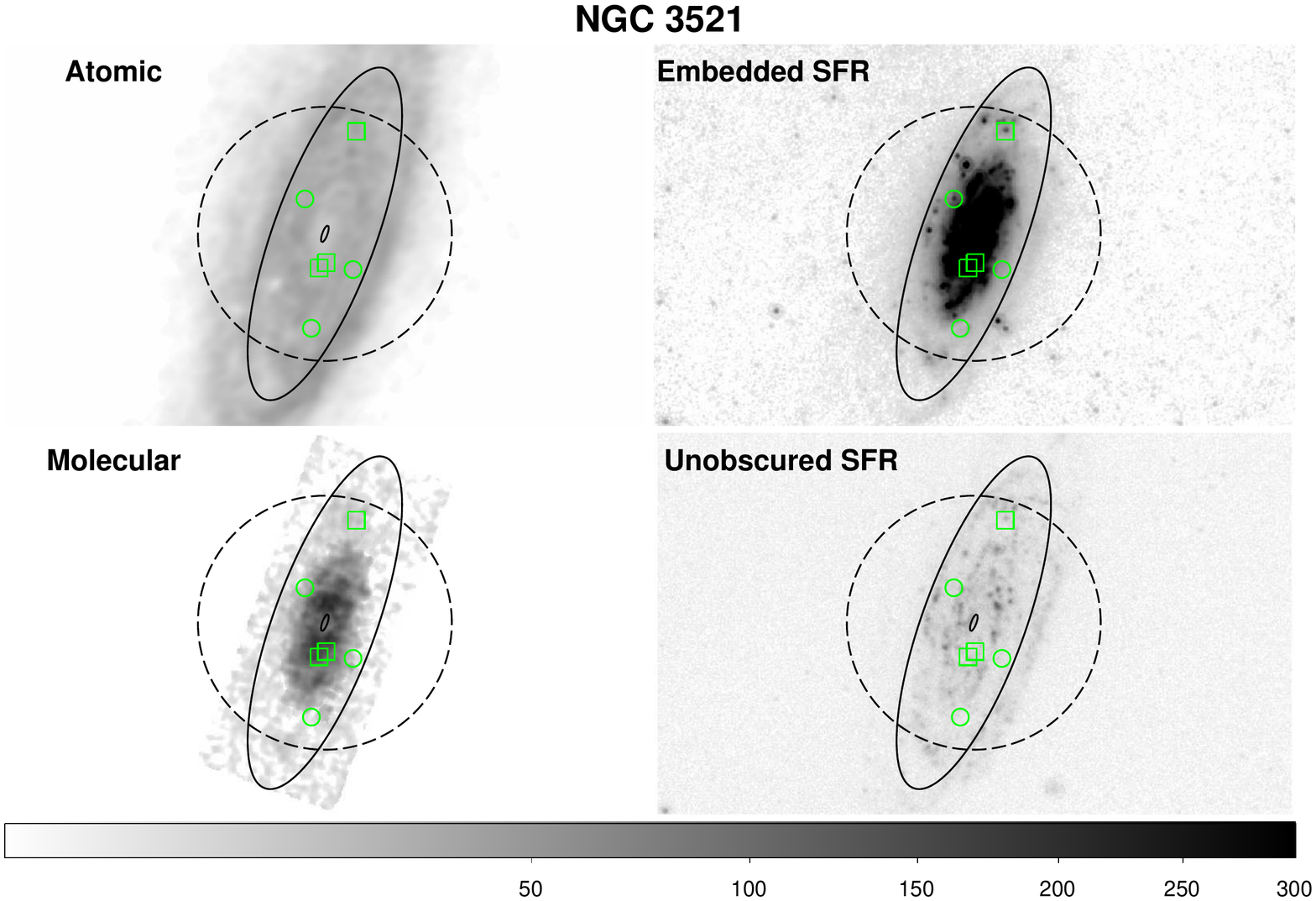}
\caption{
}
\label{fig:NGC3521_maps}
\end{figure*}

\begin{figure*}
\centering
\includegraphics[width=0.75\textwidth]{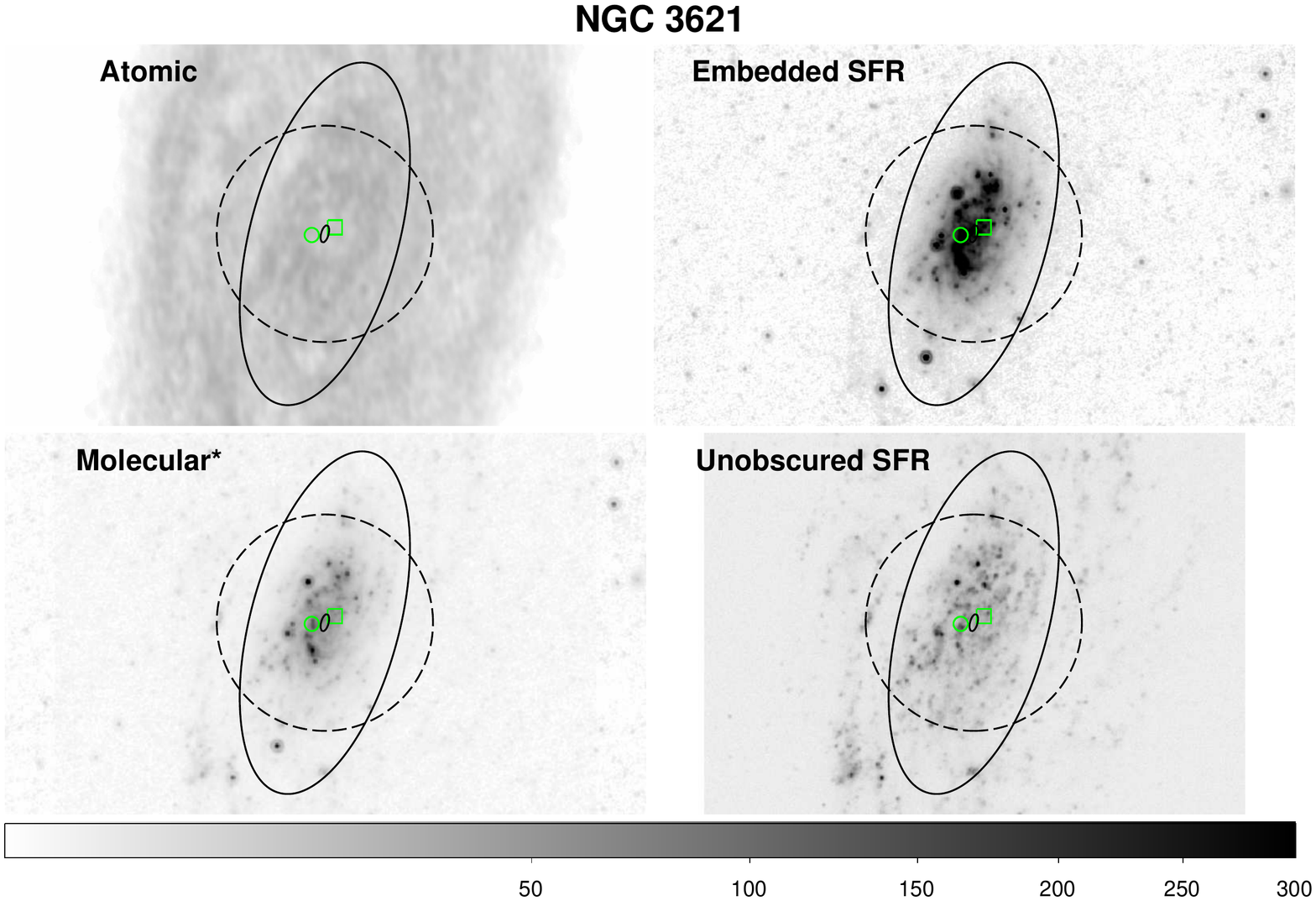}
\caption{
}
\label{fig:NGC3621_maps}
\end{figure*}

\begin{figure*}
\centering
\includegraphics[width=0.75\textwidth]{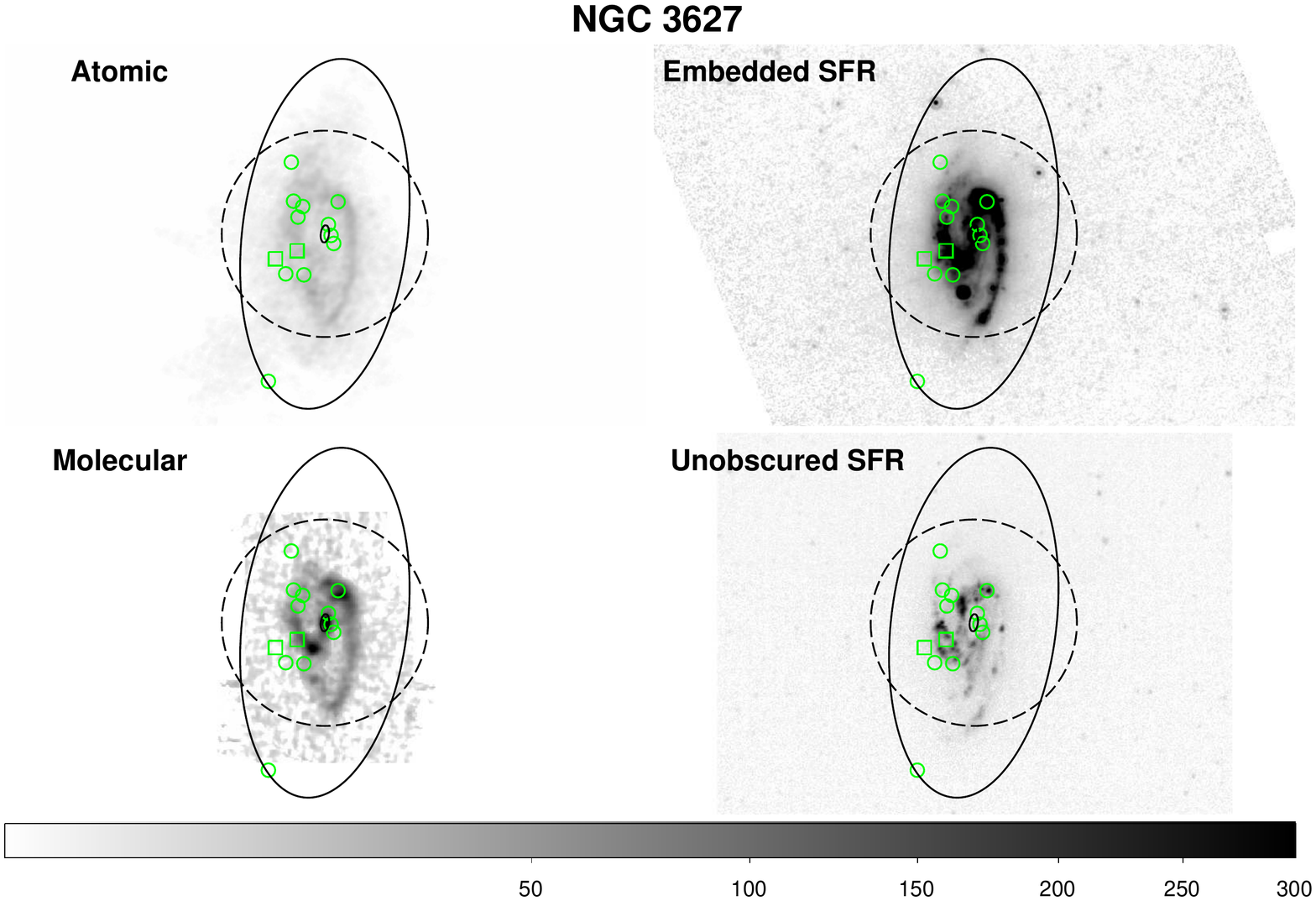}
\caption{
}
\label{fig:NGC3627_maps}
\end{figure*}

\begin{figure*}
\centering
\includegraphics[width=0.75\textwidth]{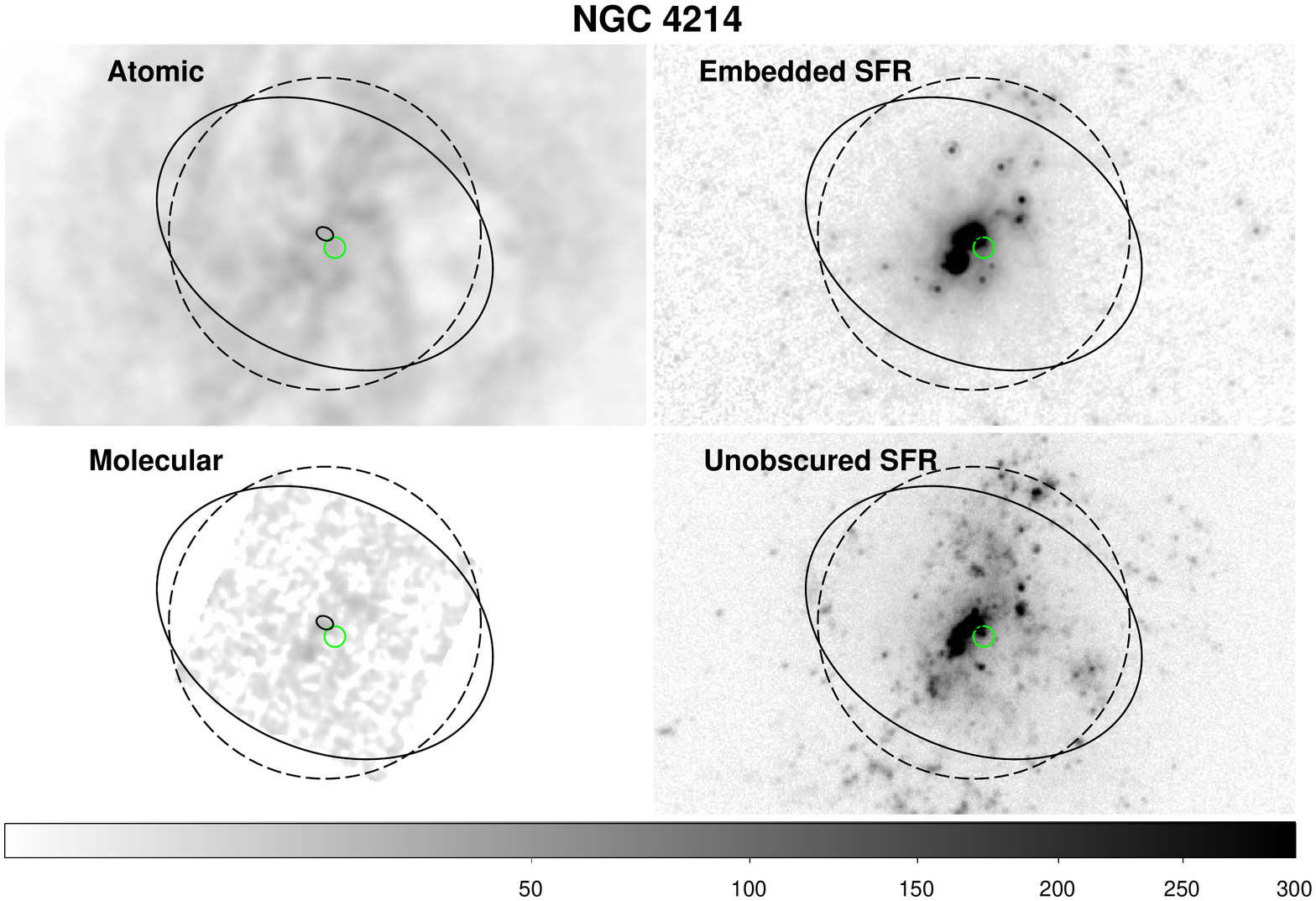}
\caption{
}
\label{fig:NGC4214_maps}
\end{figure*}

\begin{figure*}
\centering
\includegraphics[width=0.75\textwidth]{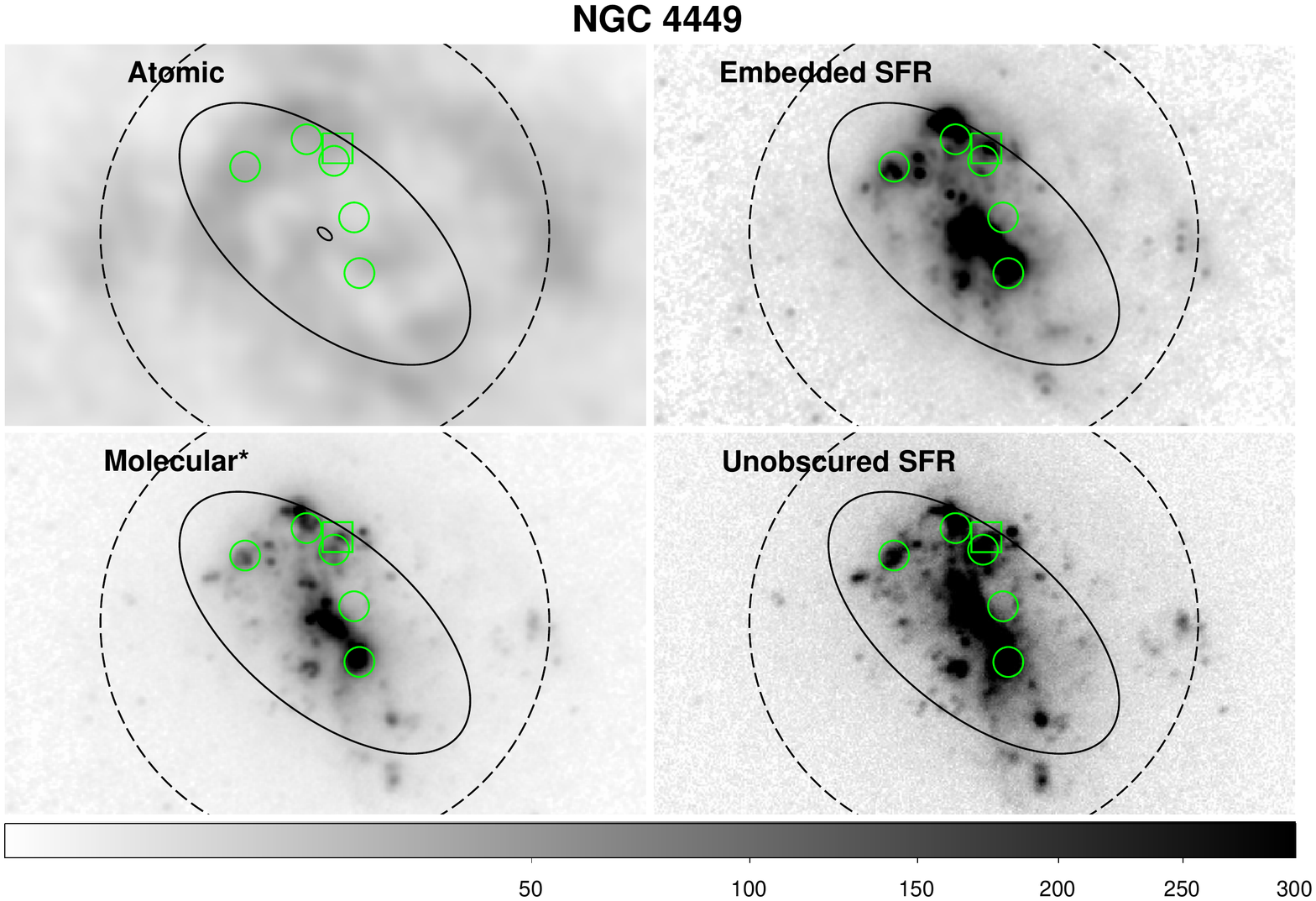}
\caption{
}
\label{fig:NGC4449_maps}
\end{figure*}

\begin{figure*}
\centering
\includegraphics[width=0.75\textwidth]{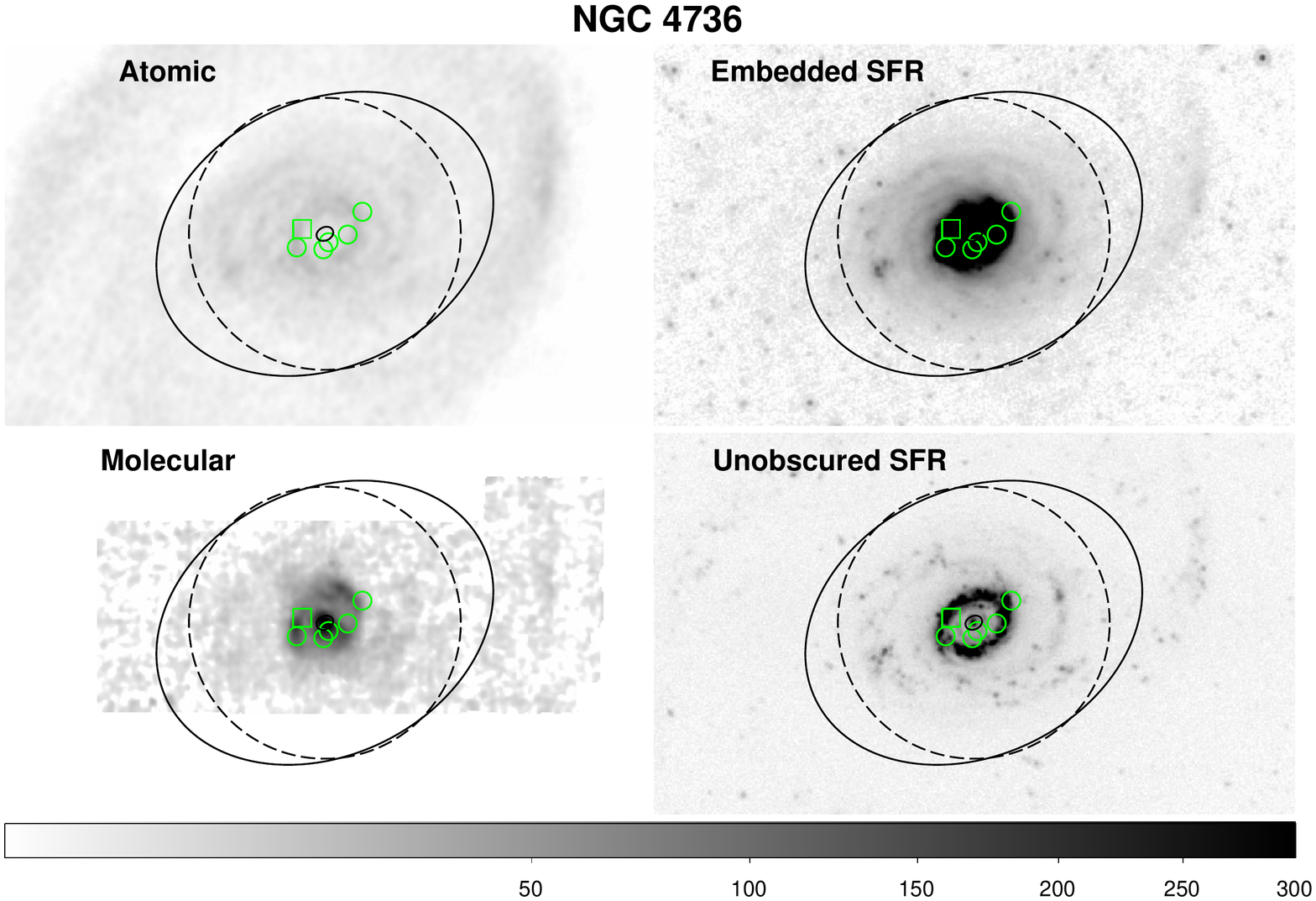}
\caption{
}
\label{fig:NGC4736_maps}
\end{figure*}

\begin{figure*}
\centering
\includegraphics[width=0.75\textwidth]{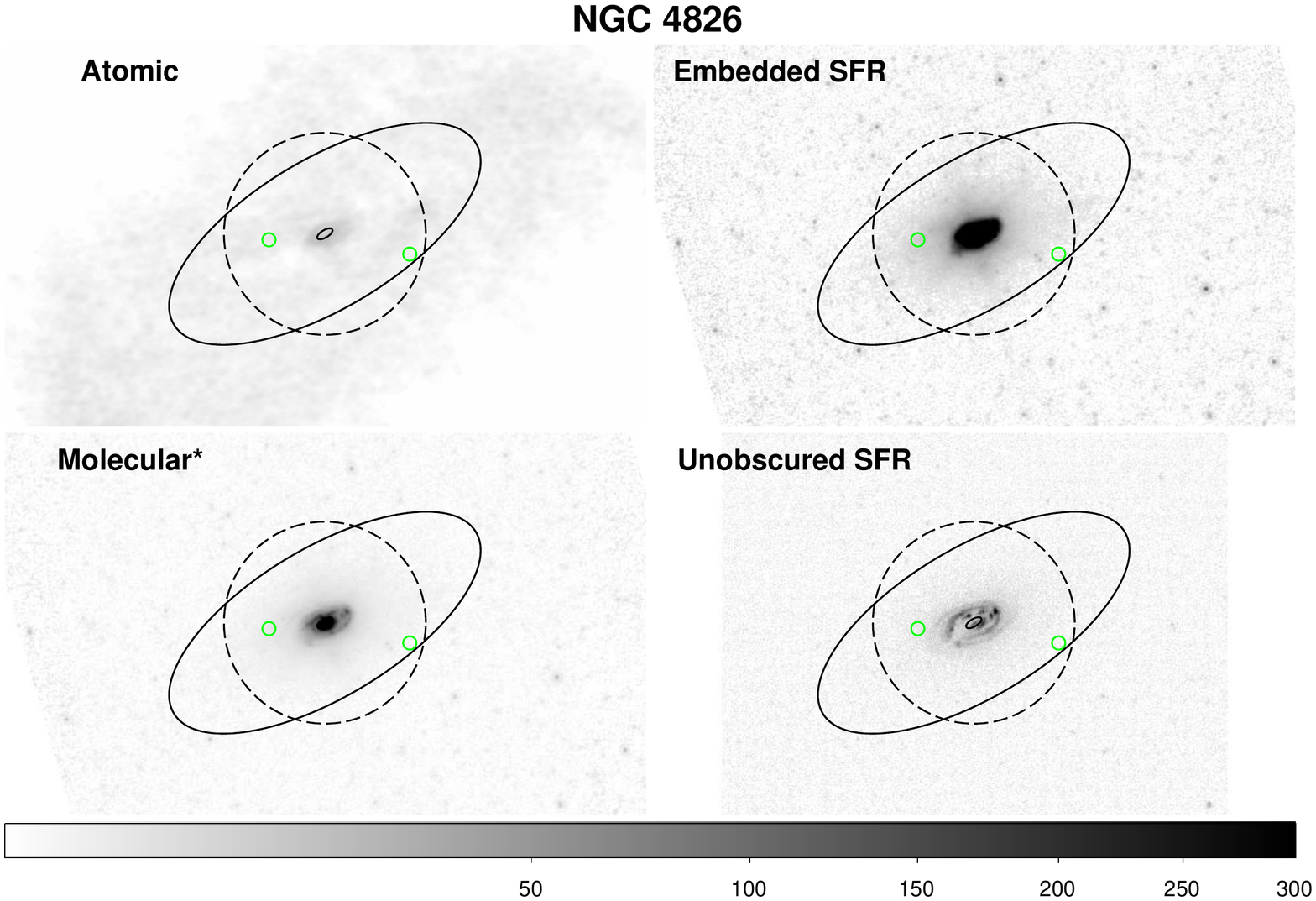}
\caption{
}
\label{fig:NGC4826_maps}
\end{figure*}

\begin{figure*}
\centering
\includegraphics[width=0.75\textwidth]{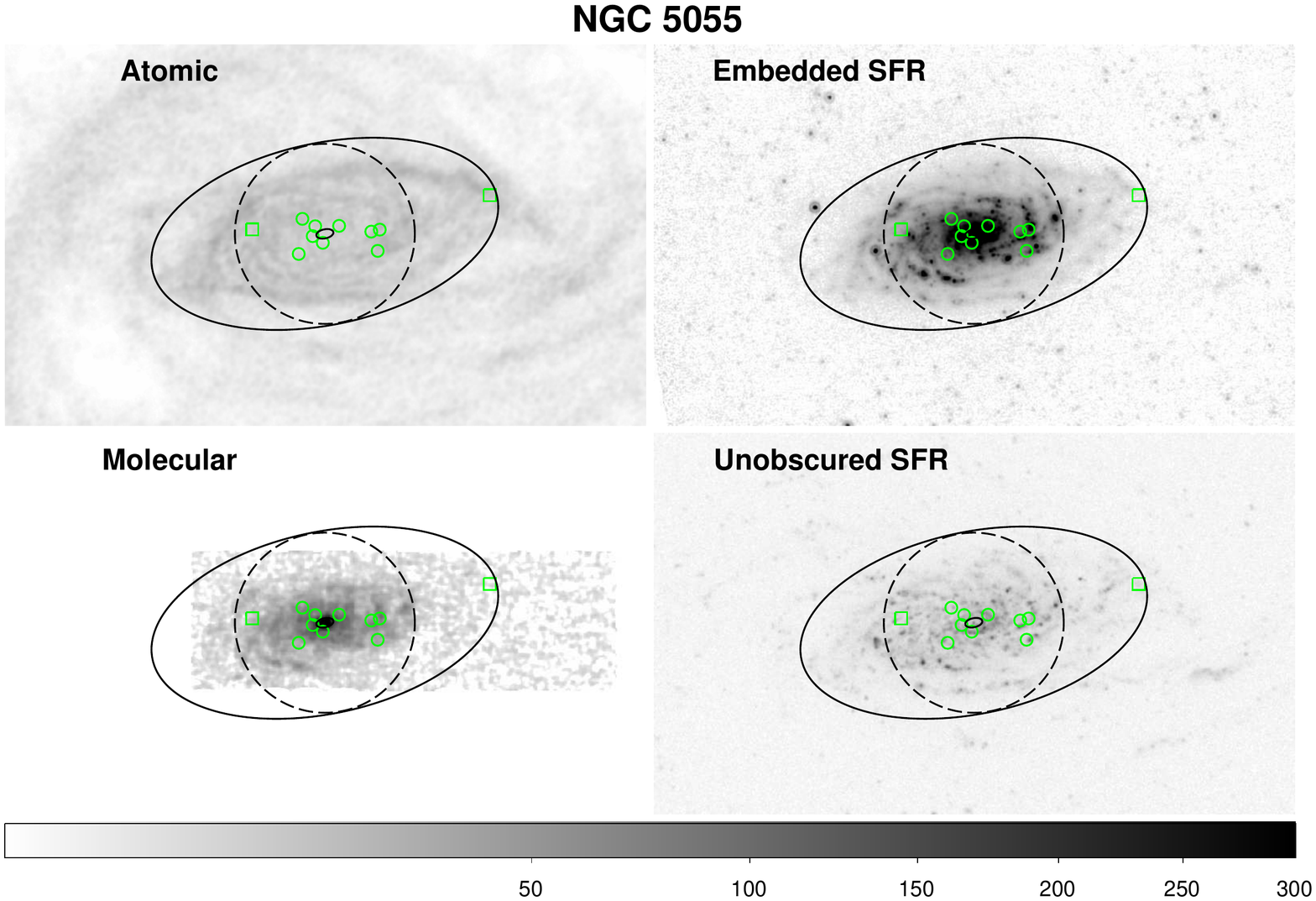}
\caption{
}
\label{fig:NGC5055_maps}
\end{figure*}

\begin{figure*}
\centering
\includegraphics[width=0.75\textwidth]{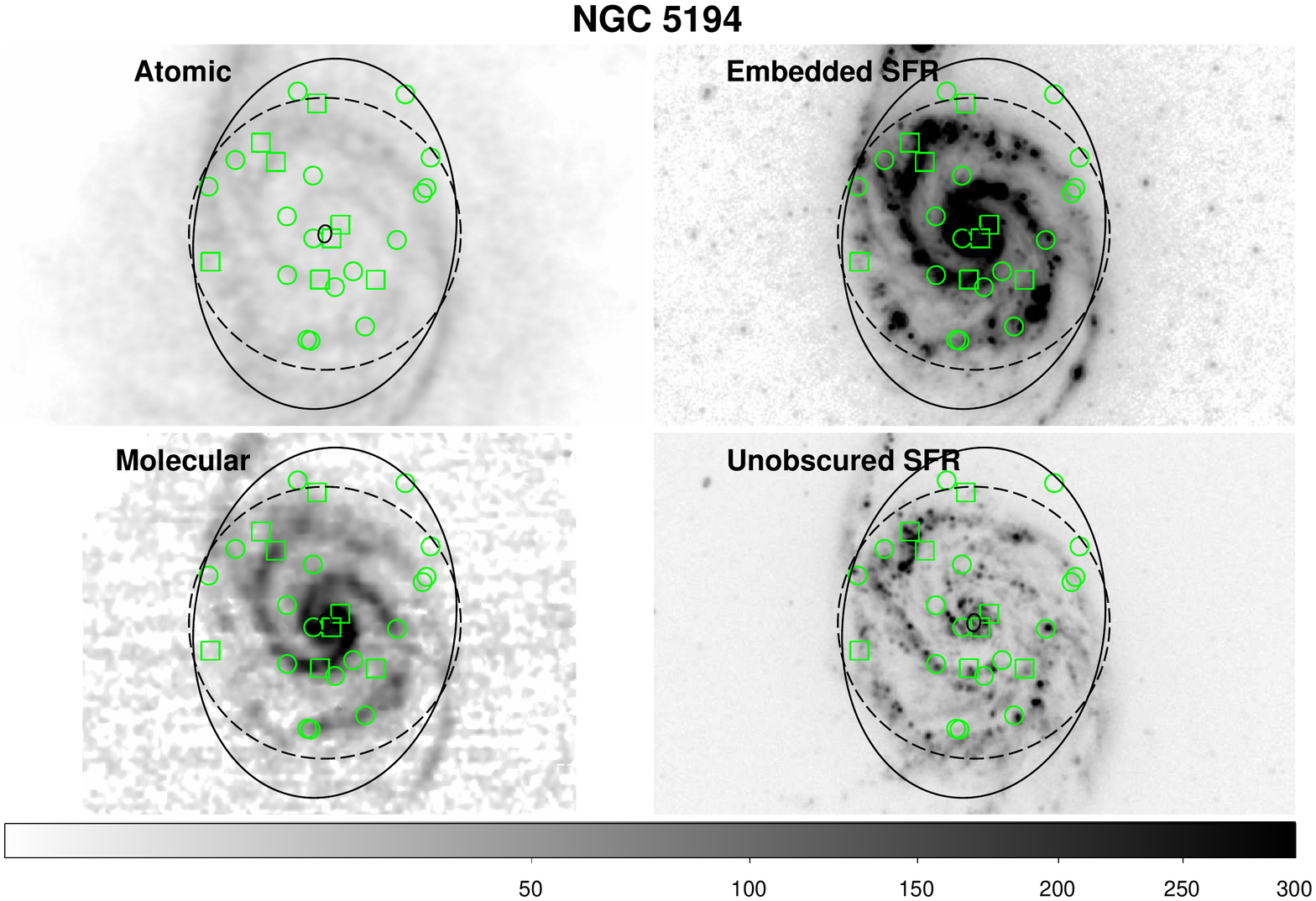}
\caption{
}
\label{fig:NGC5194_maps}
\end{figure*}

\begin{figure*}
\centering
\includegraphics[width=0.75\textwidth]{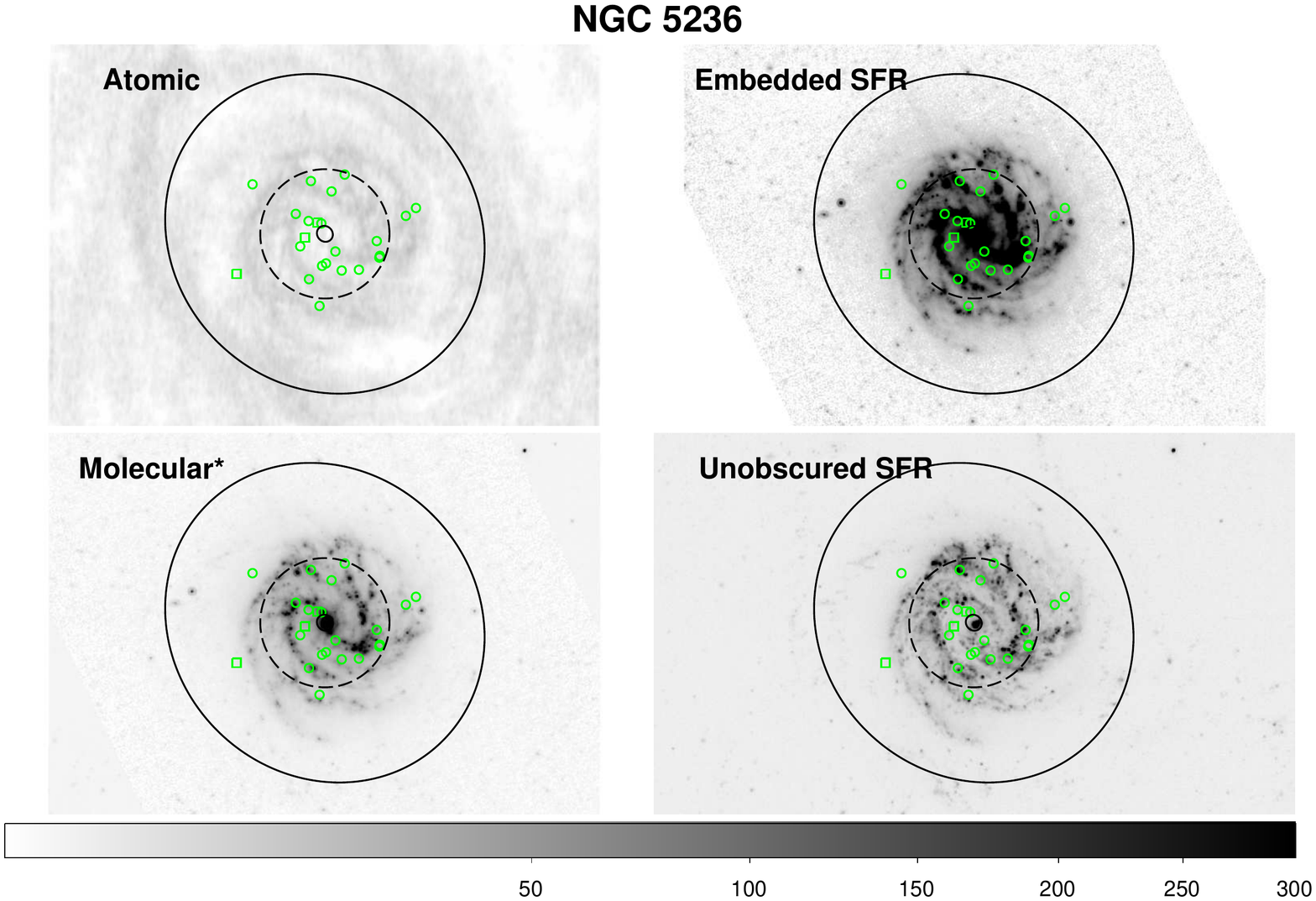}
\caption{
}
\label{fig:NGC5236_maps}
\end{figure*}

\begin{figure*}
\centering
\includegraphics[width=0.75\textwidth]{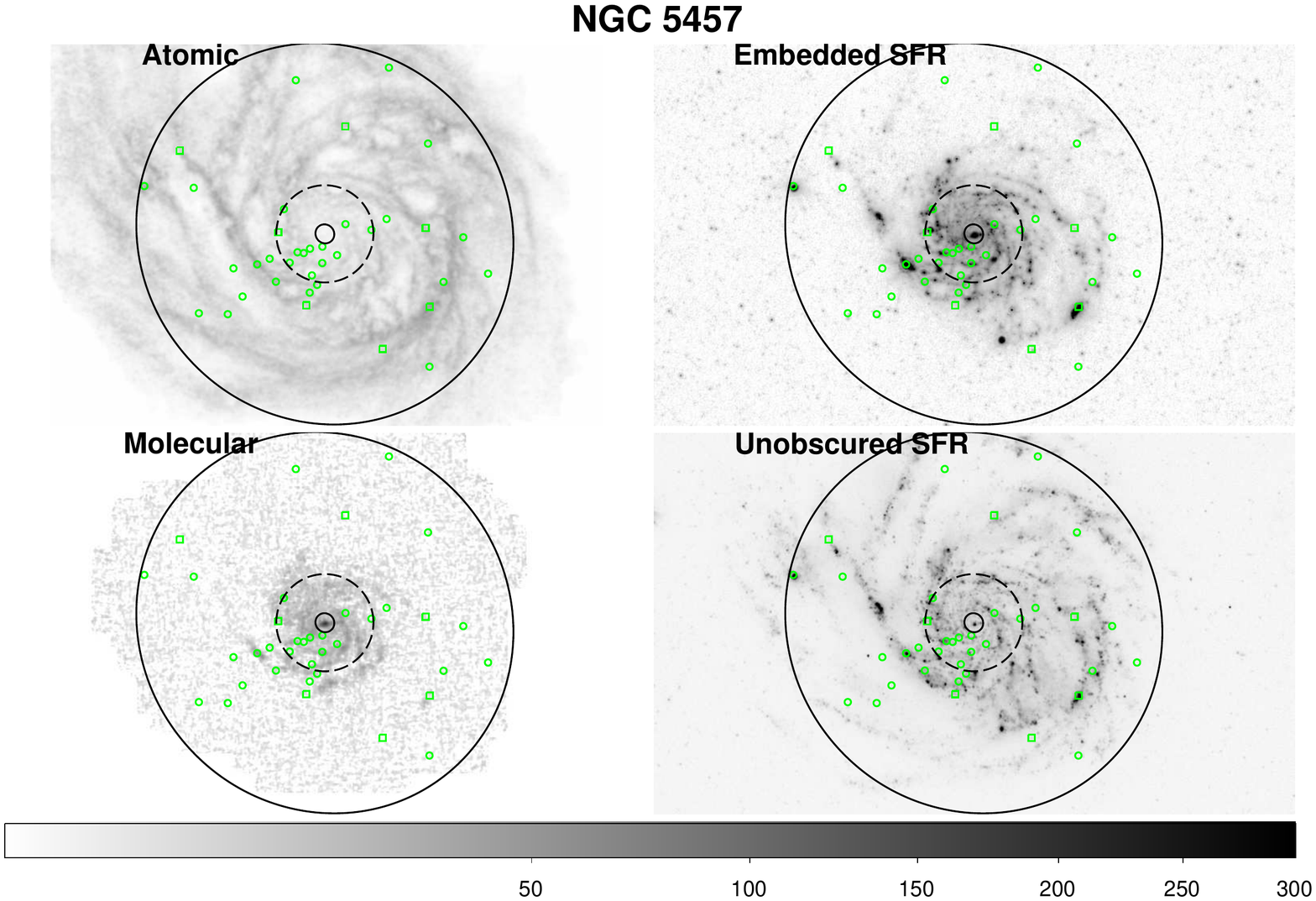}
\caption{
}
\label{fig:NGC5457_maps}
\end{figure*}

\begin{figure*}
\centering
\includegraphics[width=0.75\textwidth]{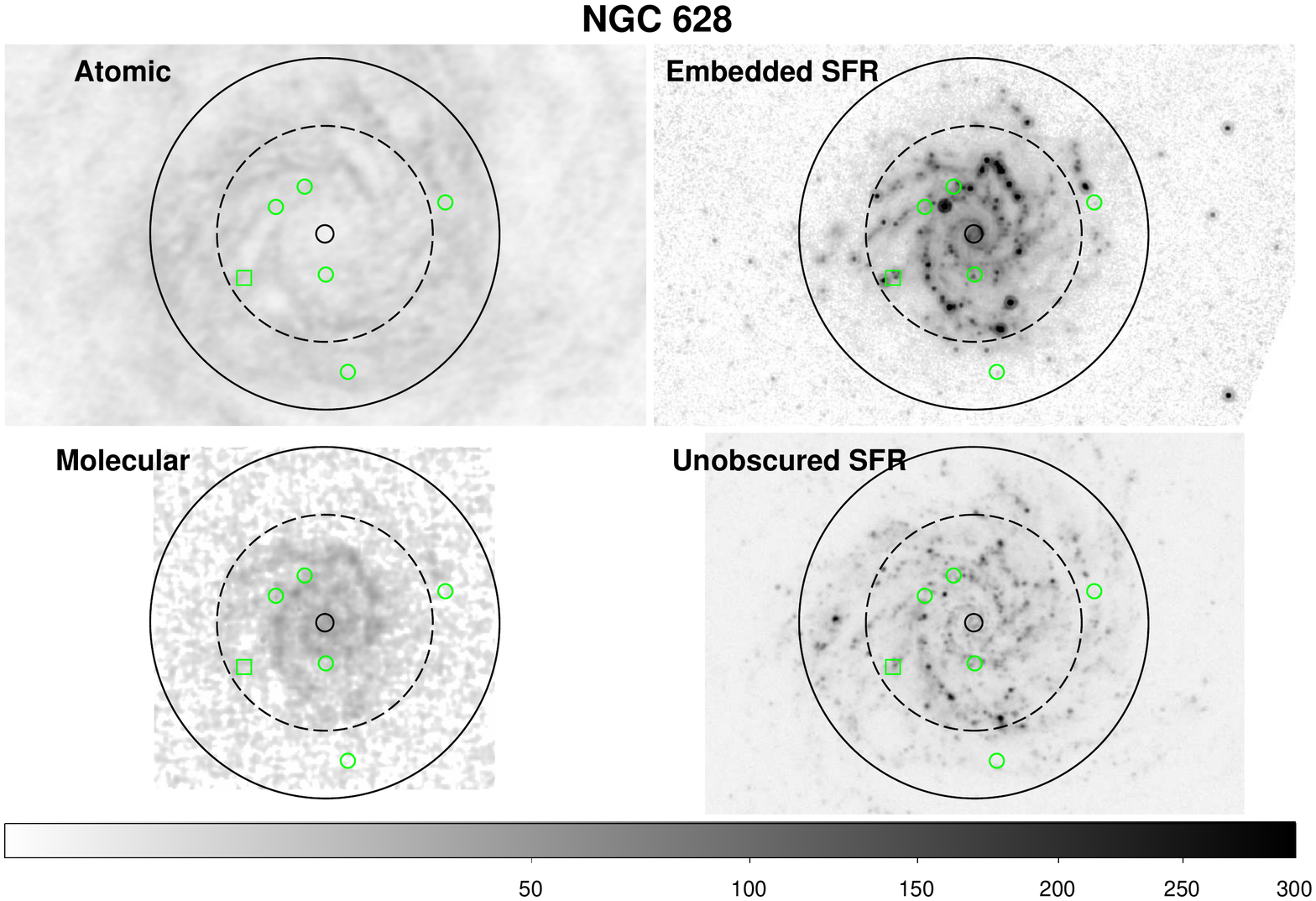}
\caption{
}
\label{fig:NGC628_maps}
\end{figure*}

\begin{figure*}
\centering
\includegraphics[width=0.75\textwidth]{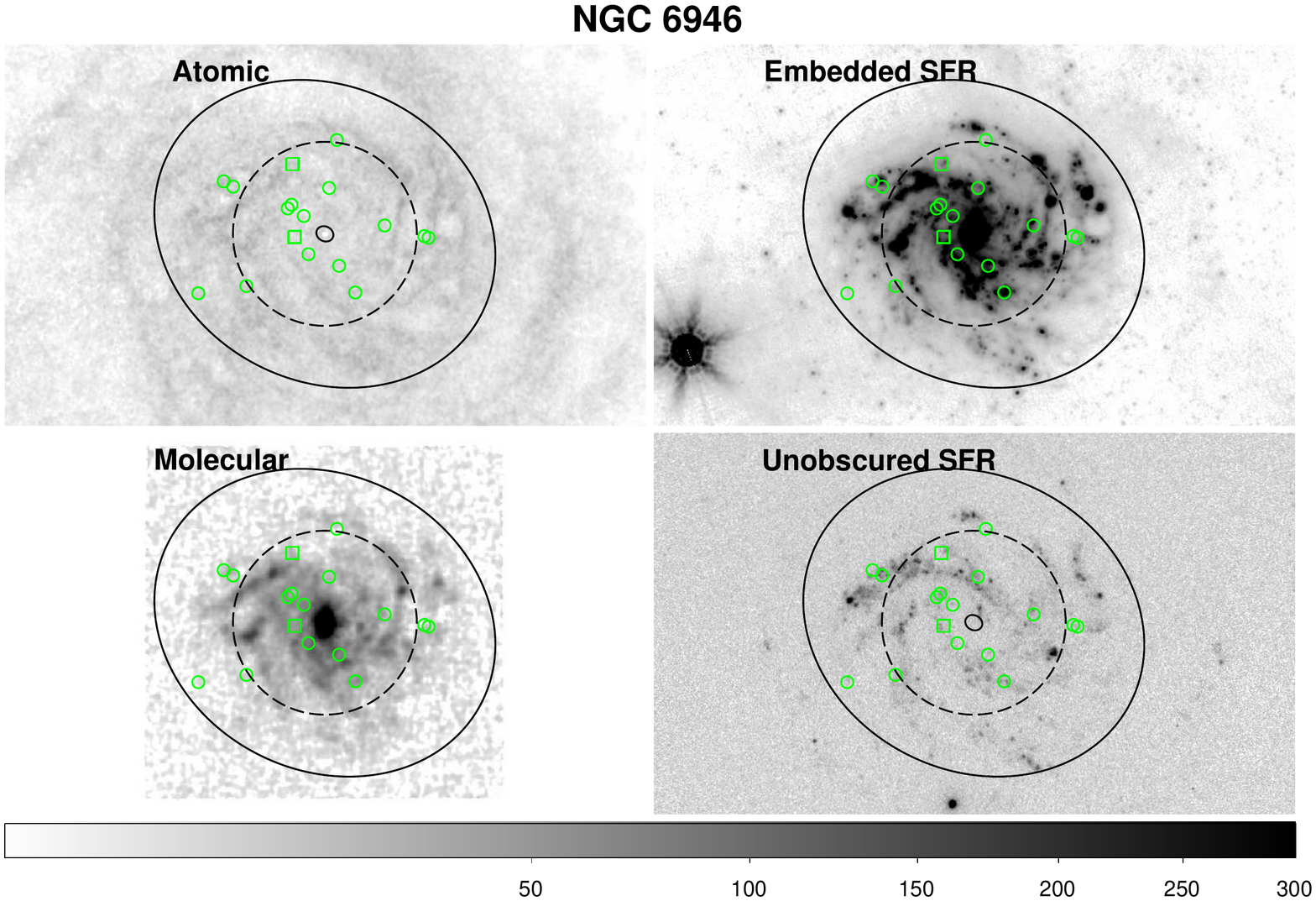}
\caption{
}
\label{fig:NGC6946_maps}
\end{figure*}

\begin{figure*}
\centering
\includegraphics[width=0.75\textwidth]{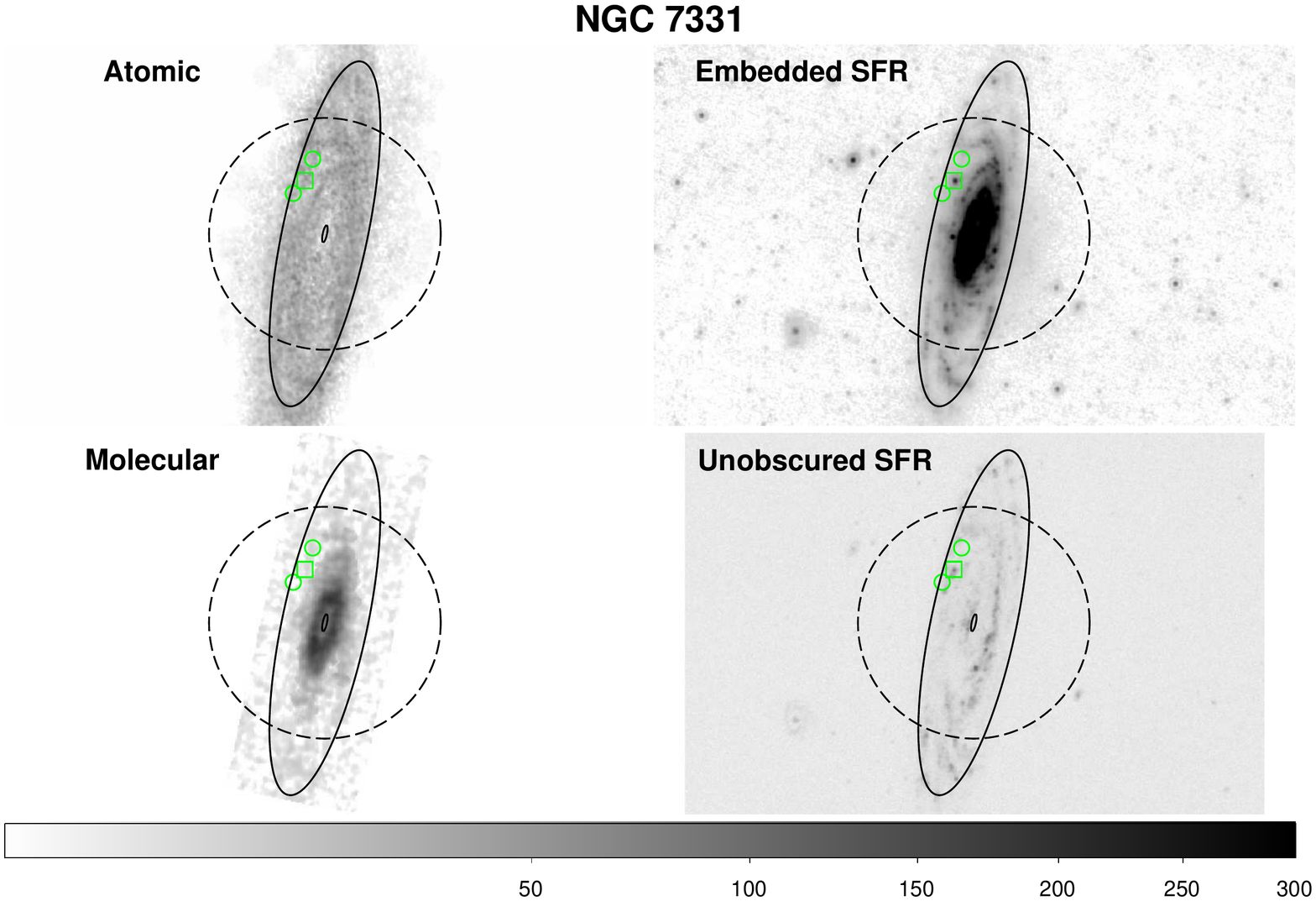}
\caption{
}
\label{fig:NGC7331_maps}
\end{figure*}

\begin{figure*}
\centering
\includegraphics[width=0.75\textwidth]{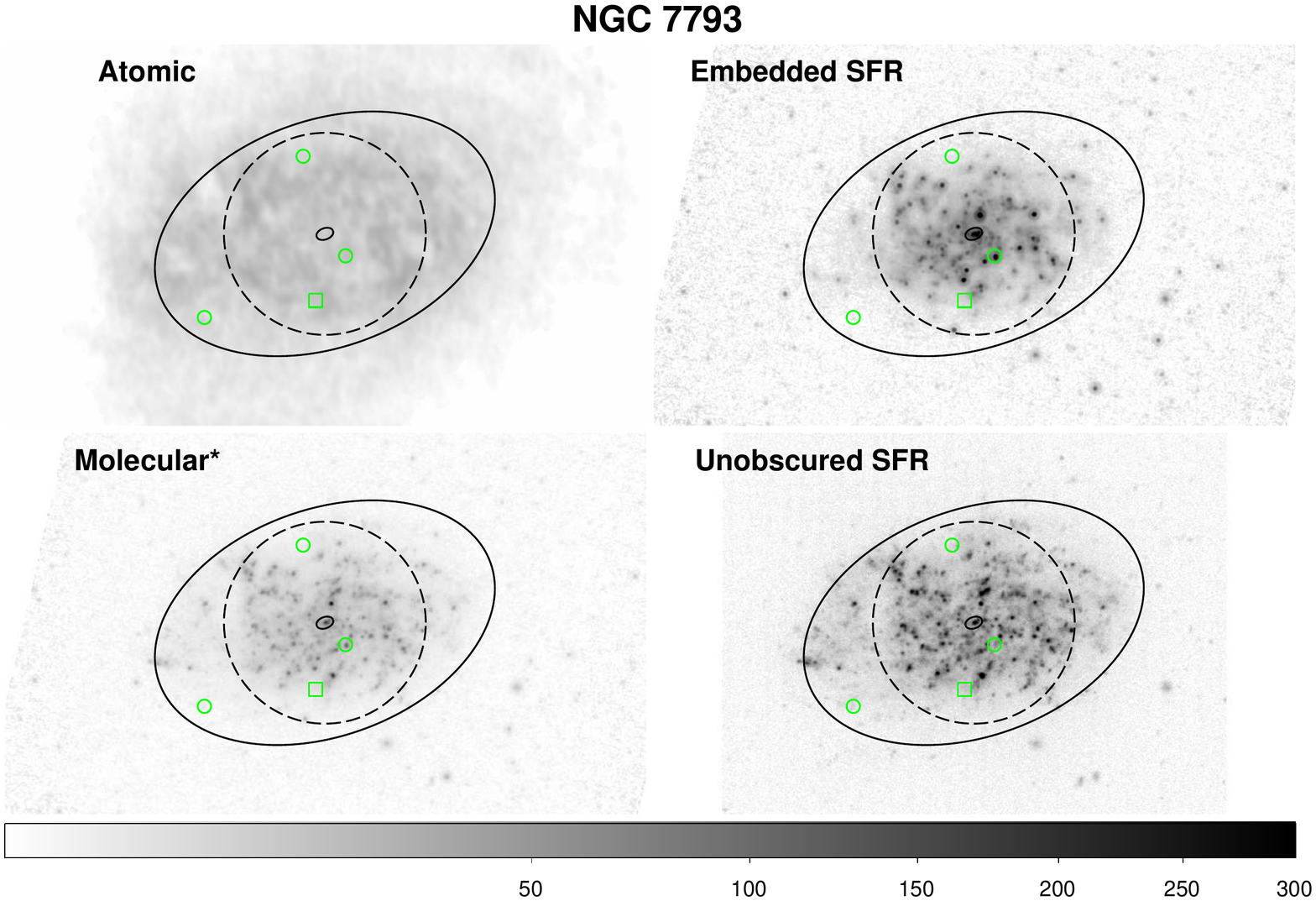}
\caption{
}
\label{fig:NGC7793_maps}
\end{figure*}

\begin{figure*}
\centering
\includegraphics[width=0.75\textwidth]{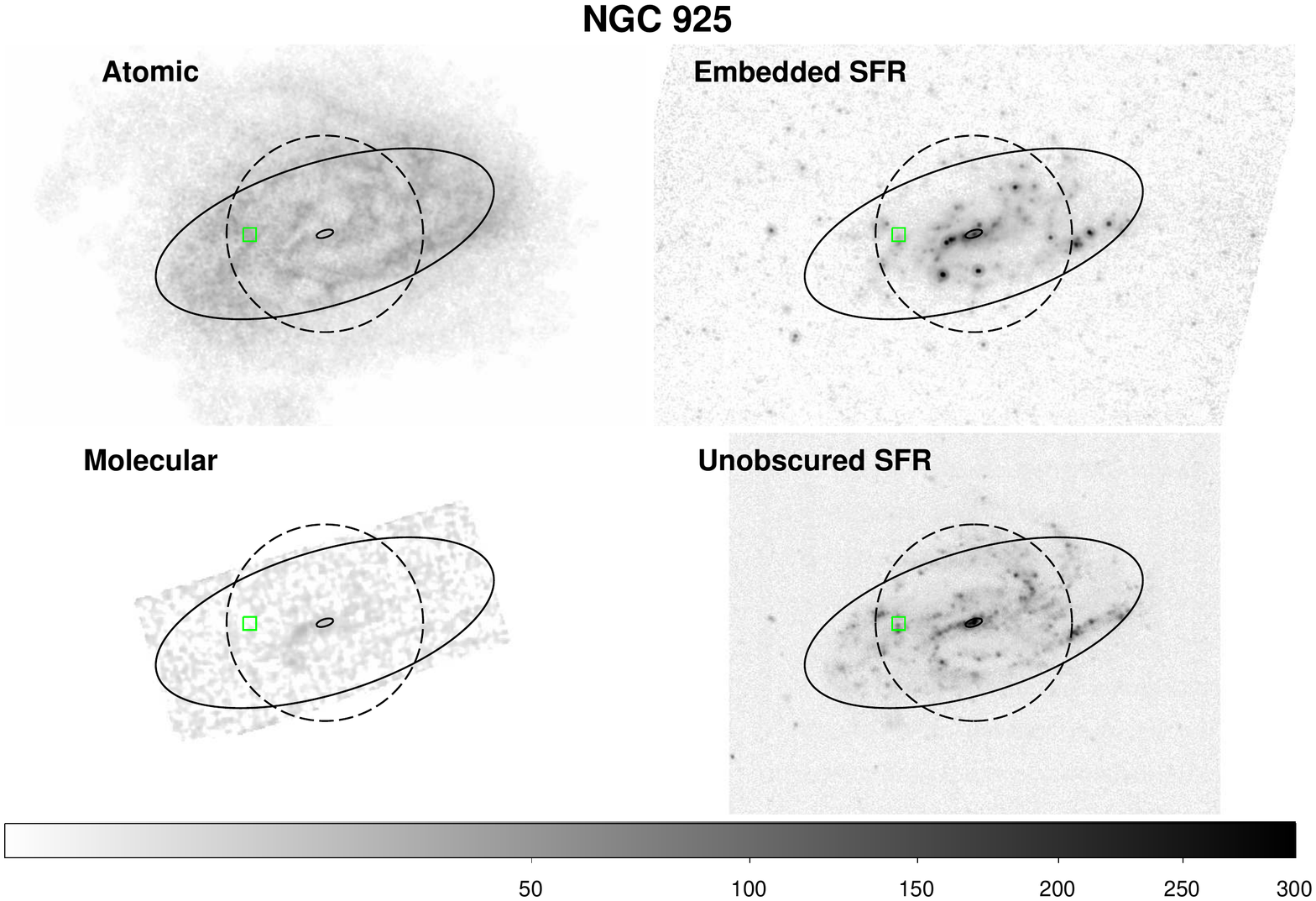}
\caption{
}
\label{fig:NGC925_maps}
\end{figure*}

Figures~\ref{fig:HOII_maps}--\ref{fig:NGC925_maps} show projected maps
of atomic and molecular gas surface densities (in units of
$10^{20}$~H~cm$^{-2}$) and of unobscured (showing in direct UV
starlight) and dust-embedded (showing in infrared re-emission) SFR (in
units of $10^{-4}$~$M_{\odot}$~kpc$^{-2}$~yr$^{-1}$) for the 27
studied galaxies.  

The maps were derived from \textit{THINGS}, \textit{HERACLES},
\textit{Spitzer}/MIPS 24~$\mu$m and \textit{GALEX} FUV intensity maps,
with the latter corrected for Galactic extinction (see further details
in Section~\ref{s:galaxy_profiles}). If no \textit{HERACLES} data are
available, synthetic H$_{2}$ maps (labeled 'Molecular$^{*}$' and
calculated from the aforementioned obscured and unobscured SFR maps
using the relation between $\SSFR$ and $\SHM$, eq.~(\ref{eq:sfr_hm}))
are shown instead. The small and large ellipses denote the
boundaries of the $0.05R_{25}$--$R_{25}$ regions used in this 
study. The dashed circle on each map has a radius of 3 arcmin. The
green circles and squares mark the positions of X-ray sources in the
clean sample with $10^{38}<\Lunabs<10^{39}$ and
$\Lunabs>10^{39}$~erg~s$^{-1}$, respectively.

\clearpage

\clearpage

\clearpage

\clearpage

\clearpage

\clearpage

\clearpage

\clearpage

\clearpage

\clearpage

\clearpage

\clearpage

\clearpage

\clearpage

\clearpage

\end{appendix}


\bsp	
\label{lastpage}
\end{document}